\documentclass[preprint2]{aastex}

\def\mm{M\'endez }

\def\lae{\lower 2pt \hbox{$\, \buildrel {\scriptstyle <}\over {\scriptstyle \sim}\,$}} 

\begin{document}

\title{The atoll source states of 4U 1608--52} 

\author{Steve van Straaten\altaffilmark{1}, Michiel van der Klis\altaffilmark{1}, Mariano M\'endez\altaffilmark{2}}

\email{straaten@astro.uva.nl}
\altaffiltext{1}{Astronomical Institute, ``Anton Pannekoek'',
University of Amsterdam, and Center for High Energy Astrophysics, 
Kruislaan 403, 1098 SJ Amsterdam, The Netherlands.}

\altaffiltext{2}{SRON, National Institute for Space Research, Sorbonnelaan 2, 
3584 CA Utrecht, The Netherlands.}

\begin{abstract}

We have studied the atoll source 4U 1608--52 
using a large data set obtained with the Rossi X-ray Timing Explorer. 
We find that the timing properties of 4U 1608--52 are almost exactly identical to those 
of the atoll sources 4U 0614+09 and 4U 1728--34 despite the fact that contrary to 
these sources 4U 1608--52 is a transient covering two orders of magnitude in luminosity. 
The frequencies of the variability components of these three sources 
follow a universal scheme when plotted versus the frequency of the upper kilohertz QPO, 
suggesting a very similar accretion flow configuration. If we plot the Z sources on this 
scheme only the lower kilohertz QPO and HBO follow identical relations.
Using the mutual relations between the frequencies of the variability components 
we tested several models; the transition layer model, 
the sonic point beat frequency model, and the relativistic precession model. None of 
these models described the data satisfactory. Recently, it has been 
suggested that the atoll sources (among them 4U 1608--52) trace out similar 
three--branch patterns as the Z sources in the color--color diagram. 
We have studied the relation between the power spectral properties and the 
position of 4U 1608--52 in the color--color diagram and conclude that the 
timing behavior is not consistent with the idea that 4U 1608-52 traces out 
a three-branched Z shape in the color-color diagram along which the timing 
properties vary gradually, as Z sources do.

\end{abstract}


\section{INTRODUCTION}

Most of the neutron star low--mass X--ray binaries can be divided into two classes, Z 
and atoll sources, based on the correlated behavior of their timing properties at low 
frequencies ($\nu < 200$ Hz) and their X--ray spectral properties \citep{hk89}. 
Both classes show quasi-periodic 
oscillations (QPOs) with frequencies ranging from a few hundred Hz to more than 1000 Hz 
(kilohertz QPOs). The low--frequency part of the power spectra of both classes is usually 
dominated by a similar broad band--limited noise component, but it is unclear if these 
are physically the same component \citep{wk99}. In addition to the band--limited noise both 
classes show several quasi-periodic oscillations below 200 Hz. In the Z sources 
these are named after the branch of the Z track in the color--color diagram (see below) 
where they mostly occur: horizontal (HBOs), normal (NBOs) and flaring branch 
oscillations (FBOs). The HBO shows a sub-- and a second harmonic at $\sim 0.5$ and $\sim 2$ 
times the frequency of the main peak, and sometimes a peak at $\sim 1.5$ times the frequency 
of the HBO \citep{jonker02}. Obviously, the most straightforward interpretation of this is that 
the sub HBO is the fundamental frequency, with second, third and fourth harmonics all 
observed \citep{jonker02}. Below 200 Hz the atoll sources show several Lorentzian components 
\citep[see e.g. ][]{vstr02}. Note that these components are called Lorentzian and not QPO; this is 
because these features are sometimes too broad (FWHM $>$ centroid frequency/2) to be classified as 
a QPO. In the atoll 
sources all components become broader as their characteristic frequency decreases 
\citep*{pbk99,vstr02}. It has been suggested that the HBO in the Z and the low--frequency 
Lorentzian in the atoll sources are the same physical components \citep{ford98,pbk99,wk99}.

The energy spectrum of neutron star low--mass X--ray binaries can be usefully 
parametrized through the use of color--color or color--intensity diagrams, where a color 
is the ratio of counts in two different energy bands. As the energy spectrum of a source 
changes, it moves through these diagrams. The timing properties of both the Z and the 
atoll sources depend on the position of the source in the color--color diagram. The Z 
sources move fast through the color--color diagram and draw up a Z track within hours to 
days \citep[see e.g. GX 340+0;][]{jonker00} whereas the typical atoll sources move 
slowly through the color--color diagram and draw up a C--shaped track within weeks to 
months \citep[see e.g. 4U 1728--34;][]{disalvo01},  
although the ``banana'' part of the track, a curved branch to the bottom and right hand sides
of the diagram, is traced out as fast as in the Z sources. The slow motion
in the island part of the diagram, to the left hand and top sides,
combined with observational windowing, tends to lead to the formation of
isolated patches of data points, which is the origin of the term ``island
state'' \citep{hk89}. Whereas the continuum power spectra of the banana state are
dominated by a power law component at low frequencies with perhaps a weak
band--limited noise component (which becomes stronger as the source
approaches the island state), the island state power spectra are dominated
by the band--limited noise. As the source moves away from the banana into
the island state the count rate drops, the X--ray spectrum becomes harder
and the band--limited noise becomes stronger while its characteristic
frequency decreases. In the most extreme island states 
(4U 1608--52; \citealt{yoshida93}; \citealt{mendez99}; 4U 0614+09; \citealt{mendez97}; 
\citealt{vstr00}; 4U 1728--34; \citealt{ford98}; \citealt{disalvo01}; 4U 1705--44; 
\citealt*{lang89}; \citealt{berg98}; \citealt*{fordetal98}; Aql X--1; \citealt{reig00}; 
KS 1731--260; \citealt{muno00}) the source
is faint and hard and the band--limited noise is strong; this
is the state in which neutron stars are most similar to BHCs in the low
hard state \citep{vdk94a,crary96,olive98,berg98}. Because of the low count rates and slow motion through
the color--color diagram the precise properties of the extreme island state 
have been hard to ascertain, although observations of 4U 1608--52 with 
Tenma \citep{mitsuda89} indicated the existence of an extended branch in this 
state that was traced out over an interval of weeks.

Recently, \citet*{muno02a} and \citet{gier02a} used large data sets
from the Rossi X--ray Timing Explorer (RXTE) to study the color--color 
diagrams of several of the Z and atoll
sources, which contain interesting additional information about, in
particular, the nature of the extreme island state in the transient atoll
sources 4U 1608--52, 4U 1705--44 and Aql X--1. They suggested that the atoll
sources trace out similar three--branch patterns as the Z sources, with
the extreme island state cast in the role of Z source horizontal branch.
However, these authors did not address the timing properties of the
sources they studied.

In this paper, we continue our work on the correlated X--ray spectral and
timing behavior of Z and atoll sources using RXTE \citep{wijnands96,wijnands97,
wijnands98a,wijnands98b,fordetal98,mendez97,mendez99,mendezvdk99,reig00,jonker00,jonker02,
vstr00,vstr01,disalvo01,homan02}
with an analysis of 4U 1608--52. In these previous analyses, we have found  
strong correlations between the behavior of the timing features and 
the position of the source in the color--color diagram in both Z and atoll sources.
4U 1608--52 is a transient source that shows outbursts with a recurrence time varying 
between 80 days and several years \citep{loch94}. low--frequency QPOs were discovered in 
the ``island'' state of 4U 1608--52 with the Ginga satellite \citep{yoshida93} and 
kilohertz QPOs were discovered with RXTE \citep{paradijs96,berg96}. The source was 
included in the samples studied by \citet{muno02a} and \citet{gier02a} and 
was one of the sources that was reported to show Z--like behavior in the 
color--color diagram. By connecting the timing with the energy spectral properties 
we can test whether 4U 1608--52 indeed behaves as a Z source. If this is true one 
would expect the power spectral properties to change smoothly along the Z track, as is 
the case in the Z sources. We find that this is not the case. In 
\S \ref{sec.disc} we also perform a more general comparison of frequencies observed in 
the Z and the atoll sources.

Parallel tracks in color--color and color--intensity diagrams were first observed in 
the Z sources \citep{hasinger90,kuulkers94} and later also in the atoll sources 4U 1636--53 
\citep*{prins97,disalvo03}, 4U 1735-44 \citep{wijnands98c} and 4U 0614+09 \citep{vstr00}. 
Recently \citet{muno02a} reported further parallel tracks 
in the color--intensity diagrams of several atoll sources. The parallel--track phenomenon
in the plot of lower kilohertz QPO frequency versus intensity was first observed in 
4U 0614+09 \citep{ford97} and 4U 1608--52 \citep{yu97} and has since been observed 
in several more atoll sources \citep[see e.g.][]{mendez00}. It has been suggested that the 
parallel tracks in the plot of intensity versus frequency of the lower kilohertz QPO and 
the parallel tracks in the color--intensity diagrams might be the same phenomena 
\citep{vdk00,vdk01,muno02a}. Van der Klis (2001) has proposed a possible explanation for this 
parallel track phenomenon in terms of a filtered response of part of the X--ray emission 
to changes in the mass accretion rate through the disk.
The frequency versus intensity parallel--track phenomenon is particularly clear in 4U 1608--52 
\citep{mendez99}. In \S \ref{sec.tracks} we investigate the relation between these frequency versus 
intensity parallel tracks and the parallel tracks in the color--intensity diagrams. 

\section{OBSERVATIONS AND DATA ANALYSIS}
\label{sec.obs}

In this analysis we use all available public data from 1996 March 3 to 2000 May 24 from RXTE's 
proportional counter array \citep[PCA; for more instrument information see][]{zhang93}. The 
data are divided into observations that consist of one to several satellite orbits. We exclude 
data for which the angle of the source above the Earth limb is less than 10 degrees or for which
the pointing offset is greater then 0.02 degrees. In our data set we found 7 type I X--ray bursts
and we exclude those ($\sim 16$ s before and $\sim 240$ s after the onset of each burst) from our analysis. 

We use the 16--s time--resolution Standard 2 mode to calculate the colors. For each of the five 
PCA detectors (PCUs) we calculate a hard color, defined as the count rate in the energy band 9.7$-$16.0 
keV divided by the rate in the energy band 6.0$-$9.7 keV, and a soft color, defined as the count 
rate in the energy band 3.5$-$6.0 keV divided by the rate in the energy band 2.0$-$3.5 keV. Per 
detector we also calculate the intensity, the count rate in the energy band 2.0$-$16.0 keV. To 
obtain the count rates in these exact energy ranges we interpolate linearly between count rates in 
the PCU channels. We calculate the colors and intensity for each time interval of 16 s. We subtract 
the background contribution in each band using the standard background model for the PCA version 2.1e.

\begin{figure}
\figurenum{1}
\epsscale{0.8}
\plotone{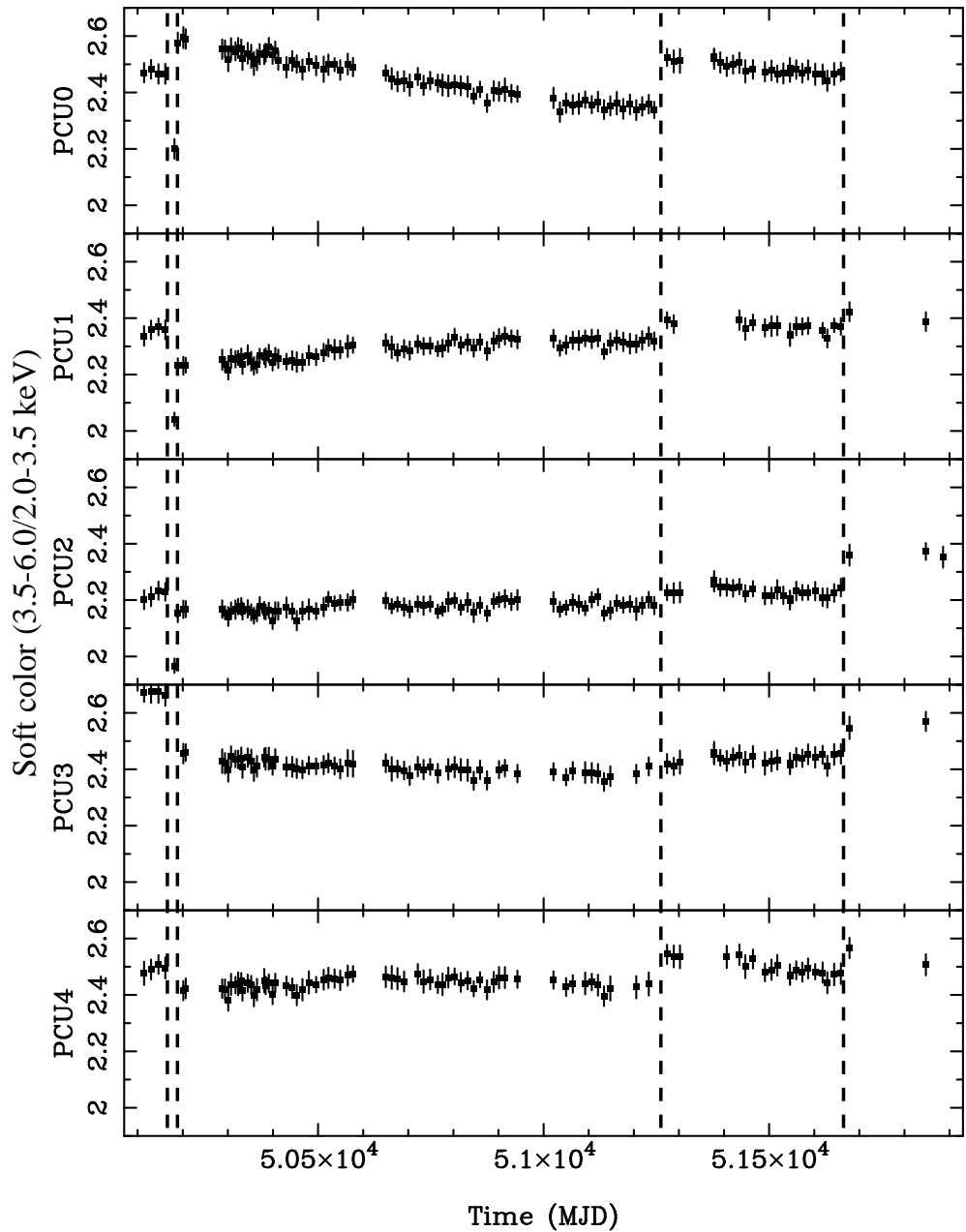}
\caption{Soft color of the Crab versus time for each PCU. Soft 
color is defined as the ratio of count rates in the energy bands 
3.5$-$6.0$/$2.0$-$3.5 keV. Dashed lines indicate the gain changes between gain epochs 1--5. 
The soft color ($\sim 1.8$) in epoch 5 for PCU0, after its propane layer was lost, is offscale.}
\label{fig.crab}
\end{figure}

In order to correct for the changes in effective area between the different gain epochs and 
for the gain drifts within those epochs as well as the differences in effective area between the 
PCUs themselves we used the method introduced by \citet{kuulkers94}: for 
each PCU we calculate the colors of the Crab, which can be supposed to be constant 
in its colors, in the same manner as for 4U 1608--52. We then average the 16 s Crab colors and intensity 
per PCU for each day. In Figure \ref{fig.crab} we can see the clear differences in the soft color trends 
of Crab between the five PCUs caused by the effects mentioned above. For each PCU we divide 
the 16 s color and intensity values obtained for 
4U 1608--52 by the corresponding Crab values that are closest in time but within the same gain epoch. 
We then average the colors and intensity over all PCUs. If we multiply the soft color by 2.36, the hard 
color by 0.56 and the intensity by 2400 c/s/PCU we approximately recover the observed, uncorrected 
values for 4U 1608--52 (quoted values are averages of the Crab colors during the observations over all 
five PCUs). To improve statistics we rebin the 16 s 
color and intensity points to 256 s and exclude all data for which the resulting relative errors are 
larger than 5\%. This led to a loss of about 40 ks of data at count rates below $\sim0.01$ Crab. For 
most of these data neither timing nor colors were of sufficient quality to determine the source state.
Starting May 12, 2000, the propane layer on PCU0, which functions 
as an anti--coincidence shield for charged particles, was lost. However, as all our data after May 12, 
2000 ($\sim10$ ks) was excluded because the relative color errors were larger than 5\% this was not 
a issue in our analysis.

For the Fourier timing analysis we use the 122$-\mu$s time--resolution Event and Single Bit modes and 
the 0.95$-\mu$s time--resolution Good Xenon modes. We only use data for which all energy channels are 
available. This led to a loss of about 9 ks of data in the lower banana state. 
In some observations there was data overflow in the timing modes due to excessive count rates 
from the source; we excluded those data. This led to a loss of about 27 ks of data at the highest intensity 
level (see also below). The power spectra were constructed using data segments of  
256 s and 1/8192 s time bins such that the lowest available frequency is 1/256 Hz and the 
Nyquist frequency 4096 Hz; the normalization of \citet{lea83} was used. To get a first 
impression of the timing properties at the different dates and positions in the color--color diagram, the 
power spectra were combined per observation. The resulting power spectra were then converted 
to squared fractional rms. We subtracted a constant Poisson noise level estimated between 2000 
and 4000 Hz where neither noise nor QPOs are known to be present. 
As fit function we use the multi--Lorentzian function; a sum of 
Lorentzian components \citep*{grove94,olive98,bpk02} plus a power law to fit the VLFN \citep[see][]{vstr02}. We only 
include those Lorentzians in the fit whose significance based on the error in the 
power integrated from 0 to $\infty$ is above 3.0 $\sigma$. 
Two to five Lorentzian components were generally needed for a good fit. For the Lorentzian that is used 
to fit the band--limited noise we fixed the centroid frequency to zero. We plot the power 
spectra in the power times frequency representation \citep[$\nu{\rm P}_{\nu}$; e.g.][]{bel97,n00}, where the 
power spectral density is multiplied by its Fourier frequency. For a multi--Lorentzian fit function 
this representation helps to visualize a characteristic frequency, $\nu_{\rm max}$, namely, the frequency 
where each Lorentzian component contributes most of its variance per logarithmic frequency interval: in 
$\nu{\rm P}_{\nu}$ the Lorentzian's maximum occurs at $\nu_{\rm max}$
\citep[$\nu_{\rm max} = \sqrt{\nu_0^2 + \Delta^2}$, where $\nu_0$ is the centroid and $\Delta$ the HWHM 
of the Lorentzian;][]{bpk02}. We represent the Lorentzian relative width by $Q$ defined as $\nu_0/2\Delta$.

\section{A TIMING TOUR THROUGH THE COLOR--COLOR DIAGRAM}
\label{sec.timingtour}

In this section we try to get an idea of the timing properties of the source as it moves through 
the color--color diagram. We step through the data in chronological 
order and look at the timing properties (per observation, see \S \ref{sec.obs}) and the position of the source 
in the color--color diagram. In this way we select continuous time 
intervals for which the power spectrum and the position in the color--color diagram remain similar.
For each continuous time interval we construct a representative power spectrum by adding up all observations within that 
interval. To get a first idea of the timing properties we made initial fits to the representative power spectra of the 
continuous time intervals using the multi--Lorentzian fit function 
described in \S \ref{sec.obs}. In most cases the power spectral features present could be directly identified with 
known components seen in other atoll sources by comparing with the power 
spectral features of 4U 0614+09 and 4U 1728--34 \citep[using Figures 1, 2 and 3 of][]{vstr02}. A more thorough fitting and 
description of the power spectral features will be presented in \S \ref{sec.comb_ps} where we obtain optimal
statistics by constructing representative intervals for all our data by adding up the continuous time intervals 
that have similar positions in the color--color diagram and show similar power spectra. 

The data on 4U 1608--52 can be 
usefully divided into 3 segments. The first segment ranges from 1996 March 3 to December 28 \citep[the decay of the 1996 
outburst, see][]{berg96}, the second segment from 1998 February 3 to September 29 \citep[the 1998 outburst, 
see][]{mendez98} and the third segment from 2000 March 6 to May 10 (persistent data). In practice most data were available for the second segment (the 1998 outburst) so we will present the results for the second segment first. These
results can then serve as a template for the rest of the data.

\subsection{Segment 2 (the 1998 outburst)}
\label{sec.segment2}

\begin{figure}
\figurenum{2}
\epsscale{0.7}
\plotone{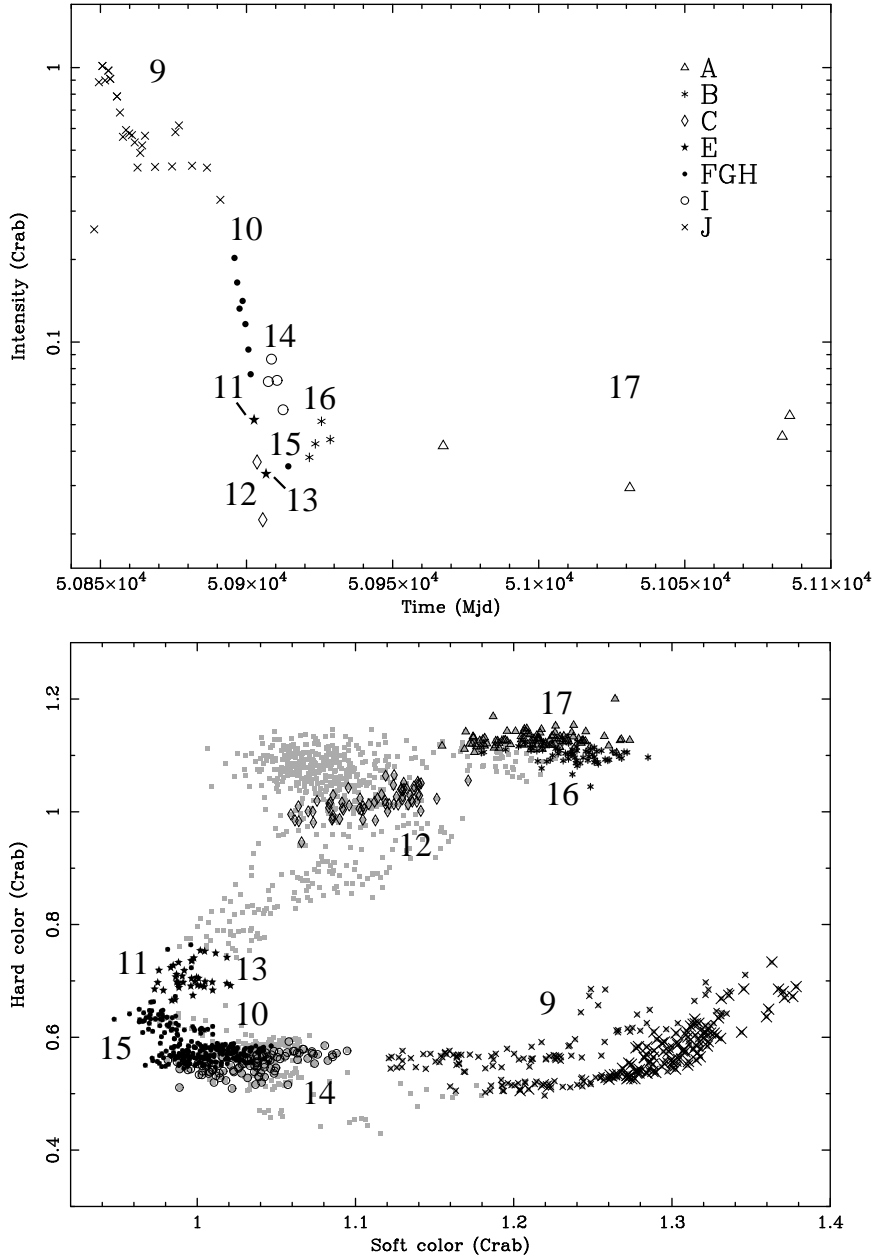}
\caption{\footnotesize Lightcurve (top) and color--color diagram (bottom) for the decay of the 1998 outburst. 
In the bottom frame, the black points correspond to the 1998 outburst and the grey points to the full data set. 
Intensity is defined as the count rate in the energy band 2.0$-$16.0 keV; hard and soft colors  
are the count rate ratios in the 9.7$-$16.0$/$6.0$-$9.7 keV and 3.5$-$6.0$/$2.0$-$3.5 keV bands respectively. 
Intensity and colors are in units of Crab (see \S \ref{sec.obs}). 
Each point in the color--color diagram spans 256 s. For clarity no errors are shown, but  
relative errors are always below 5\% (see \S \ref{sec.obs}). Each point in the 
lightcurve corresponds to a one--day average. The numbers indicated in the plot correspond to 
continuous time intervals for which the power spectra and the position in the color--color diagram remain 
similar. The symbols indicate positions in the color--color diagram for which the power spectra are 
similar; a legend linking the symbols to the classification of \S \ref{sec.comb_ps} is provided in the top 
frame. The enlarged crosses at the peak of the outburst and the corresponding enlarged crosses in interval 
9 of the color--color diagram indicate data for which no power spectra could be computed due to 
data overflow (see \S \ref{sec.obs}).}
\label{fig.cc_deel2}
\end{figure}

In Figure \ref{fig.cc_deel2} we show as black points the lightcurve and color--color diagram 
for the second segment of the data (the 1998 outburst). The grey points in the lower frame represent the overall color--color 
diagram including all the data. Intensity and colors are normalized to the Crab 
(see \S \ref{sec.obs}). The numbers 9--17 (1--8 are reserved for the first segment of the lightcurve) indicate the 
continuous time intervals for which the power spectra and the position in the color--color diagram remain similar (see 
\S \ref{sec.timingtour}). The data for which no power spectra could be computed due to 
data overflow (see also \S \ref{sec.obs}) are indicated by the larger crosses at the peak of the outburst and 
in interval 9 of the color--color diagram. In Table \ref{tbl.timingtour} we present the duration of each interval, 
the time until the next interval and the 2--16 keV intensity during each interval. We identify several power spectral 
features and classify the continuous time intervals into color intervals A--J; these will be discussed in \S \ref{sec.comb_ps} where we put 
all information together. For the 1998 outburst we find 7 different classes, i.e., the source returns twice to similar 
positions in the color--color diagram where it displays similar power spectral shapes. Each class is marked with a 
different symbol in the color--color diagram of Figure \ref{fig.cc_deel2}.

For the 1998 outburst we confirm the result of \citet{mendez99} that the color--color diagram 
shows the classical atoll shape \citep[see][]{hk89}. The characteristic frequencies 
of most of the power spectral components increase along the track starting at the open triangles and
ending at the crosses in Figure \ref{fig.cc_deel2} (for a more detailed discussion of the power spectral behavior 
with respect to the position in the color--color diagram, see \S \ref{sec.move_cc}).  

\subsection{Segment 1 (the decay of the 1996 outburst)}
\label{sec.segment1}

\begin{figure}
\figurenum{3}
\epsscale{0.7}
\plotone{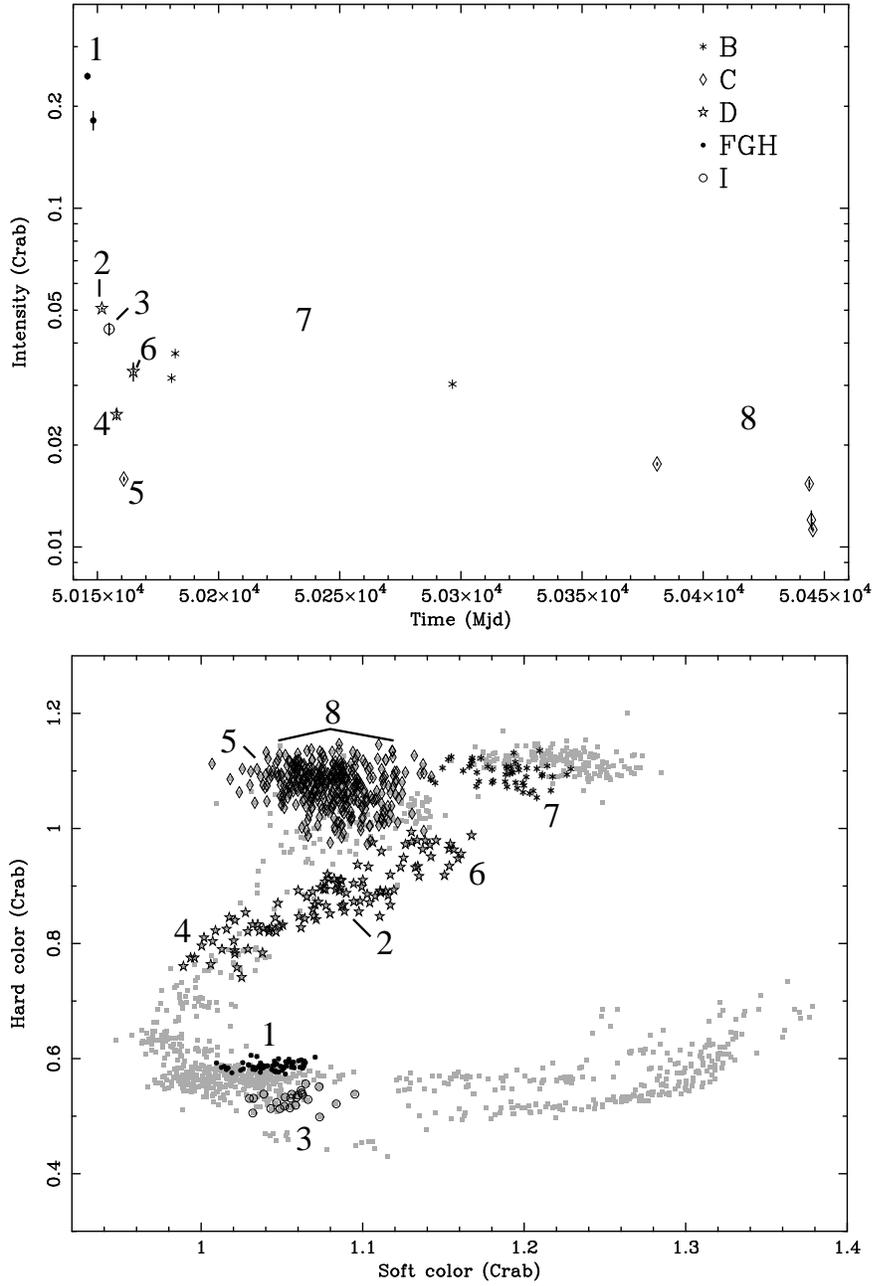}
\caption{As Figure \ref{fig.cc_deel2}, but for the 1996 outburst.}
\label{fig.cc_deel1}
\end{figure}

In Figure \ref{fig.cc_deel1} we show as black points the lightcurve and color--color diagram 
for the first segment of the data (the decay of the 1996 outburst). The grey points represent the overall color--color 
diagram including all the data. The numbers 1--8 again indicate 
continuous time intervals for which the power spectra and the position in the color--color diagram remain 
similar. The symbols and entries in Table \ref{tbl.timingtour} are as described in \S \ref{sec.segment2}.

We find one class additional to those observed during intervals 9--17. This class is 
composed of continuous time intervals 2, 4 and 6, fills up the region between the open diamonds and the 
filled stars in the color--color diagram, and is marked with open stars in Figure \ref{fig.cc_deel1}. 
Note further that part of the open diamonds here are in a slightly different position in the color--color diagram, at a 
higher hard color and a lower soft color, than where they were during the second segment of the lightcurve. This is 
accompanied by a lower intensity. 
If we sort the classes by characteristic frequency as measured in the power spectra, 
the characteristic frequencies change along an ``$\epsilon$--shaped'' track in the 
color--color diagram (see also \S \ref{sec.move_cc}).

\subsection{Segment 3}
\label{sec.segment3}

\begin{figure}
\figurenum{4}
\epsscale{0.7}
\plotone{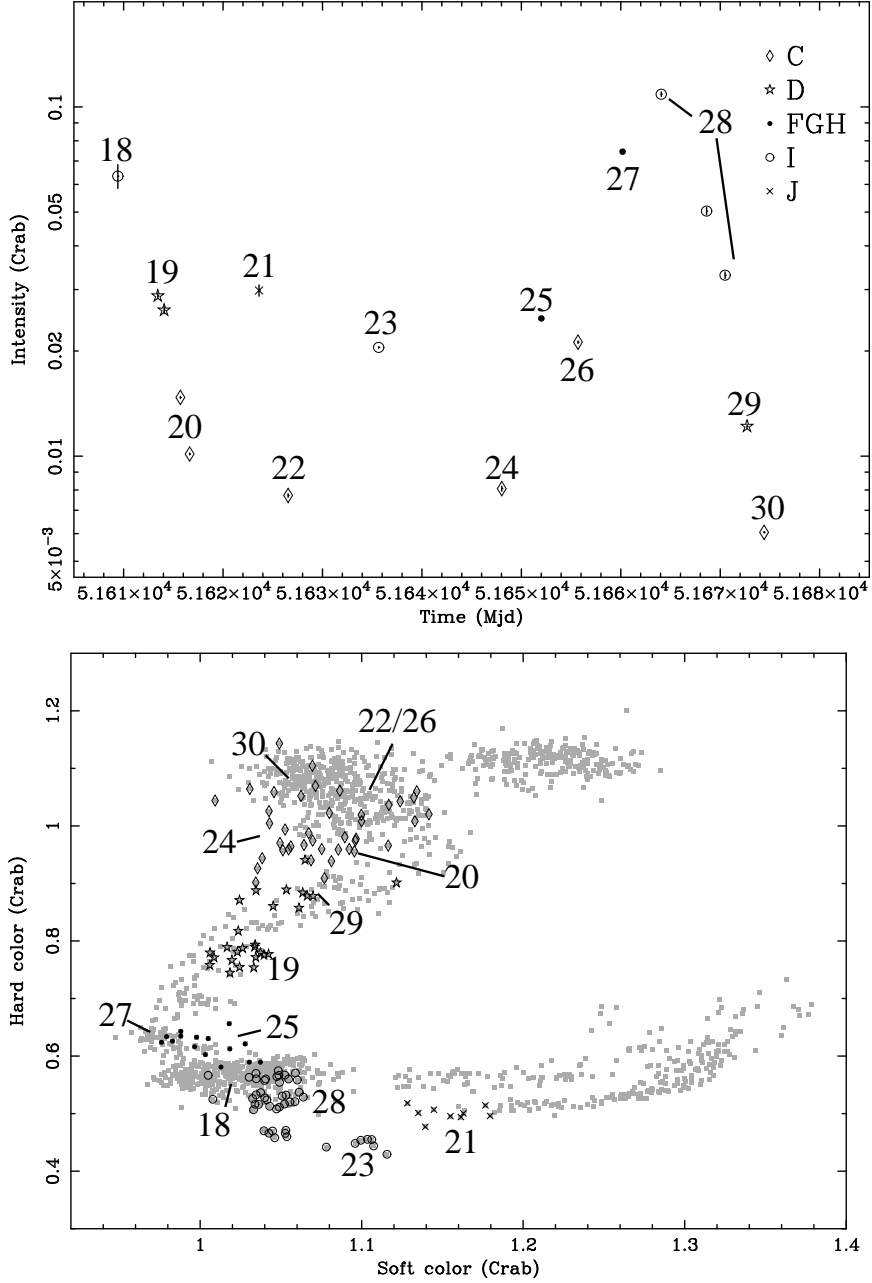}
\caption{As Figure \ref{fig.cc_deel2}, but for the third segment of the data (persistent data). 
In this case, the classification marked by the symbols is based on the 
position in the color--color diagram only (see \S \ref{sec.segment3}).}
\label{fig.cc_deel3}
\end{figure}

In the third segment of the data the source count rates were low and in most observations one or more detectors were switched 
off. This, together with the short durations of the observations (only $\sim500-2500$ s per observation) led to 
bad statistics. In most cases it was impossible to identify any power spectral features, therefore the classification 
for this segment of the data was solely done on position in the color--color diagram. If any power spectral 
features were detected they were always consistent with this color--based classification. 
As before, in Figure \ref{fig.cc_deel3} black points represent the lightcurve and color--color diagram 
for the third segment of the data and grey points the overall color--color diagram. The numbers 18--30 now indicate 
continuous time intervals for which the position in the color--color diagram remained 
similar, and the symbols mark the classification, here based on the 
position in the color--color diagram only; entries in Table \ref{tbl.timingtour} are as described in \S \ref{sec.segment2}.
Note that the source is in the class marked with the crosses for only the second time 
during continuous time interval 21, but now at much lower intensity (0.03 Crab) than it was the first time in continuous 
time interval 9 (0.24--1.30 Crab) at the peak of the 1998 outburst.

\subsection{The combined power spectra}
\label{sec.comb_ps}

To improve the statistics we average the power spectra in each of the 8 classes. In the class marked with the filled circles 
the lower kilohertz QPO is extremely narrow and varies in frequency over several 
hundreds of Hz. This leads to multiple peaks in the power spectrum. Therefore we split up this class into three parts 
depending on lower kilohertz QPO frequency; 
the first with the lower kilohertz QPO ranging from 540 to 640 Hz, the second with the lower kilohertz QPO ranging from 
640 to 710 Hz and the third with the lower kilohertz QPO ranging from 710 to 900 Hz. A finer division would have compromised
the statistics at low frequencies. The observations in this class where no lower (or 
upper) kilohertz QPO was detected were added based on position in the color--color diagram. In Figure \ref{fig.cc_int} 
we show the resulting 10 intervals in the color--color diagram and in Figure \ref{fig.hid_sid_int} we show the corresponding 
hard color and soft color vs. intensity diagrams. Note that the data for which no power spectra could be computed due to 
data overflow (see \S \ref{sec.obs} and \ref{sec.segment2}) are not included in Figures \ref{fig.cc_int} and 
\ref{fig.hid_sid_int}. We mark the different classes in order of increasing 
characteristic frequencies from A to J. We use letters here to avoid confusion with the numbered continuous time 
intervals of \S \ref{sec.timingtour}. 

We fit each interval with the multi--Lorentzian fit function described in \S 2. For the intervals where 
the kilohertz QPOs have sufficiently high frequencies not to interfere with the low--frequency features 
and vice versa, we fit the kilohertz QPOs between 500 and 2048 Hz and then fix the kilohertz 
QPO parameters when we fit the whole power spectra. This is for computational reasons only; the results are the same 
as those obtained with all parameters free. If the $Q$ value of a Lorentzian 
becomes negative in the fit, we fix it to zero. No significantly negative Q's occured.
The fits have a $\chi^2$/dof below 1.4 for intervals A--G. For interval H the 
$\chi^2$/dof is high (5.43); this is caused by the motion of the lower kilohertz QPO from 710 to 900 Hz (see above). For intervals 
I and J the $\chi^2$/dof are 2.5 (dof = 100) and 1.8 (dof = 97). This is caused by deviations of the VLFN from the power law used to fit it (see appendix \ref{sec.detailed_ps}).

We now specify the terminology that we will use for the various power spectral components. 
Typical power spectral components observed for the atoll sources are described for the 
two sources 4U 0614+09 and 4U 1728--34 in \citet{vstr02}. 
As in \citet{bpk02} we call the upper kilohertz QPO L$_u$ and its characteristic frequency 
$\nu_u$. The lower kilohertz QPO has been linked to a broad bump at $\sim10$ to $\sim25$ 
Hz found in the low luminosity bursters 1E 1724--3045, GS 1826--24 and SLX 1735--269 
\citep{bpk02} and in the atoll source 4U 0614+09 at its lowest characteristic frequencies 
\citep{vstr02}. Although we emphasize that this identification is very tentative (see 
\S \ref{comp:timing:atoll}) we will, as was done in \citet{bpk02}, call both the lower kilohertz QPO and this 
10--25 Hz bump (which do not occur simultaneously), L$_\ell$ (characteristic frequency $\nu_\ell$). 
No standard terminology yet existed for the hectohertz Lorentzian \citep{ford98}; here we will call it L$_{hHz}$ 
(characteristic frequency $\nu_{hHz}$). The behavior of the band--limited noise and QPOs in the 0.1--50 Hz 
range in the atoll sources 4U 1728--34 and 4U 0614+09 is complex \citep[see also][]{disalvo01,vstr02}, we 
describe it as a function of $\nu_u$ or position in the color--color diagram (see Figure \ref{fig.freq_freq}). 
First, the characteristic frequency of 
the band--limited noise increases from $\sim0.3$ to $\sim15$ Hz. In this phase the band--limited 
noise is broad and usually fitted with a zero--centered Lorentzian or a broken power law. Then at 
$\sim15$ Hz this noise component appears to ``transform'' into  
a narrow QPO \citep[called very low--frequency Lorentzian in][]{vstr02} 
whose frequency smoothly continues to increase up to $\sim50$ Hz, while 
what appears to be another band--limited noise component appears at lower characteristic 
frequencies. It is unclear exactly how these three components are related. Here, to 
remain true to the naming scheme of \citet{bpk02} we call the band--limited noise component that 
becomes a QPO as well as the QPO it becomes L$_{b}$ (characteristic frequency $\nu_{b}$) and the ``new'' 
broad band--limited noise appearing at lower frequency L$_{b2}$ (characteristic frequency $\nu_{b2}$), but 
please note the uncertainties in the interpretation underlying this nomenclature. Finally, in the banana branch 
of 4U 1728--34 and 4U 0614+09 (where usually no kilohertz QPOs are detected)
a broad Lorentzian is also present at $\sim20-35$ Hz for which it is unclear whether it is L$_{b}$, L$_{b2}$ or a new 
component; here we shall list it as L$_{b2}$. 
The atoll sources 4U 1728--34 and 4U 0614+09 both additionally show a $\sim1-50$ Hz Lorentzian at frequencies above L$_{b}$, called 
the low--frequency Lorentzian by \citet{vstr02}. At high characteristic frequencies ($\nu_{\rm max} > 25$ Hz) 
this Lorentzian appears as a narrow QPO. At low characteristic frequencies ($\nu_{\rm max} < 10$ Hz) this Lorentzian 
is broad and can be identified with the component in the low luminosity bursters which \citet{bpk02} call L$_h$ ('hump').
For this reason, we also call this low--frequency Lorentzian L$_h$ (characteristic frequency $\nu_h$). 
In some observations the low luminosity bursters show a narrow low--frequency QPO \citep[called L$_{\rm LF}$ in][]{bpk02}
simultaneously with L$_h$. The $\nu_{\rm max}$ of L$_{\rm LF}$ is slightly lower than that of L$_h$. For 1E 1724--3045 the 
Lorentzian centroid frequencies of L$_{\rm LF}$ and L$_h$ coincided but this was not the case for GS 1826--24 \citep{bpk02}.
L$_{\rm LF}$ was not observed in 4U 1728--34 or 4U 0614+09 and we also do not detect it in 4U 1608--52. 

A detailed description of the behaviour of the power spectral components of 4U 1608--52 is given in
appendix \ref{sec.detailed_ps}.
In Figure \ref{fig.powspec_1608} we show the averaged power spectra and best fit functions of the intervals A--J. 
The fit parameters for the low--frequency part of the power--spectra are listed in Table 
\ref{tbl.lowfitpar}, the fit parameters for the high--frequency part are listed in Table 
\ref{tbl.highfitpar}. 
We obtain 95\% confidence upper limits for the fractional rms amplitude 
of L$_{hHz}$ in intervals B, C and E and of L$_h$ in 
intervals B, C and E using $\Delta\chi^{2} = 2.71$, fixing 
$Q$ to 0.2 and allowing $\nu_{hHz}$ to run between 100 and 200 Hz. For setting upper limits to L$_h$
we fix $Q$ to 1.5, 3.0 and 3.5 and let $\nu_h$ run 
between 22--32 Hz, 34--44 Hz and 42--50 Hz respectively (the values expected from the results 
for 4U 1728--34 and 4U 0614+09). 

In Figure \ref{fig.freq_freq} we plot the characteristic 
frequencies of 4U 1608--52 versus $\nu_u$, together 
with the results of \citet{vstr02} for 4U 1728--34 and 4U 0614+09. The black points mark the results 
for 4U 1608--52, the grey points the results for 4U 1728--34 and 4U 0614+09. The different symbols indicate 
different power spectral components. The results for 4U 1608--52 mostly fall on the relations
established for 4U 1728--34 and 4U 0614+09. Also the $Q$ values of the various components 
versus $\nu_u$ are similar to those for 4U 1728--34 and 4U 0614+09. In Figure \ref{fig.rms_all} we plot the rms 
fractional amplitude of all components versus $\nu_u$. The general trends in this plot for 4U 1608--52 are 
similar to those of 4U 1728--34 and 4U 0614+09 \citep{vstr02}, but there is an offset in rms between the relations for the three sources \citep*{mendez01}. 
For the two kilohertz QPOs the rms fractional amplitudes in 4U 1608--52 are similar to those in 4U 0614+09 but larger than 
in 4U 1728--34 \citep{disalvo01}. For all other features; L$_{hHz}$, L$_h$, L$_{b}$, L$_{b2}$ and at low frequencies L$_\ell$   
 the rms fractional amplitudes for 4U 0614+09 
are the largest followed by 4U 1728--34 whereas the rms fractional amplitudes for 4U 1608--52 are the lowest. 
As we describe in appendix \ref{sec.detailed_ps} in more detail, both the relations in Figures \ref{fig.freq_freq} 
and \ref{fig.rms_all} as well as a direct comparison with the power spectra of 4U 1728--34 and 4U 0614+09 
\citep[Figures 1 and 2 in][]{vstr02} allow us to identify the power spectral components of 4U 1608--52 
in the different intervals within the identification scheme described above.

\subsection{How does 4U 1608--52 move through the color--color diagram?}
\label{sec.move_cc}

Now that we have described and identified all power spectral components in 
4U 1608--52, we can link the timing properties (Tables \ref{tbl.lowfitpar} and 
\ref{tbl.highfitpar}) to the position in the 
color--color diagram. We can also link the timing properties and X--ray spectral 
properties to the classical island and banana states described by \citet{hk89}. 
Intervals A--D are occurrences of the island state (strong broadband noise with low 
$\nu_{\rm max}$, no VLFN) in which intervals A--C can be classified as the 
extreme island state. The extreme island state shows L$_{b}$, L$_h$, L$_\ell$ and 
L$_u$ all at low characteristic frequencies. The power spectral components are all 
broad and strong. Interval E forms a transition between the island state 
and the lower banana, it still has relatively strong broad band noise but also shows a pair of 
narrow kilohertz QPOs that are typical for what \citet{hk89} called the lower left banana state.
So, intervals F--H are all lower left banana state. Here the 
VLFN appears, the kilohertz QPOs are double, L$_{b}$ transforms from a band--limited 
noise component into a QPO and L$_{b2}$ appears (see also \S \ref{sec.comb_ps}). 
Intervals I and most of interval J are also still in what \citet{hk89} called the lower banana state based on their 
position in the color--color diagram; only the upper right part in the color--color diagram 
of interval J is in the upper banana state based on color--color position. The VLFN 
is strong and the broad band noise is weak in these intervals. 
Intervals I and J show L$_u$ at the highest frequencies, L$_{b2}$, 
and a strong VLFN component. The characteristic 
frequencies increase in order A--J and form an ``$\epsilon$--shaped'' track in the 
color--color diagram.

Recently, \citet{muno02a} and \citet{gier02a} studied the color--color diagrams of several 
of the Z and atoll sources, including 4U 1608--52, and suggested that the atoll sources 
trace out similar three--branch patterns as the Z sources. We observe the same shape of 
the color--color diagram for 4U 1608--52 as \citet{muno02a} and \citet{gier02a} did 
(Fig. \ref{fig.cc_int}). Interval C in Figure \ref{fig.cc_int} represents a deviation from 
the classical atoll shape. According to the interpretation of \citet{muno02a} and \citet{gier02a} 
the source would have to move in the Z--track order D--AB--C in Figure \ref{fig.cc_int}. 
Interval C would then correspond to the horizontal branch of the Z sources. We observe a 
transition from D to C and back to D with gaps of three days during the decay of the 1996 
outburst (Fig. \ref{fig.cc_deel1}). We also observe a transition from E to C and back to E 
with gaps of only one day during the decay of the 1998 outburst (Fig. \ref{fig.cc_deel2}). 
The source is first in E for an interval of $\sim0.1$ day during which the hard color increases 
by $\sim0.04$ (so $v_{\rm hard~color} \approx 0.4$ day$^{-1}$), then there is a gap of 0.9 
day after which the source appears in C with an increase in hard color of $\sim0.3$ (if the source 
moved directly from E to C $v_{\rm hard~color} \approx 0.3$ day$^{-1}$). Then in C there 
is an interval of $\sim0.16$ day where the hard color increases by $\sim0.04$ 
($v_{\rm hard~color} \approx 0.3$ day$^{-1}$). So this is consistent with the source moving 
directly from E to C at approximately constant speed (probably through D) and not through A or B 
as required if 
4U 1608--52 behaved as a Z source. In C there is a gap of $\sim1.8$ day after which the 
source appears again in C but at a slightly ($\sim0.03$) lower hard color. For an $\sim0.1$ 
day interval the source stays in C while the hard color decreases by $\sim0.03$ 
($v_{\rm hard~color} \approx -0.3$ day$^{-1}$). Then there is a gap of $\sim1.0$ day and after 
that the source has returned to E with a hard color of $\sim0.3$ less ($v_{\rm hard~color} 
\approx -0.3$ day$^{-1}$). This is consistent with the source moving directly back from C to E, 
again at constant speed, and not through A or B. Transitions from and to states 
A and B had gaps of 7 and more days, so it is impossible to draw conclusions from these. The 
characteristic frequencies of the timing features decrease in the order D--C--B--A 
(appendix \ref{sec.detailed_ps}). This is also not consistent with the idea that 4U 1608-52 behaves as 
a Z source, as in Z sources the characteristic frequencies of the timing features increase 
along the Z starting at the horizontal branch (i.e., this would predict frequencies decreasing 
in the order DBAC). One might say that 4U 1608--52 draws up an approximate ``$\epsilon$'' 
shape in the color--color diagram along which the characteristic frequencies of the timing features 
change smoothly. 

\section{A DETAILED STUDY OF THE PARALLEL TRACK PHENOMENA}
\label{sec.tracks}

Parallel tracks in the intensity versus lower kilohertz QPO 
frequency in 4U 1608--52 were discovered by \citet{yu97}, and extensively studied by \citet{mendez99} and 
\citet{mendez01}. The hard color and soft color vs. intensity diagrams of 4U 1608--52 (see Fig. \ref{fig.hid_sid_int}) 
and those of several other atoll sources also show narrow parallel tracks \citep{prins97,disalvo03,muno02a}. Based 
on the appearance of the diagrams, it has been suggested that there is a relation between these two types of parallel 
tracks \citep{vdk00,vdk01,muno02a}.

Based on our new analysis it is now possible to directly link the parallel tracks in the color intensity diagrams to
those in the QPO frequency--intensity diagrams using the frequency of the lower kilohertz QPO as obtained by \citet{mendez01}. 
Similarly to \citet{mendez99} and \citet{mendez01} we only include data for which both kilohertz QPOs are detected 
simultaneously and therefore the lower kilohertz QPO is 
identified unambiguously. We rebin our 16 s colors in such a way that they match the 64--448 s data intervals of \citet{mendez01}.
In Figure \ref{fig.mendez_rate} we plot the frequency of the lower kilohertz QPO and the hard and soft color versus the 2.0--16.0 
keV intensity. The alternating black/grey symbols represent the parallel tracks in the intensity versus the 
frequency diagram; the tracks contain data that are continuous in time, with only the $\sim2500$ s gaps due to Earth occultations. 
Note that the parallel track phenomenon in the lower kilohertz QPO frequency vs. intensity diagram is only observed in 
a small region of the color intensity diagrams (see Figure \ref{fig.hid_sid_int}) where the lower kilohertz QPO is strong and
narrow enough to be accurately traced on 64--448 s timescales.

With this analysis, we can investigate how the parallel tracks in the frequency vs. intensity diagram relate to those 
in the color intensity diagrams. Interestingly, it turns out that the frequency--intensity tracks can not, as previously 
thought, be identified with the narrow vertical tracks visible in the color intensity diagrams of Figures \ref{fig.hid_sid_int} 
and \ref{fig.mendez_rate}. Each of these latter tracks corresponds to a single satellite orbit. 
The parallel tracks in the frequency vs. intensity diagram are only identifiable as (rather fuzzy) tracks in the color intensity 
diagrams after several satellite orbits have elapsed and thus turn out to be composed of several of those narrow 
vertical tracks. The parallel tracks in the frequency versus intensity diagram could also identified in the hard 
color--intensity diagram of 4U 1636--53, where two banana tracks shifted by 20 \% in intensity were observed in the hard 
color--intensity diagram \citep{disalvo03}, and exactly the same 20 \% shift was observed between the corresponding 
two parallel lines in the frequency versus intensity diagram. In 4U 1608--52 we do not observe complete banana tracks 
shifted in intensity as in 4U 1636--53. The  drift in X--ray flux in 4U 1608--52 already occurs on timescales of several days 
\citep{mendez99}, too short for a complete track to form, where in 4U 1636--53 the source can stay on one track for several 
months \citep{disalvo03}. These fast changes in intensity in 4U 1608--52 are due to its transient character, where 4U 1636--53 
is a persistent source for which changes in intensity occur much more gradually.
Note, that within each of the two parallel lines in the frequency versus intensity diagram of 4U 1636--53 there is no correlation
observed between frequency and intensity.

The narrow vertical tracks in the color intensity diagrams are caused by variations in color that for the most part can 
be explained by the scatter due to counting statistics. 
The parallel tracks in the frequency versus intensity diagram, although composed of data covering several consecutive satellite 
orbits, already show up within individual orbits due to a short term correlation between the lower kilohertz QPO frequency and 
intensity in combination with small errors in these quantities but persist over several consecutive orbits.
The correlation between colors and intensity (or kilohertz QPO frequency) in individual orbits is veiled 
by the limited counting statistics, so in the color intensity diagrams the parallel QPO tracks can only be identified on the 
longer timescales of several satellite orbits. 

To further illustrate this point in Figure
\ref{fig.mendez_sc} we plot soft color versus the frequency of the lower kilohertz QPO; lines connect the points of individual
satellite orbits, and for clarity four representative individual orbits are highlighted in black. On orbit time scales, the 
lower kilohertz QPO frequency and soft color seem uncorrelated due to the counting statistics errors on the color, 
whereas on a longer time scale a clear correlation emerges.
We note that for some parallel tracks in Figure \ref{fig.hid_sid_int} there is a correlation between color and 
intensity on short timescales, especially in interval J. In those cases the counting statistics are better due to the higher count rates and 
thus the parallel track phenomenon in the color diagrams is more similar to 
that in the kilohertz QPO frequency versus intensity diagram \citep{muno02a}. However, in our interval J 
there is no timing feature present that is trackable on short timescales. 

In conclusion we can now identify the parallel QPO tracks in the color intensity diagrams.
Bad counting statistics that shows up in the form of the narrow nearly vertical parallel tracks 
in the color intensity diagrams cause the $\nu-$color correlation to be veiled on short timescales (less than hours) 
so in the color intensity diagrams the parallel QPO tracks can only be found for the longer timescales 
of several satellite orbits.

\section{DISCUSSION}
\label{sec.disc}

We have studied the color diagrams and the power spectral behavior of 
the atoll source 4U 1608--52. We found that the timing behavior of 4U 1608--52 is almost identical 
to that of the other atoll sources 4U 1728--34 and 4U 0614+09. If we plot the characteristic frequencies 
of the timing features versus the characteristic frequency of the upper kilohertz QPO, together with the 
results of \citet{vstr02} for 4U 1728--34 and 4U 0614+09, the three sources follow the same relations 
(see Fig. \ref{fig.freq_freq}). Also the behavior of the $Q$ value is the same for the three sources. 
The general trends in rms fractional amplitude for 4U 1608--52 are also similar to those of 4U 1728--34 
and 4U 0614+09, but there is an offset between the relations for the three sources (see Fig. \ref{fig.rms_all}). 
We connected the timing behavior with the position of the source in the color--color diagram 
and found that the timing behavior is not consistent with the idea that 4U 1608-52 traces out a 
three-branched Z shape in the color-color diagram along which the timing properties vary gradually as 
is the case in in Z sources. Instead, the power spectral properties change along an ``$\epsilon$--shaped'' track. 
Finally, our measurements for the colors together with the precise measurements of lower kilohertz QPO 
frequency of \citet{mendez01} gave us an opportunity to link the parallel tracks in the intensity versus 
color diagrams with the parallel tracks in the intensity versus lower kilohertz QPO frequency diagrams. 
We found that the parallel tracks in the frequency versus intensity diagram can be found back in the 
color--intensity diagrams as fuzzy structures; they should not be confused with the narrow vertical 
parallel tracks visible in the intensity versus color diagrams which are mostly due to the errors in the colors.

\subsection{Comparison with other sources; colors}
\label{sec.comp_col}

The power spectral properties in 4U 1608--52 change along an ``$\epsilon$--shaped'' track. 
The question now arises whether the ``$\epsilon$--shaped'' track we find in the color--color 
diagram of 4U 1608--52 is universal for atoll sources. We can compare our results with those 
of \citet*{olive03} who studied the X--ray color and timing properties of a well sampled state 
transition of 4U 1705--44 where the source moves from the lower banana to the extreme island state and 
back. This transition was also included in \citet{muno02a} who studied a larger dataset of 4U 1705--44.
In the color--color diagram the source moves from the lower banana, to the left of the extreme island state
as the count rate decreases. Then both the soft color and the count rate increase, whereas 
the hard color remains approximately constant, and the source traces out a horizontal track in the color--color 
diagram. Then, while the count rate continues to increase, the source moves back 
from the right of the extreme island state to the lower banana \citep{muno02a,barret02,olive03}. 
In the whole horizontal track the power spectra remained the same \citep{olive03}.

In this extreme island state 4U 1705--44 did not reach characteristic frequencies as low as those in interval 
A of 4U 1608--52 (it is very similar to interval B). If we look at 
the overall light and color curves of 4U 1705--44 presented in \citet{muno02a} we see that apart from 
the extreme island state just described, which took place in the forty days around MJD 51230, there is 
another extreme island state, with much sparser sampling, in which a higher hard color is reached. 
If we take a quick look at the timing of this MJD 51380 data we find that when the source reaches this 
higher hard color, the source shows a power spectrum very similar to that of interval A of 4U 1608--52.
As for the MJD 51230 extreme island state, during this MJD 51380 extreme island state the count rate and the 
soft color increase simultaneously and in the color--color diagram another horizontal track is drawn up 
above the MJD 51230 one. 

Returning now to 4U 1608--52 we note that in the state transition of 4U 1608--52 during the 1998 outburst 
(see Figure \ref{fig.cc_deel2}) we observe the same phenomenon of a horizontal track being traced out in the 
extreme island state. The source is in interval C (continuous time 
interval 12 in Fig. \ref{fig.cc_deel2}) for two observations, the first of which has a count rate about a factor 
two higher than the second. This higher count rate is accompanied by a 5 \% higher soft color in the first observation. 
The power spectrum is the same in both observations.
So, like the case of 4U 1705--44, the change in count rate is accompanied by a correlated change in soft color which 
traces out a small almost (the hard color changes slightly, see above) horizontal track in the color--color 
diagram (continuous time interval 12 in Fig. \ref{fig.cc_deel2}). 

A possible explanation for this behavior of atoll sources in the extreme island state is the 
so--called ``secular motion'' in the color--color diagram. This phenomenon was first observed in Z sources, in which 
the Z--shaped track in the color--color and color--intensity diagrams is traced out within 
several hours up to a day. On longer timescales the whole Z track shifts both in soft color and count rate 
\citep[e.g.][]{hasinger90,kuulkers94,jonker00,jonker02,homan02,muno02a}. 
The timing properties remain mostly unaffected by these shifts and are primarily determined by the position along the Z track 
\citep[e.g.][]{kuulkers94,jonker00,jonker02,homan02}. The same phenomenon has been observed in the 
banana state of the atoll source 4U 1636--53 \citep{prins97,disalvo03}. The horizontal tracks in the extreme island 
state of the atoll sources may be entirely caused by a secular motion, 
similar to that in the Z sources. Because the sources remain in one particular island state (similar timing properties 
and hard color) for a long time (weeks to months), the slow process of secular motion has time to draw up a horizontal branch. In this horizontal 
branch the timing remains similar, as is the case for Z source secular motion. As noted by \citet{olive03}, this aspect 
of the behavior can be explained by the scenario that \citet{vdk01} proposed to explain the parallel track phenomenon in the 
intensity versus lower kilohertz QPO frequency diagram. 

It seems that the Z shape in the color--color diagram of some of the atoll sources is caused by 
transitions between the banana and the extreme island states that occur at different 
soft color. Figure \ref{fig.schema} provides a schematic of what in our interpretation
occurs in these sources. Several different horizontally extended extreme island branches appear 
above each other in the color--color diagram at different hard color values. These are traced out 
at different epochs. To first order only the hard color determines the timing properties. 
Within each extreme island state the soft color changes in correlation with intensity. During state transitions the 
extreme island state is entered or left at a soft color value depending on intensity. 
A possible explanation for the fact that the timing remains similar while the intensity (and thus soft color) changes 
and that the state transitions can occur at different intensities (soft colors) is 
the scenario of \citet{vdk01} where the truncation radius of the disk, which determines the timing 
properties and therefore the state, is not set by the accretion rate through the disk but by the ratio 
of this accretion rate over its long--term average \citep[c.f.][]{olive03}. 
So, if a source shows a state transition to an extreme island state at a minimum intensity which increases after that
\citep[as happened in the MJD 51230 state transition of 4U 1705--44][]{muno02a,barret02,olive03}, the source will 
enter the extreme island state at a lower soft color and move to a higher soft color as the intensity increases.
If a state transition to an extreme island state takes place during a decay after an outburst of a transient source, 
we would expect that after the extreme island state is entered at a particular soft color, the source then moves to 
a lower soft color as it fades \citep*[this and the reverse state transition from an extreme island to the banana 
state during the outburst rise were observed in Aql X--1, see][]{reig03}.
In our interpretation, the shape we observe in the color--color diagram of 4U 1608--52 
is caused by observations of the source in extreme island states at different intensities 
and therefore at different soft colors. Extreme island states A and B are observed at
higher intensities than extreme island state C, causing C to have a lower soft color.
Contrary to the case in 4U 1705--44 and Aql X--1 we do not observe a large change in soft color (or intensity)
within each extreme island state and it is unclear whether this is because the source 
was not observed sufficiently long or dense enough in these extreme island states, or 
whether this is a property of 4U 1608--52.

\subsection{Comparison with other sources; timing}

\subsubsection{Atoll sources, low luminosity bursters, and black hole candidates}
\label{comp:timing:atoll}

The three atoll sources, 4U 1608--52, 4U 1608--52 and 4U 0614+09, we have studied up to now \citep[][this paper]{vstr02} 
show a similar and very distinct timing behaviour (Figures \ref{fig.freq_freq} and \ref{fig.rms_all}). 
4U 1608--52 differs from 4U 0614+09 and 4U 1728--34 in that 4U 1608--52 is a transient source whose intensity 
changes by about 2 orders of magnitude where this is only about 1 order of magnitude for 4U 0614+09 and 
4U 1728--34. Although 4U 1608--52 has a much larger range in intensity ($\sim0.006-1.0$ Crab in the 2.0--16.0 
keV range) than 4U 0614+09 ($\sim0.017-0.1$ Crab), 4U 1608--52 and 4U 0614+09 
show a similar range in characteristic frequencies for the different power spectral components.
For 4U 1608--52 $\nu_u$ ranges from 216 to 1061 Hz (excluding intervals I and J for which the identification 
of $\nu_u$ is uncertain see appendix \ref{sec.detailed_ps}). This range in $\nu_u$ is reached with a minimum intensity 
range of $\sim0.05-0.1$ Crab (see Fig. \ref{fig.hid_sid_int}). For 4U 0614+09 a similar $\nu_u$ range of 233 to 1067 Hz
is reached with a minimum intensity range of $\sim0.03-0.05$ Crab. 
So, both sources reach similar frequency ranges within a similar change in intensity.
4U 1728--34 has not yet shown a similar range in characteristic 
frequencies as 4U 0614+09 and 4U 1608--52 \citep[until now $\nu_u > 400$ Hz;][]{disalvo01,vstr02}. Also no
2.0--16.0 keV intensities were published for the intervals used in \citet{disalvo01} and \citet{vstr02}.

There are also some differences in the timing behaviour; in particular in L$_{b}$ when it appears as a QPO.
As a function of $\nu_u$, $\nu_b$ is slightly higher in 4U 1608--52 than in 4U 1728--34 and 4U 0614+09
but it covers the same range. The $Q$ values for L$_{b}$ here are similar for the three sources. 
The relation between $\nu_b$ and $\nu_u$ in Figure \ref{fig.freq_freq} has a turnover 
at 30 Hz in 4U 0614+09 \citep{vstr00}; 
there might be a turnover at 45 Hz in 4U 1728--34, but in 4U 1608--52 no turnover has been observed.
The plot of rms fractional amplitude of all components versus $\nu_u$ (Fig. \ref{fig.rms_all}) 
shows an offset between the relations for the three atoll sources 4U 1608--52, 4U 0614+09, and 4U 1728--34.
This offset is different for the low--frequency features, L$_{hHz}$, L$_h$, L$_{b}$, L$_{b2}$, and at 
low frequencies L$_\ell$, than for the two kilohertz QPOs (see also \S \ref{sec.comb_ps}). 
It seems that there are two groups, one containing the high and the other the low--frequency features. 
The difference between the two groups might be due to geometrical effects.
The different offsets in Figure \ref{fig.rms_all} hint towards the Lorentzian at $\sim200$ Hz, found in the 
lowest frequency states of 4U 1608--52 (interval A) and 4U 0614+09, being L$_u$ (see appendix \ref{sec.detailed_ps}), 
whereas it could have been identified as either L$_u$ or L$_{hHz}$ based on 
its frequency (see appendix \ref{sec.detailed_ps}). The identification of the $\sim200$ Hz Lorentzian is important because the power spectra 
of 4U 1608--52 and 4U 0614+09 at the lowest inferred mass accretion rate closely resemble those of the low luminosity bursters
1E 1724--3045, GS 1826--24, and SLX 1735--269 \citep{bpk02} and the millisecond X--ray pulsar SAX J1808.4-3658 \citep{wk98}. 
These sources all seem to be atoll sources that are only observed at low mass accretion rate. If the 
$\sim200$ Hz Lorentzian is the upper kilohertz QPO then the high--frequency peaks in those sources are probably also upper 
kilohertz QPOs. 
Note that although in 4U 1608--52 L$_h$ already becomes undetectable at $\sim20$ Hz whereas this happens 
at $\sim45$ Hz in 4U 1728--34 and 4U 0614+09, this is not a significant difference as the upper limits on the 
fractional rms of L$_h$ in 4U 1608--52 are consistent with this component still being present at those frequencies. 

The broad L$_\ell$ component found in intervals A and B was previously found in 4U 1608--52 with Ginga by \citet{yoshida93} 
and has also been identified in the atoll source 4U 0614+09 \citep{vstr02}. This component has also been found 
in the low--luminosity bursters 1E 1724--3045, GS 1826--24, and SLX 1735--269 \citep{bpk02}. The Lorentzian has been tentatively 
identified with the lower kilohertz QPO based on extrapolations of frequency--frequency relations 
\citep[see][]{pbk99,bpk02,vstr02}. The rms fractional amplitude of the broad L$_\ell$ is much higher than that of the 
lower kilohertz QPO (see Fig. \ref{fig.rms_all}), however, a similar increase in rms is visible in that Figure for L$_u$.
But note that L$_u$ is detected in all intervals where L$_\ell$ is present as the lower kilohertz QPO in intervals H--E then 
dissapears in intervals D and C and only shows up again as the broad Lorentzian in intervals B and A. 

Interval B of 4U 1608--52 shows a narrow QPO with a characteristic frequency (2.458 Hz) between those of 
L$_{b}$ and L$_{h}$. This QPO is probably the same as the one discovered in the island state 
of 4U 1608--52 by \citet{yoshida93}: their power spectrum of the 1989 August 25--26 interval is very similar to our interval B, 
it has L$_{b}$ with a characteristic frequency of $\sim1$ Hz, L$_h$ at 
$\sim6$ Hz and a narrow QPO at $\sim2$ Hz \citep[see Figure 4 of][]{yoshida93}. The low--luminosity bursters 
1E 1724--3045, GS 1826--24 \citep{bpk02} as well as the BHCs GX 339--4 \citep*{n02} and Cyg X--1 \citep{pott02} also show 
narrow QPOs with a characteristic frequency between those of 
L$_{b}$ and that of L$_h$. For 1E 1724--3045 the centroid frequency of the narrow QPO 
coincides with the centroid frequency of L$_h$, but for 4U 1608--52 this is not the case; here 
the centroid frequency of the narrow QPO ($2.45\pm0.04$ Hz) is close to half the centroid frequency of L$_h$ 
($4.58\pm0.28$ Hz). No similar narrow Lorentzians were fitted by \citet{vstr02} for 4U 1728--34 or 4U 0614+09, but 4U 1728--34 
does show narrow residuals between the characteristic frequency of L$_{b}$ and that of  L$_h$. 
In Figure \ref{fig.narrowqpo} we plot $\nu_h$ versus the 
$\nu_{\rm max}$ of these narrow QPOs. For Cyg X--1 many narrow QPOs were fitted by \citet{pott02}; we 
only plot those QPOs that have a characteristic frequency between $\nu_b$ and $\nu_h$.
The points of the low--luminosity bursters and most of those of GX 339--4 line up in Figure \ref{fig.narrowqpo}.
These QPOs were all labeled as L$_{\rm LF}$ by \citet{bpk02}. 
The two points of GX 339--4 that deviate from the line are from two observations where GX 339--4 showed two 
simultaneous QPOs; the highest frequency QPO falls on the relation in Figure \ref{fig.narrowqpo} 
while the lowest frequency QPO falls below it. 
If we fit the points of the low--luminosity bursters and GX 339--4 with a power law our result as well as those of \citet{yoshida93} 
for 4U 1608--52 fall below this relation. So we can not identify this narrow QPO in 4U 1608--52 with L$_{\rm LF}$.
In the two observations where GX 339--4 showed two narrow simultaneous QPOs, the characteristic 
frequencies of the lowest frequency QPO as well as the results for Cyg X--1 also fall well below the power law. It might be that 
these QPOs and those of 4U 1608--52 are related. These points fall close to a line that indicates half the low--luminosity burster 
relation which is also plotted in Figure \ref{fig.narrowqpo}. Note that if we use centroid frequency instead of $\nu_{\rm max}$ 
the relations in Figure \ref{fig.narrowqpo} worsen.

\subsubsection{Z sources}

In addition to comparing the color--color diagram tracks and the associated power spectra, 
we can also compare the timing properties of Z sources with those of atoll sources by making similar 
plots of $\nu_{\rm max}$ of the different power spectral components versus $\nu_u$
as we did for the atoll sources 4U 1608--52, 4U 0614+09 and, 
4U 1728--34 in Figure \ref{fig.freq_freq}. In Figure \ref{fig.atoll_z} we do this for the Z sources GX 5-1 \citep{jonker02}, 
GX 340+0 \citep{jonker00}, GX 17+2 \citep{homan02} and Cyg X--2 \citep{kuz02}. We include the lower kilohertz QPO, 
the Low--Frequency Noise (LFN), the Horizontal Branch Oscillations (HBO), and the harmonic and sub--harmonic of the HBO, 
and plot these versus $\nu_u$. The grey symbols are the results of the atoll sources 
also displayed in Figure \ref{fig.freq_freq}, the black symbols represent the Z sources.
Note that the broad--band noise in these Z sources was not fitted with a zero--centered 
Lorentzian but with a cutoff power law, $P(\nu)\propto \nu^{-\alpha} e^{-\nu/\nu_{cut}}$, for GX 5-1, GX 340+0, and GX 
17+2 and with a smooth 
broken power law, $P(\nu)\propto \nu^{-\alpha} [1 + (\nu/\nu_{b})^{\beta}]^{-1}$, for Cyg X--2.
This leads to expressions for $\nu_{\rm max}$, defined as the frequency of maximum power density in $\nu{\rm P}_{\nu}$, 
of $(1-\alpha)\nu_{cut}$ for the cutoff power law, and 
$[(1-\alpha)/(\beta + \alpha -1)]^{1/\beta} \nu_{b}$ for the smooth broken power law.

From Figure \ref{fig.atoll_z} we confirm the identification of the HBO in the Z sources with L$_h$ in the atoll sources made by 
\citet{pbk99}. However, the suggestions that either the LFN \citep{vdk94b,jonker00} or the sub--HBO \citep{wk99,jonker00} in the 
Z sources might be similar to the classical broad--band noise in the island state of the atoll sources seems not to be 
supported by the relation of the characteristic frequencies of these components with $\nu_u$. 
At $\nu_u \approx 750$ Hz the band--limited noise component in the L$_{b}$--L$_u$ relation is replaced by a QPO 
(see \S \ref{sec.comb_ps}) and the band--limited noise component L$_{b2}$ appears and follows a new relation with $\nu_u$.
The LFN points fall below the L$_{b}$---L$_u$ relation but seem to line up with L$_{b2}$. The sub--HBO points 
fall above the L$_{b}$---L$_u$ relation until $\nu_u \approx 750$ Hz where the QPO takes over 
from the band--limited noise component (see above). 

Based on this comparison of frequencies, the LFN might still be 
associated with L$_{b2}$ and the sub--HBO might be related with L$_{b}$ when it is a QPO, although for 4U 0614+09 and 
4U 1728--34 no harmonic relation was present between L$_h$ and L$_{b}$ when it is a QPO \citep{vstr02}. 
Note also that the LFN in GX 17+2 follows a different relation compared to
that in the other Z sources. Other differences between GX 17+2 and the other Z sources in Figure \ref{fig.atoll_z} are that the 
LFN in GX 17+2 is peaked, the harmonic of the HBO is relatively strong and it shows a flaring branch oscillation (FBO) whereas the 
other sources show a flat LFN, a relatively weak harmonic of the HBO and no FBO \citep{jonker02}.
We note that none of these differences with the other Z sources serve to make GX 17+2 more similar to the other atoll sources.

\subsubsection{Summary}

The results presented in sections \ref{sec.comb_ps}, appendix \ref{sec.detailed_ps}, and this section show that the low--frequency part of the power spectra 
of both the atoll and the Z sources behaves in a very complex manner. For the atoll sources many different Lorentzian components 
appear, disappear or change from broad Lorentzians into narrow QPOs. Although this behavior is complex, the power spectral 
components of the three atoll sources for which we have performed a multi--Lorentzian timing study show remarkable 
similarities. By using frequency vs. frequency plots such as the one in Figure \ref{fig.freq_freq}, and by comparing the power spectra of 
different sources directly with each other it is possible to identify all these components within a single classification. We do this 
here for 4U 1608--52 (section \ref{sec.comb_ps}). 
Note however, that sometimes additional narrow features become significant in the power spectra (see \ref{comp:timing:atoll}). 
As another example, a recent RXTE observation of 4U 1608--52 during an outburst in 2001, shows 
a power spectrum (Figure \ref{fig.nupnu_2001}) very similar to that of interval B (Fig. \ref{fig.powspec_1608}), 
except that instead of the broad peak at $\sim 25$ Hz, a narrow ($Q = 5.5$) QPO with an rms fractional amplitude of $4.3\pm0.6 \%$ 
appears at 30 Hz. Also several, marginally significant, features are visible between 2 and 5 Hz.

If we compare the low--frequency features of the atoll sources with those 
found in the Z sources, in a frequency vs. frequency plot we see that only the HBO of the Z sources can be 
unambiguously identified with the L$_{h}$ of the atoll sources. The frequency vs. $\nu_u$ relations of the other low--frequency features partly 
overlap in Figure \ref{fig.freq_freq} but are not identical, and the relation of the characteristic frequency of the LFN with 
$\nu_u$ is not even the same for all Z sources. It appears that for some components the frequency vs. $\nu_u$ relations are not 
universal. There might be small physical differences between the atoll sources and Z sources (and within the Z sources themselves) 
that affect these relations.
So, either these features are different phenomena in different types of sources, or something else (e.g. neutron star mass, 
magnetic field strength) affects the relations. If this is so, this hidden parameter does not change much from source to 
source, or the relations do not depend strongly on this parameter as the relations are similar, as the relations are very similar 
within each source type.  

\subsection{QPO models}

Three different characteristic frequencies can be used to test the 
frequency relations predicted by the several QPO models; $\nu_{\rm max}$, $\nu_0$ and $\nu_{\rm damped}$. 
$\nu_{\rm max}$ is the frequency at which a Lorentzian contributes most of its variance 
per log frequency \citep{bpk02} and is the one used in this paper. $\nu_0$ is the centroid frequency of
the Lorentzian and can be obtained from our fit parameters $\nu_{\rm max}$ and $Q$ as
$\nu_0 = 2 \nu_{\rm max} Q/\sqrt{4 Q^2 + 1}$. 
According to \citet{titarchuk02} the $\nu_0$ of the Lorentzian is shifted with respect to the eigenfrequency 
of the oscillation due to damping. This shift depends on the damping rate which can be estimated from the
width of the Lorentzian \citep{titarchuk02}. We can calculate this eigenfrequency, $\nu_{\rm damped}$,
from $\nu_{\rm max}$ and $Q$ as $\nu_{\rm damped} = \nu_{\rm max} \sqrt{1 + 1/(4 Q^2 + 1)}$. In this section we 
use our fit parameter $\nu_{\rm max}$ unless stated otherwise.

The transition layer model \citep[TLM][]{ot99,to99} associates the lower 
kilohertz QPO with the Keplerian frequency of the inner disk edge ($\nu_K$). 
Between the neutron star and the Keplerian disk a transition layer is present.
In the TLM the upper kilohertz QPO is produced by radial oscillations of a blob 
thrown out of the transition layer into a magnetosphere. This radial eigenmode or hybrid 
frequency ($\nu_{hybrid}$) relates to $\nu_K$ as 
$\nu_{hybrid} = (\nu_K^2 + 4 \Omega^2)^{1/2}$, where $\Omega$ is the rotational frequency of the star's magnetosphere 
near the equatorial plane. 
This implies that $\nu_{hybrid}$ should always exceed $2 \Omega$. For 4U 1608--52
\citet{ot99} found that $\Omega$ is always larger than 300 Hz and thus $\nu_{hybrid}  > 600$ Hz should apply. 
For 4U 1608--52 we find that when there is a pair of kilohertz QPOs present 
(intervals E--H) the upper kilohertz QPO frequency ranges from 830 to 1062 Hz. But the 200--682 Hz single kilohertz QPO 
of intervals A--D, which we identify as the upper kilohertz QPO based on frequency--color \citep{mendez99,vstr00,disalvo01}, 
frequency--frequency (Fig. \ref{fig.freq_freq}) and rms--frequency \citep[see][and Fig. \ref{fig.rms_all}]{mendez01} 
correlations, fall well below the 600 Hz predicted by the TLM model. 
Note that neither the use of $\nu_0$, nor $\nu_{\rm damped}$, shifts these frequencies above the 600 Hz limit. If these single kilohertz 
QPOs were not the upper, but the lower kilohertz QPOs, this would lead to the unlikely scenario that the presence of an upper 
kilohertz QPO would lead to a completely different power spectrum for a similar lower kilohertz QPO frequency (compare, 
e.g., C with E or D with G in Fig. \ref{fig.powspec_1608}). 

The $\nu_h$ vs. $\nu_\ell$ relation for neutron stars and BHCs (see \citealt{pbk99} for the relation in the 
$\nu_0$ representation, and \citealt{bpk02} for the relation in the $\nu_{\rm max}$ representation) 
was recently extended towards 
lower frequencies by including 17 white dwarf sources \citep[see][and references therein]{mauche02,warner03}. 
The points of 4U 0614+09, 4U 1728--34, and 4U 1608--52 fall on the $\nu_h$ vs. $\nu_\ell$ relation; only the point 
from interval B of 4U 1608--52 deviates.
The $\nu_h$ vs. $\nu_\ell$ relation can be fitted with a power law with an index close to 1 \citep{pbk99,mauche02}. 
The TLM explains this by assuming that $\nu_\ell$ is the Keplerian frequency of the inner disk (see above), and $\nu_h$ represents 
the frequency of magnetoacoustic oscillations, $\nu_{\rm MA}$, in the disk transition layer. The relation between $\nu_h$ and 
$\nu_\ell$ is then a result of a global relation between $\nu_{\rm MA}$ of the transition layer and $\nu_K$ at the 
adjustment radius \citep{tit_wood02}. 
For each individual source the TLM predicts an index for the $\nu_{\rm MA}$ vs. $\nu_K$ relation that is steeper than
one and should be studied separately \citep{tit_brad_wood01}. Note, however, that the atoll sources cover a large 
range of the $\nu_h$ vs. $\nu_\ell$ relation, as there are points around $\nu_\ell = 20$ Hz and around $\nu_\ell = 550$ Hz.
While 4U 1608--52 only contributes points around $\nu_\ell = 20$ Hz and 4U 1728--34 only around $\nu_\ell = 550$ Hz,
4U 0614+09 contributes both.
This suggests that the \citet{pbk99} relation is not just a global relation between different sources as suggested by
\citet{tit_wood02} but can also be found within individual sources.
Note also that although the TLM explains the \citet{pbk99} relation as above in \citet{tit_wood02}, 
the HBOs of the Z sources, which are part of \citet{pbk99} relation, are explained differently, namely 
as the vertical eigenmode of a blob rotating with a Keplerian frequency thrown into the magnetosphere 
(see above) in other TLM papers \citep[e.g.][]{titarchuk02}.

In the sonic point beat frequency model \citep*[SPBFM;][]{miller98} the upper kilohertz QPO represents the 
Keplerian frequency at the inner disk edge. The lower kilohertz QPO then arises from a beat of the Keplerian 
frequency at the inner disk edge with the neutron star spin frequency. This led to an early prediction 
that $\Delta \nu$, the frequency difference between the upper and the lower kilohertz QPO, is equal to the spin 
frequency and should therefore be constant. Observations of several sources, among which 4U 1728--34 \citep{mendezvdk99} 
and 4U 1608--52 \citep{mendez98}, showed a significant decrease in $\Delta \nu$ when the upper kilohertz QPO frequency 
increased. Further refinements of the SPBFM by \citet{lamb01} could explain this decrease in $\Delta \nu$. Neither the 
use of $\nu_{\rm max}$ nor $\nu_{\rm damped}$ instead of $\nu_0$ makes $\Delta \nu$ constant. Note that our measurements
of the high--frequency kilohertz QPOs ($\nu_\ell > 500$ Hz) are not the best to use here as the addition 
of a lot of data leads to an artificial broadening of the Lorentzian (see \S \ref{sec.comb_ps}).
Therefore here we used the precise measurements of these high--frequency kilohertz QPOs made by \citet{mendezvdk99} 
for 4U 1728--34 and by \citet{mendez98} for 4U 1608--52 to calculate $\nu_{\rm max}$ and $\nu_{\rm damped}$.
For both sources $\Delta \nu$ increases by less than 1.5 \% if $\nu_{\rm max}$ is used instead of $\nu_0$ and 
by less than 3.1 \% if $\nu_{\rm damped}$ is used instead of $\nu_0$.
Note that, just as for the TLM, the low frequencies we find for the upper kilohertz QPO (see above) 
are a problem for the SPBFM. 
In the SPBFM the lower bound on the upper kilohertz QPO frequency is set by
the Keplerian frequency at the maximal radius where the radiation coming from the neutron star surface can still 
remove sufficient angular momentum from the gas in the disk so that it falls supersonically to the neutron 
star surface \citep{miller98}. This radius is about $3R_{ms}$, where $R_{ms}$ is the radius 
of the marginally stable orbit.
So, to reach $\nu_u < 300$ Hz the mass of the neutron star has to exceed  $\sim2.5$ $M_{\odot}$.

The relativistic precession model \citep[RPM;][]{stella98,stella99} assumes that the upper kilohertz QPO represents the 
Keplerian frequency of the inner disk. The lower kilohertz QPO frequency represents the periastron precession frequency 
of the accretion disk, which is assumed to contain slightly elliptical orbits.
This leads to the prediction that the frequency difference ($\Delta \nu$) between the two kilohertz QPOs decreases both 
at low and high kilohertz QPO frequencies \citep[see Figure 1 of][]{stella99}. If the broad L$_\ell$ at low frequencies 
\citep[intervals A and B of 4U 1608--52 and interval 1 of 4U 0614+09 in][]{vstr02} is the lower kilohertz QPO 
(but see \S \ref{comp:timing:atoll}), then this predicted decrease in $\Delta \nu$ at low kilohertz QPO frequencies is 
observed in 4U 0614+09 and 4U 1608--52. 
Note that the point of interval B of 4U 1608--52 falls above the curves of Figure 1 in \citet{stella99} by $3\sigma$. 
More complicated modelling within the framework of the RPM as is done for Sco X--1 in \citet{stella99} is beyond the scopes 
of this paper.

The RPM also predicts a low--frequency QPO at the Lense--Thirring precession frequency of 
$\nu_{LT} = 4.4\times10^{-8} I_{45} M^{-1} \nu_s \nu_K^2$ Hz \citep{stella98}, where $I_{45}$ is the moment of inertia 
in units of $10^{45}$ g cm$^2$, $M$ is the mass of the neutron star in $M_{\odot}$, and $\nu_s$ is the neutron star 
spin frequency. We searched our data for a power law with an index of 2 as predicted by the RPM by fitting the 
frequency vs. $\nu_u$ relations for the low 
frequency features (Fig. \ref{fig.freq_freq}) for the three sources 4U 1728--34, 
4U 0614+09 and 4U 1608--52 with power laws. We used $\nu_{\rm max}$, 
$\nu_0$, and $\nu_{\rm damped}$ for the characteristic frequencies (see above). 
We find that the $\nu_h$ vs. $\nu_u$ relation has an index that is the closest to the 2 predicted by the RPM. 
If we use $\nu_{\rm max}$ or $\nu_{\rm damped}$ for the frequency the relations clearly deviate from a power law.
The points fall below the fitted power law at low and high frequencies and above at intermediate frequencies, leading to 
very large $\chi^2$/dof of 4605/21 ($\nu_{\rm max}$) and 7179/21 ($\nu_{\rm damped}$).
If we use $\nu_0$ for the frequency we find that the $\nu_h$ vs. $\nu_u$ relation can be described
by a power law of the form $\nu = 5.4 (\pm0.4) \times 10^{-5}\nu_u^{2.01(\pm0.02)}$ (see Figure \ref{fig.powerlawfit}).
The $\chi^2$/dof of this fit is still high (373/20) but this is not due to systematic deviations from a power law.
The averaged relative scatter around the fitted power law is only 6 \% on the $\nu_h$ axis and only 3 \%
on the $\nu_u$ axis. The high $\chi^2$/dof is probably a result of not including any systematic errors 
when determining the errors of the power spectral fit parameters \citep[see][]{ford98}.
Note, that if we use $\nu_0$ for the frequency we exclude 
those points that have $\nu_0 = 0$ (from fits with a zero--centered Lorentzian).
The result that the $\nu_h$ vs. $\nu_u$ relation can be described by a power law with an index close to 2 was found 
previously for a more limited frequency range ($\nu_u > 400$ Hz) in 4U 1728--34 \citep[index 2.11;][]{ford98} and 
4U 0614+09 \citep[index 2.46;][]{vstr00}. It is remarkable that now that we have extended the $\nu_h$ vs. $\nu_u$ relation
towards $\nu_u \approx 200$ Hz, the relation can still be described by a power law and that the power law index is found 
to be even closer to 2.
We can get an indication of the spin frequency by looking at the burst oscillation frequency 
which is 363 Hz for 4U 1728--34 \citep{stroh96} and 620 Hz for 4U 1608--52 \citep{muno02b}. 
If these two frequencies are the spin frequency of the neutron star in each case 
\citep[the most likely scenario, see][]{chakrabarty03}, then it is difficult to understand 
why the $\nu_h$ vs. $\nu_u$ relations of these sources coincide. If we assume that we see the 
spin frequency in 4U 1728--34 and twice the spin frequency in 4U 1608--52, we find an average spin frequency of about 
335 Hz for these sources. For the $\nu_0$ representation fit, this leads to $I_{45}/M = 3.7$ which is too large for proposed 
equations of state \citep[the acceptable range is $0.5<I_{45}/M<2$;][]{stella98}. If we assume that we see the spin frequency 
in 4U 1608--52 and half the spin frequency in 4U 1728--34, the average spin frequency is about 
670 Hz for these sources. This leads to $I_{45}/M = 1.8$ which is acceptable.

\subsection{The energy spectrum of 4U 1608--52}

Recently, \citet{gier02b} studied the X--ray energy spectra of 4U 1608--52 as a function of position in the color--color 
diagram. They find that, similar to other atoll sources, the spectra of the island state (our intervals A, B and C) of 
4U 1608--52 are dominated by a hard power law spectrum. There is also a soft component present which in their view is probably 
due to the neutron star surface rather than the accretion disk. The banana state (our intervals E--J) spectrum is soft and 
it is uncertain whether the soft component is due to the neutron star surface or the disk. To explain the spectral behavior 
of 4U 1608--52 \citet{gier02b} use a scenario where in the island state the disk is far from the surface of the neutron 
star. The large radius of the inner disk prevents the disk from being observed directly in the PCA spectra.
In the island/banana state transition the inner accretion flow, which was geometrically thick and optically thin in the 
island state, collapses into a  geometrically thin and optically thick disk. The inner edge of the disk is now close to the 
neutron star, and it is observed as the soft component and increases the soft color. From the DISKBB model, \citet{gier02b} 
estimate an inner disk radius in the banana state of $\sim30$ km. The inner accretion flow becomes more optically thick 
lowering the hard color. We can compare this scenario with the results of our timing study. If we assume that $\nu_u$ represents 
the Keplerian frequency at the inner disk edge \citep[e.g.][]{miller98} we find, assuming a neutron star mass of 
1.4 M$_\odot$ (for a different neutron mass the results change by a factor (M$_{NS}/1.4)^{1/3}$, which is 1.2
for a 2.4 M$_\odot$ neutron star), 
that the inner disk terminates at $47-28$ km for the island state (intervals A, 
B and C). For the banana state (intervals E--J) we find an inner disk radius of $\sim19-15$ km and decreasing towards 
higher soft color. Indeed we find that the inner disk radius is closer to the neutron star in the banana than in the island 
state, but the values in both the banana and the island state are close to the $\sim30$ km estimated by \citet{gier02b} for 
the banana state. So, the difference in radius between island and banana state seems too small to make the disk disappear 
from the PCA spectrum in the island state as  proposed in the scenario of \citet{gier02b} described above.
If we assume that instead of $\nu_u$, $\nu_{\ell}$ represents the Keplerian frequency at the inner 
disk edge \citep[e.g.][]{to99}, this would lead to much larger differences in inner disk radius between the 
island ($280-201$ km) and the banana state ($26-20$ km), so this interpretation could be consistent with the 
\citet{gier02b} scenario. The identification of $\nu_\ell$ as the Keplerian frequency at the inner 
disk edge is also suggested by the recently proposed  extension of the PBK relation \citep{pbk99,bpk02} towards 
lower frequencies by including 17 white dwarf sources \citep[see][and references therein]{warner03}. For the 
white dwarf sources a QPO is plotted versus the dwarf nova oscillation, DNO, the DNOs are thought to occur at 
the Keplerian frequency of the inner disk edge \citep[][and references therein]{warner02}. 

The behavior in the soft color vs. intensity diagram of 4U 1608--52 and 4U 1705--44 in 
their extreme island states can also be explained within the scenario of \citet{gier02b}. Let us assume that $\nu_{\ell}$ 
represents the Keplerian frequency of the inner disk edge. In the extreme island states the disk is far out at 200 km and 
can not be seen in the PCA energy spectrum (see above), the soft color should therefore only depend on the spectral properties 
of the neutron star. If we look at the extreme island states A and B of 4U 1608--52 and 4U 1705--44 in the soft color versus 
intensity diagram, there seems to be a one to one relation between soft color and intensity (see Fig. \ref{fig.hid_sid_int} 
for 4U 1608--52). We can fit this relation with a power law with an index of about 0.1. 
We propose that in the extreme island state as the accretion rate (intensity) increases, the neutron star surface 
gets more heated and the soft color increases. 
Outside the extreme island states the soft color is determined by the spectral properties of both the neutron star 
surface and the disk and its dependence on accretion rate is therefore more complicated.

\section{CONCLUSIONS}

The main conclusions of this paper can be summarized as follows:

\begin{itemize} 
 
\item[1]	The timing behavior of 4U 1608--52 is almost identical to that of the atoll sources 4U 1728--34 and 
		4U 0614+09, and can be described in terms of a well defined set of components 
		(L$_u$, L$_\ell$, L$_{hHz}$, L$_h$, L$_{b}$, L$_{b2}$) which all vary in characteristic frequency 
		together (except for L$_{hHz}$ for which $\nu_{\rm max}$ is constant), sharing the same frequency correlations.

\item[2]	The timing behavior is not consistent with the idea that 4U 1608-52 traces out 
		a three-branched Z shape in the color-color diagram along which the timing properties vary gradually as
		in Z sources. Instead, the power spectral properties change smoothly along 
		an ``$\epsilon$--shaped'' track, which is probably composed of three, partially observed, 
		extended extreme island branches.

\item[3]	4U 1608--52, at its lowest frequencies, shows 
	        a power spectrum almost identical  to that of the low luminosity bursters 1E 1724--3045, GS 1826--24, 
		and SLX 1735--269, and the millisecond X--ray pulsar SAX J1808.4-3658. The low luminosity bursters
		and the millisecond X--ray pulsar
		all seem to be atoll sources at low mass accretion rate.
		A similar conclusion was reached previously based on 4U 0614+09 \citep{vstr02}.

\item[4]	The high--frequency peak at about 200 Hz in interval A of 4U 1608--52 is more likely to be the upper 
		kilohertz QPO than the hectohertz Lorentzian. This means that the corresponding $\sim 200$ Hz 
		peaks in 4U 0614+09, 1E 1724--3045, GS 1826--24, SLX 1735--269 and SAX J1808.4-3658
		also most likely represent the upper kilohertz QPO.

\item[5]	The low--frequency part of the power spectra of both the atoll and the Z sources behaves in a very complex 
		manner. Nevertheless, for the atoll sources it is possible to identify all these components within a single 
		classification. If we compare the low--frequency features of the atoll sources with those found in the Z 
		sources we find that the HBO of the Z sources can be identified with L$_h$ of the atoll sources. Based 
		upon the frequency relations with the upper kilohertz QPO neither the LFN nor the 
		sub--harmonic of the HBO of the Z sources can straightforwardly 
		be identified with the band--limited noise of the atoll sources. 

\item[6]	The parallel tracks in the frequency versus intensity diagram can be identified in the intensity versus 
		color diagrams. However, the statistical spread in the colors, which shows up as the narrow nearly vertical 
		parallel tracks in the color intensity diagrams, cause the $\nu-$color correlation to be veiled on short 
		timescales (less than hours) so in the color intensity diagrams parallel tracks similar to the parallel QPO 
		tracks can only be identified on longer timescales.

\item[7]	We have tested the transition layer model, the sonic point beat frequency model, and the relativistic 
		precession model using our results for the three atoll sources 4U 0614+09, 4U 1728--34, and 4U 1608--52.
		Neither the transition layer model nor the sonic point beat frequency model can explain the 
		upper kilohertz QPO frequency range we find in these sources. The $\nu_h$ vs. $\nu_u$ can be described 
		by a power law with an index of 2 as predicted by the relativistic precession model, however it is very likely 
		that the burst oscillation frequencies found for 4U 1728--34 (363 Hz), and 4U 1608--52 (620 Hz) represent the 
		spin frequency of the neutron star in each case. Then it is difficult to understand why the $\nu_h$ vs. $\nu_u$ 
		relations of these sources coincide.

\item[8]	Our timing and color results are consistent with the scenario  proposed by \citet{gier02b} to explain the
		results of their X--ray energy spectral study of 4U 1608--52, only if $\nu_{\ell}$, and not $\nu_u$
		represents the Keplerian frequency of the inner disk edge. In this scenario the disk is close to the 
		neutron star in the banana state, and far from the neutron star in the island state. The large radius of the 
		inner disk in the island state prevents the disk from being observed directly in the PCA spectra. 

\end{itemize}

\section{ACKNOWLEDGEMENTS}

This work was supported by NWO SPINOZA grant 08--0 to E.P.J. van den Heuvel, 
by the Netherlands Organization for Scientific Research (NWO), and by 
the Netherlands Research School for Astronomy (NOVA). 
This research has made use of data obtained through
the High Energy Astrophysics Science Archive Research Center Online Service, 
provided by the NASA/Goddard Space Flight Center. We would like to thank 
Peter Jonker for providing us with tables for the GX 5--1 and GX 340+0 data.

\appendix
\section{DETAILED DESCRIPTION OF THE POWER SPECTRA}
\label{sec.detailed_ps}

\begin{figure}
\figurenum{5}
\epsscale{0.9}
\plotone{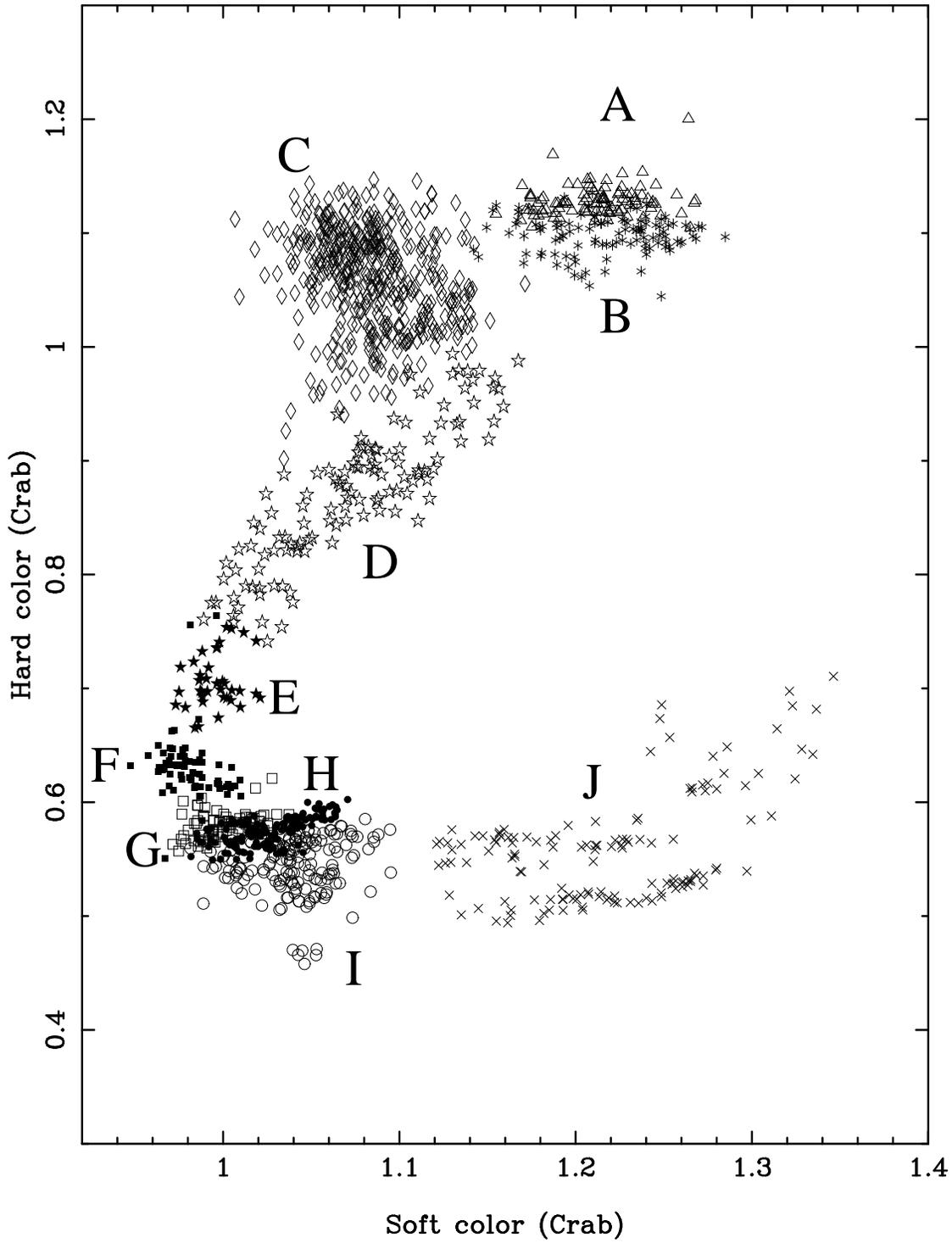}
\caption{Color--color diagram divided into 10 representative intervals marked from A to J. Colors are in units of 
Crab (see \S \ref{sec.obs}).}
\label{fig.cc_int}
\end{figure}

\begin{figure}
\figurenum{6}
\epsscale{0.9}
\plotone{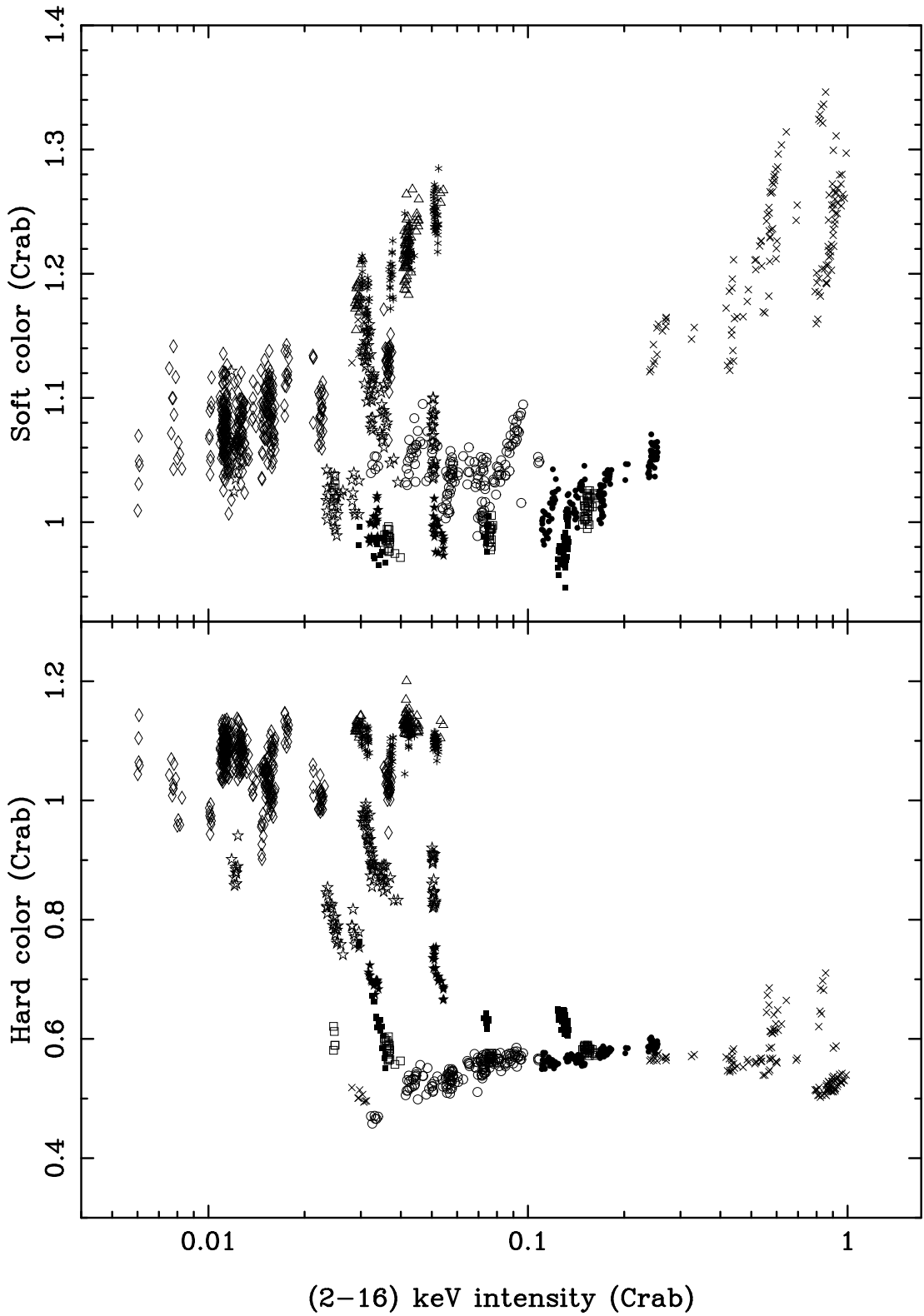}
\caption{Soft color (top panel) and hard color (bottom panel) plotted versus intensity. The symbols correspond to the 10 
representative intervals of Figure \ref{fig.cc_int} (see \S \ref{sec.comb_ps}). Intensity and colors are in units of Crab (see 
\S \ref{sec.obs}).}
\label{fig.hid_sid_int}
\end{figure}

We now provide a detailed description of the components found in the combined power spectra of 4U 1608--52 
and compare them to previous results on 4U 1728--34, 4U 0614+09 of \citet{vstr02} and the low luminosity 
bursters discussed by \citet{bpk02}. The combined power spectrum of {\bf interval A} (Fig. \ref{fig.powspec_1608}) 
is very similar to interval 1 of 4U 0614+09 
\citep[see Figure 2 in][]{vstr02}. It shows L$_{b}$ and L$_h$ 
at the lowest characteristic frequencies observed in atoll sources. Figure 
\ref{fig.freq_freq} confirms the identification of these components. As was the case in 4U 0614+09, 
interval A shows L$_\ell$ at $\sim15$ Hz and a broad Lorentzian at $\sim200$ Hz. 
The Lorentzian at $\sim200$ Hz found in 4U 1608--52 (interval A)
 and 4U 0614+09 can be identified as either L$_u$ or L$_{hHz}$ based on its 
frequency. However, this Lorentzian has a similar rms fractional amplitude for both 4U 1608--52 and 4U 0614+09. 
The upper kilohertz QPOs of 4U 1608--52 and 4U 0614+09 always have a similar fractional rms, whereas for L$_{hHz}$ 
the fractional rms is always about 5 percentage points less in 4U 1608--52 than in 4U 0614+09 
(see Fig. \ref{fig.rms_all}). This hints towards this component being L$_u$. 
{\bf Interval B} is similar to interval A, it shows L$_{b}$, L$_h$, L$_\ell$ and L$_u$, but all at
 higher characteristic frequencies. Interval B also shows a narrow QPO at 
2.458 Hz which will be discussed in \S \ref{comp:timing:atoll}. Intervals A and B both show a component designated L$_\ell$, 
at $\sim15$ and $\sim
25$ Hz respectively. The broad L$_\ell$ at these low frequencies can be tentatively identified with the 
lower kilohertz QPO based on extrapolations of frequency--frequency relations \citep[see][]{pbk99,bpk02,vstr02}. 
However, this Lorentzian might also be an unrelated feature (see \S \ref{comp:timing:atoll}). {\bf Interval C} shows only 
three components, 
L$_{b}$, L$_h$ and L$_u$ again all at higher frequencies than in the previous interval. {\bf Interval D} shows L$_{b}$, L$_{hHz}$ 
and L$_u$. Compared to interval C, L$_h$ is no longer present and L$_{hHz}$ has appeared.
Based on the similarities with 4U 1728--34 and 4U 0614+09, L$_h$ would be expected to be more narrow than in intervals A--C 
(expected $Q$ about 1--2) and have a characteristic frequency of about 30 Hz. Figure \ref{fig.powspec_1608} shows a 
non--significant narrow peak at $\sim21$ Hz in interval D which can probably be attributed to L$_h$. 
As noted above, L$_{hHz}$, a broad Lorentzian with a constant $\nu_{\rm max}$ of $\sim150$ Hz, appears in interval D for the 
first time. In interval A L$_{hHz}$ would lie on top of L$_u$ (see above) and in intervals B and C L$_{hHz}$ would lie in the flank of L$_u$.
Therefore it is hard to detect L$_{hHz}$ in intervals A--C. Note that there is some excess power at the expected location of 150 Hz 
in interval C. In interval D L$_u$ has moved above this frequency range so that L$_{hHz}$ is more easily 
detectable. Both the frequency 
and the fractional rms of L$_{b}$ in interval D are above the relations in Figures \ref{fig.freq_freq} and 
\ref{fig.rms_all}. Including the little peak at 
$\sim21$ Hz in the fit hardly affects L$_{b}$.
These deviations from the relations in Figures \ref{fig.freq_freq} and \ref{fig.rms_all}
are most likely due to frequency shifts inside interval D, which covers 
a large range in the color--color diagram (see Fig. \ref{fig.cc_int}). 
Splitting this interval further up would compromise the statistics.
{\bf Interval E} shows L$_{b}$, L$_{b2}$ and a pair of kilohertz QPOs. 
Here the character of L$_{b}$ has changed from a band--limited noise component (fitted with a zero--centered Lorentzian)
to a QPO. L$_{hHz}$  is visible in the power spectrum but is not 
significant (2.4 $\sigma$ single trial) in this interval. Because leaving it out of the fit makes L$_{b2}$ fit power up 
to hundreds of Hz leading to an overestimated $\nu_{b2}$ and rms fractional amplitude, we include
a L$_{hHz}$ at 150 Hz (the average value in 4U 1728--34 and 4U 0614+09) with $Q = 0.2$ (see above) in the
fit. In Table \ref{tbl.highfitpar} and in Figure \ref{fig.rms_all} we only show the upper limit for this 
L$_{hHz}$. The characteristic frequency of L$_{b2}$ in interval E is still high compared to the general relation between 
$\nu_{b2}$ and $\nu_u$ (see Fig. \ref{fig.freq_freq}). This is similar to what was observed in 4U 1728--34 where in this range 
there are also two points where $\nu_{b2}$ is significantly above the relation. 
These deviations all occur close to the L$_{b}$ ``transformation'' and when the L$_{b}$ QPO is still weak compared to the 
band--limited noise component L$_{b2}$ (see Fig. \ref{fig.rms_all}), so one possible explanation is that in these transition 
power spectra a narrow and a broad L$_{b}$ are merged together.
The power spectra of {\bf intervals F, G and H} all show very low--frequency noise (VLFN), and L$_{b}$, L$_{b2}$, 
L$_{hHz}$ and the pair of kilohertz QPOs all increasing in frequency. {\bf Intervals I and J} both show 
VLFN, L$_{hHz}$ and a broad kilohertz QPO. It is unclear whether the kilohertz peak represents 
the upper, the lower or even a blend of both kilohertz peaks. There were no other power spectral 
features present which could be used to identify this component with the help of the correlations of 
Figure \ref{fig.freq_freq}. 
If these peaks, with characteristic frequencies of about 1050 and 1250 Hz, would be the lower kilohertz 
QPO they would be among the highest frequency lower kilohertz QPOs found in any source up to date 
\citep[see e.g.][]{vdk00}. 
So to be slightly conservative and also because  it is more convenient, as in Figures \ref{fig.freq_freq} and \ref{fig.rms_all}
we use the characteristic frequency of L$_u$ to plot against, we list these kilohertz QPOs as L$_u$.
The $Q$ values for the kilohertz QPOs in intervals I and J are much lower than those in 4U 1728--34 and 
4U 0614+09. This is probably an artificial broadening due to the movement of the kilohertz QPO during the 
intervals. A finer subdivision of intervals I and J does not lead to observing L$_u$ and L$_\ell$ separately.
Note that no kilohertz QPOs were found in the individual observations of these intervals \citep{mendez98}.
Interval J shows 
an additional broad bump which is also seen in 4U 1728--34 
\citep[interval 18 and 19 of Figure 1 in][]{vstr02} and 4U 0614+09 \citep[interval 9 of Figure 2 in][]{vstr02}; 
this Lorentzian can be identified as either L$_{b}$, L$_{b2}$ or a new type of Lorentzian. Here we list it as L$_{b2}$.

The VLFN in 4U 1608--52 shows a similar behavior to that in other atoll sources. The power law index is about constant around 
1.5, with the exception of interval F where it is 2.5. The rms fractional amplitude generally increases with position 
in the color--color diagram. In intervals I and J the VLFN clearly deviates from just a power law. This leads to high 
$\chi^2$/dof values of 2.5 and 1.8.  Adding Lorentzians in interval I and J to fit the deviations leads to significantly 
better $\chi^2$/dof of 1.3 and 1.2. The $\nu_{\rm max}$ of the Lorentzian in interval I is then 0.02 Hz and in interval J 0.3 Hz
In interval I the Lorentzian does not only fit the deviation but almost the entire VLFN. The inclusion of the Lorentzians 
lowers the rms fractional amplitude of the power law to 1.59 \% for interval I and 2.27 \% for interval J.

\clearpage

\begin{center}
\begin{deluxetable}{lcccccc}
\tabletypesize{\small}
\tablewidth{0pt}
\tablecaption{\label{tbl.timingtour}} 
\tablehead{ 
\colhead{Interval} & \colhead{Duration} & \colhead{$\Delta t$} & 
\colhead{Intensity} & \colhead{Classification} \\
\colhead{Number} & \colhead{(days)} & \colhead{(days)} & 
\colhead{(Crab)} & \colhead{} \\ 
}

\startdata
\multicolumn{5}{c}{segment 1 of the lightcurve}\\
\hline
1 & 3 & 3 & 0.16--0.25 & H	\\
2 & 1 & 3 & 0.05 & D	\\ 
3 & 1 & 3 & 0.04--0.05 & I	\\
4 & 1 & 3 & 0.02--0.03 & D	\\
5 & 1 & 4 & 0.02 & C	\\
6 & 1 & 16 & 0.03--0.04 & D	\\
7 & 116\tablenotemark{1} & 84 & 0.03--0.04 & B	\\
8 & 64\tablenotemark{1} & 402 & 0.01--0.02 & C	\\
\hline
\multicolumn{5}{c}{segment 2 of the lightcurve}\\
\hline
9 & 43 & 5 & 0.24--1.20 & J	\\
10 & 6 & 1 & 0.08--0.21 & FGH	\\					  
11 & 1 & 1 & 0.05 & E	\\					  
12 & 2 & 1 & 0.02--0.04 & C	\\					  
13 & 1 & 1 & 0.03 & E	\\ 					  
14 & 5 & 2 & 0.05--0.10 & I	\\ 					  
15 & 1 & 7 & 0.03--0.04 & FG	\\					  
16 & 7 & 38 & 0.04--0.05 & B	\\					  
17 & 119\tablenotemark{1} & 523 & 0.03--0.05 & A	\\ 
\hline							  
\multicolumn{5}{c}{segment 3 of the lightcurve}\\		  
\hline							  
18 & 1 & 4 & 0.06--0.07 & I	\\ 
19 & 1 & 2 & 0.03 & D	\\
20 & 1 & 7 & 0.01 & C \\
21 & 1 & 3 & 0.03 & J	\\
22 & 1 & 9 & 0.01 & C	\\
23 & 1 & 12 & 0.02 & I	\\
24 & 1 & 4 & 0.01 & C	\\ 
25 & 1 & 4 & 0.02--0.03 & G	\\
26 & 1 & 4 & 0.02 & C	\\
27 & 1 & 4 & 0.07--0.08 & F	\\
28 & 6 & 2 & 0.03--0.11 & I	\\
29 & 1 & 2 & 0.01 & D	\\
30 & 1 & -- & 0.01 & C	\\

\enddata

\tablenotetext{1}{Large time gaps occured between observations (see Figs. \ref{fig.cc_deel2} 
and \ref{fig.cc_deel1}).}
\tablecomments{
For each continuous time interval as defined in \S 3 we list the duration in days (with a minimum 
value of 1 day), the number of days until the next interval ($\Delta t$), the intensity range in the 2--16 keV energy 
band, and a classification based on the power spectrum and the position of the source in the color--color 
diagram (see \S \ref{sec.comb_ps}). For the duration, 
$\Delta t$ and the intensity range the 256 s lightcurve obtained in \S 2 was used.}

\end{deluxetable}
\end{center}

\oddsidemargin -1.7cm

\begin{center}
\begin{deluxetable}{lcccccccccccccc}
\tabletypesize{\tiny}
\tablewidth{0pt}
\tablecaption{Low--frequency fit parameters of the multi--Lorentzian fit for 
4U 1608--52\label{tbl.lowfitpar}} 
\tablehead{ 
\colhead{Interval} & \multicolumn{2}{c}{VLFN} & \multicolumn{3}{c}{L$_{b}$} & 
\multicolumn{3}{c}{L$_{b2}$} & \multicolumn{3}{c}{L$_h$} \\
\colhead{Letter} & \colhead{rms} & \colhead{$\alpha$} & 
\colhead{$\nu_{\rm max}$} & \colhead{Q} & \colhead{rms} &
\colhead{$\nu_{\rm max}$} & \colhead{Q} &  \colhead{rms} &
\colhead{$\nu_{\rm max}$} & \colhead{Q} &  \colhead{rms} \\
\colhead{} & \colhead{(\%)} & \colhead{} & 
\colhead{(Hz)} &            & \colhead{(\%)} &
\colhead{(Hz)} & \colhead{} &  \colhead{(\%)} &
\colhead{(Hz)} & \colhead{} &  \colhead{(\%)} 
}
\startdata
A & -- & -- & 0.247$\pm$0.012 & 0 (fixed) & 11.03$\pm$0.36 & -- & -- & -- & 1.613$\pm$0.098 & 0.156$\pm$0.091 & 12.02$\pm$0.91 \\
B & -- & -- & 1.165$\pm$0.045 & 0 (fixed) & 9.97$\pm$0.19 & -- & -- & -- & 6.17$\pm$0.27 & 0.56$\pm$0.11 & 9.73$\pm$0.96 \\
C & -- & -- & 3.22$\pm$0.32 & 0 (fixed) & 9.62$\pm$0.67 & -- & -- & -- & 19.8$\pm$1.5 & 0.24$\pm$0.14 & 11.79$\pm$0.86 \\
D & -- & -- & 22.54$\pm$0.94 & 0 (fixed) & 15.09$\pm$0.25 & -- & -- & -- & -- & -- & $<6.1$\tablenotemark{1} \\
\hline
E & -- & -- & 22.77$\pm$0.52 & 3.8$^{+2.1}_{-1.4}$ & 4.46$^{+0.92}_{-0.64}$ & 30.9$\pm$4.9 &  0 (fixed) & 10.4$\pm$1.0 & -- & -- & $<2.0$\tablenotemark{1} \\
F & 0.840$\pm$0.044 & 2.49$\pm$0.16 & 32.11$\pm$0.76 & 1.15$\pm$0.13 & 5.99$\pm$0.28 & 9.9$\pm$2.8 & 0 (fixed) & 3.16$\pm$0.42 & -- & -- & $<2.8$\tablenotemark{1} \\
G & 0.740$\pm$0.078 & 1.41$\pm$0.16 & 37.1$\pm$1.1 & 1.37$\pm$0.22 & 4.83$\pm$0.31 & 10.6$^{+6.5}_{-3.7}$ & 0 (fixed) & 2.44$^{+0.59}_{-0.39}$ & -- & -- & -- \\
H & 1.180$\pm$0.025 & 1.537$\pm$0.041 & 47.4$\pm$1.1 & 2.08$\pm$0.35 & 2.68$\pm$0.18 & 15.4$^{+6.4}_{-4.1}$ & 0 (fixed) & 1.98$\pm$0.26 & -- & -- & -- \\
I & 3.587$\pm$0.039 & 1.436$\pm$0.016 & -- & -- & -- & -- & -- & -- & -- & -- & --\\
\hline
J & 2.446$\pm$0.019 & 1.317$\pm$0.010 & -- & -- & -- & 17.3$\pm$1.6  & 0.37$\pm$0.23  & 1.66$\pm$0.17  & -- & -- & -- \\
\enddata

\tablenotetext{1}{95 \% confidence upper limits (see \S \ref{sec.comb_ps}).}
\tablecomments{The fit parameters of the low--frequency part of the power spectra. Listed are the 
parameters of the very Low--Frequency noise component (VLFN), which is fitted with a power law, L$_{b}$, L$_{b2}$, and L$_h$ 
which are fitted with Lorentzians. The horizontal line between intervals D and E marks the ``transfer'' 
of L$_{b}$ from band--limited noise component into a QPO and the simultaneous appearance of L$_{b2}$
(see \S \ref{sec.comb_ps}). 
The horizontal line between intervals I and J marks the broad Lorentzian in interval J for which it is unclear
whether it is L$_{b}$, L$_{b2}$ or a new component (see \S \ref{sec.detailed_ps}).  
L$_{b}$ and L$_{b2}$ were fitted with a zero--centered Lorentzian ($Q = 0$)
when they showed their band--limited noise like behavior. 
For the power--law $\alpha$ is the power law indices and the fractional rms is integrated from 0.01 to 1 Hz. For the 
Lorentzians we show the characteristic frequencies ($\equiv 
\nu_{\rm max}$), $Q$ values ($\equiv \nu_{\rm 0}/2\Delta$) and the integrated fractional rms (from 0 to $\infty$ Hz, 
over the full PCA energy band). 
Not listed is the narrow Lorentzian found in interval B, this Lorentzian has $\nu_{\rm max} = 2.458\pm 0.037$, 
$Q = 7.5^{+4.2}_{-2.1}$ and rms $= 1.83\pm0.27$. The quoted errors use $\Delta\chi^{2}$ = 1.0.\\
}

\end{deluxetable}
\end{center}

\begin{center}
\begin{deluxetable}{lccccccccc}
\tabletypesize{\tiny}
\tablewidth{0pt}
\tablecaption{High--frequency parameters of the multi--Lorentzian fit for 
4U 1608--52\label{tbl.highfitpar}} 
\tablehead{
\colhead{Interval} &
\multicolumn{3}{c}{L$_{hHz}$} & 
\multicolumn{3}{c}{L$_\ell$} & \multicolumn{3}{c}{L$_u$} \\
\colhead{Letter} &
\colhead{$\nu_{\rm max}$} & \colhead{Q} &  \colhead{rms} &
\colhead{$\nu_{\rm max}$} & \colhead{Q} &  \colhead{rms} &
\colhead{$\nu_{\rm max}$} & \colhead{Q} &  \colhead{rms} \\
\colhead{} &
\colhead{(Hz)} & \colhead{} &  \colhead{(\%)} &
\colhead{(Hz)} & \colhead{} &  \colhead{(\%)} &
\colhead{(Hz)} & \colhead{} &  \colhead{(\%)}  
}
\startdata
A & -- & -- & -- & 14.74$\pm$0.82 & 0.05$\pm$0.10 & 15.76$\pm$0.87 & 216$\pm$34 & 0 (fixed) & 14.10$\pm$0.58\\
B & -- & -- & $<9.1$\tablenotemark{2} & 24.2$\pm$3.9 & 0 (fixed) & 12.48$\pm$0.55 & 309$\pm$13 & 0.93$\pm$0.14 & 12.29$\pm$0.52\\
\hline
C & -- & -- & $<12.6$\tablenotemark{2} & -- & -- & -- & 474$\pm$21 & 0.77$\pm$0.14 & 18.50$\pm$0.73\\
D & 189$\pm$23 & 0.90$\pm$0.29 & 8.91$\pm$0.94 & -- & -- & -- & 682.5$\pm$7.3 & 2.96$\pm$0.27 & 16.95$\pm$0.54\\
E & -- & -- & $<10.3$\tablenotemark{2} & 531$^{+11}_{-17}$ & 6.1$^{+3.6}_{-2.4}$ & 6.9$\pm$1.0 & 830.3$\pm$5.8 & 6.11$\pm$0.67 & 13.64$\pm$0.54\\
F & 134.1$\pm$6.1 & 1.32$\pm$0.27 & 5.36$\pm$0.34 & 602.2$\pm$1.3 & 19$^{+14}_{-5}$ & 8.3$^{+1.5}_{-0.4}$ & 896.0$\pm$3.0 & 8.30$\pm$0.61 & 8.57$\pm$0.22\\
G & 157$\pm$18 & 1.17$\pm$0.49 & 3.94$\pm$0.49 & 668.1$\pm$2.3 & 16.8$^{+4.0}_{-2.1}$ & 9.91$^{+0.36}_{-0.20}$ & 973.1$\pm$6.3 & 10.2$\pm$1.6 & 5.53$\pm$0.32\\
H & 158.9$\pm$5.8 & 3.0$^{+1.4}_{-0.9}$ & 2.13$\pm$0.25 & 784.35$\pm$0.91 & 8.04$\pm$0.17 & 9.549$\pm$0.077 & 1061.9$\pm$6.3 & 12.5$\pm$2.8 & 3.16$\pm$0.23\\
I & 93.1$\pm$5.0 & 1.00$\pm$0.22 & 5.25$\pm$0.28 & -- & -- & -- &  1046$\pm$45\tablenotemark{1} & 2.83$\pm$0.75\tablenotemark{1} & 5.98$\pm$0.57\tablenotemark{1}\\
J & 123$\pm$25 & 0.37$\pm$0.23 & 1.66$\pm$0.17 & -- & -- & -- & 1252$\pm$95\tablenotemark{1} & 0.77$\pm$0.22\tablenotemark{1} & 3.61$\pm$0.26\tablenotemark{1}\\
\enddata

\tablenotetext{1}{The identification of this Lorentzian is not certain.}
\tablenotetext{2}{95 \% confidence upper limits (see \S \ref{sec.comb_ps}).}
\tablecomments{
The fit parameters of the high--frequency part of the power spectra. Listed are the parameters of L$_{hHz}$, L$_\ell$, 
and L$_u$ which are all fitted with Lorentzians. The horizontal line separates intervals A and B where 
L$_\ell$ is a broad low--frequency Lorentzian (see \S \ref{sec.comb_ps}). In intervals E--H L$_\ell$ is the lower kilohertz QPO.
Note however, that the identification of L$_\ell$ in intervals A and B as the lower kilohertz QPO is very tentative 
(see \S \ref{comp:timing:atoll}). For each Lorentzian we show 
the characteristic frequencies ($\equiv \nu_{\rm max}$), $Q$ values ($\equiv \nu_{\rm 0}/2\Delta$) and the 
integrated fractional rms (from 0 to $\infty$ Hz, over the full PCA energy band). The quoted errors 
use $\Delta\chi^{2}$ = 1.0.}

\end{deluxetable}
\end{center}

\clearpage

\onecolumn

\begin{figure}
\figurenum{7}
\epsscale{0.45}
\begin{tabular}{ccc}
\plotone{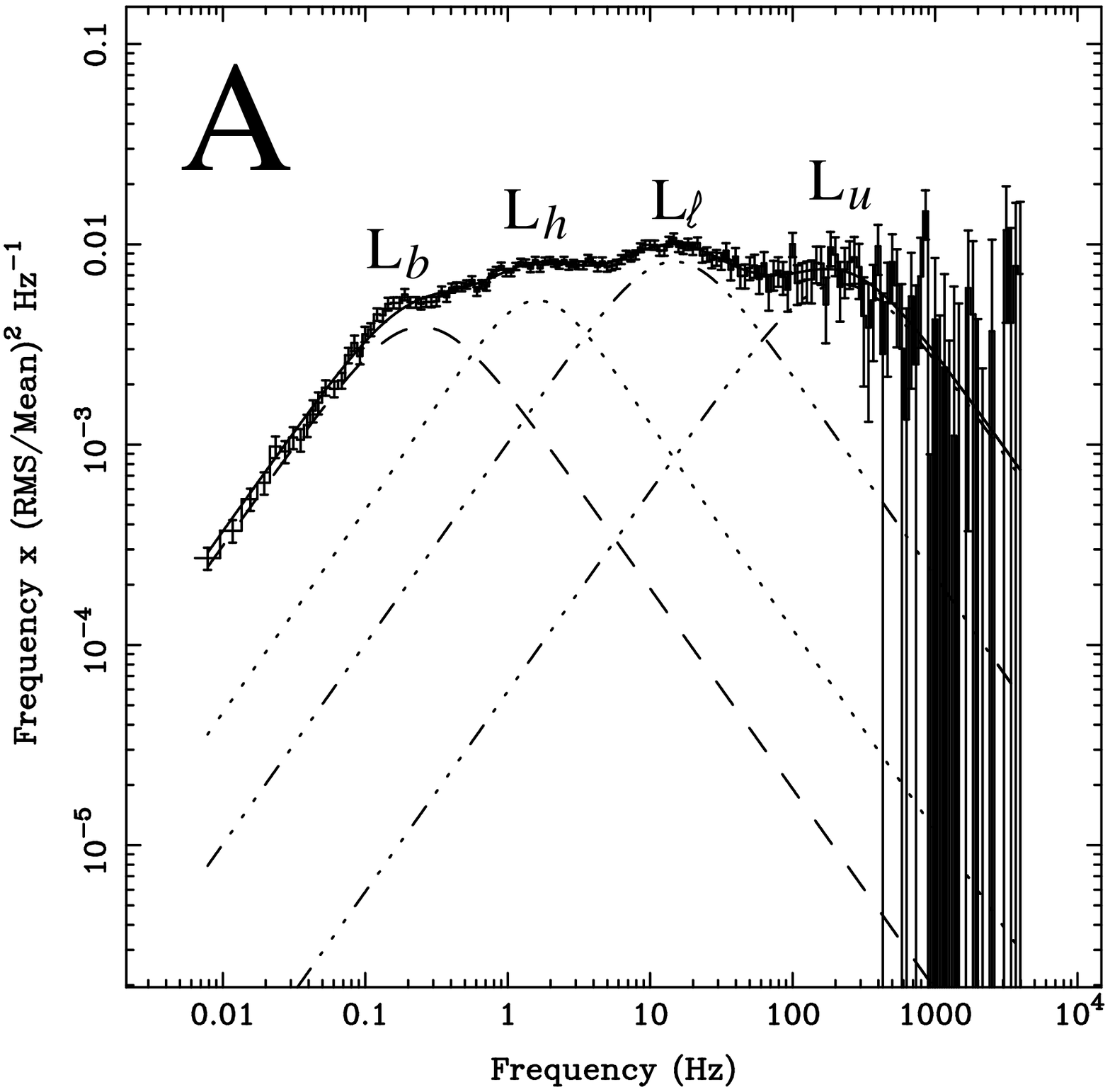} & 
\plotone{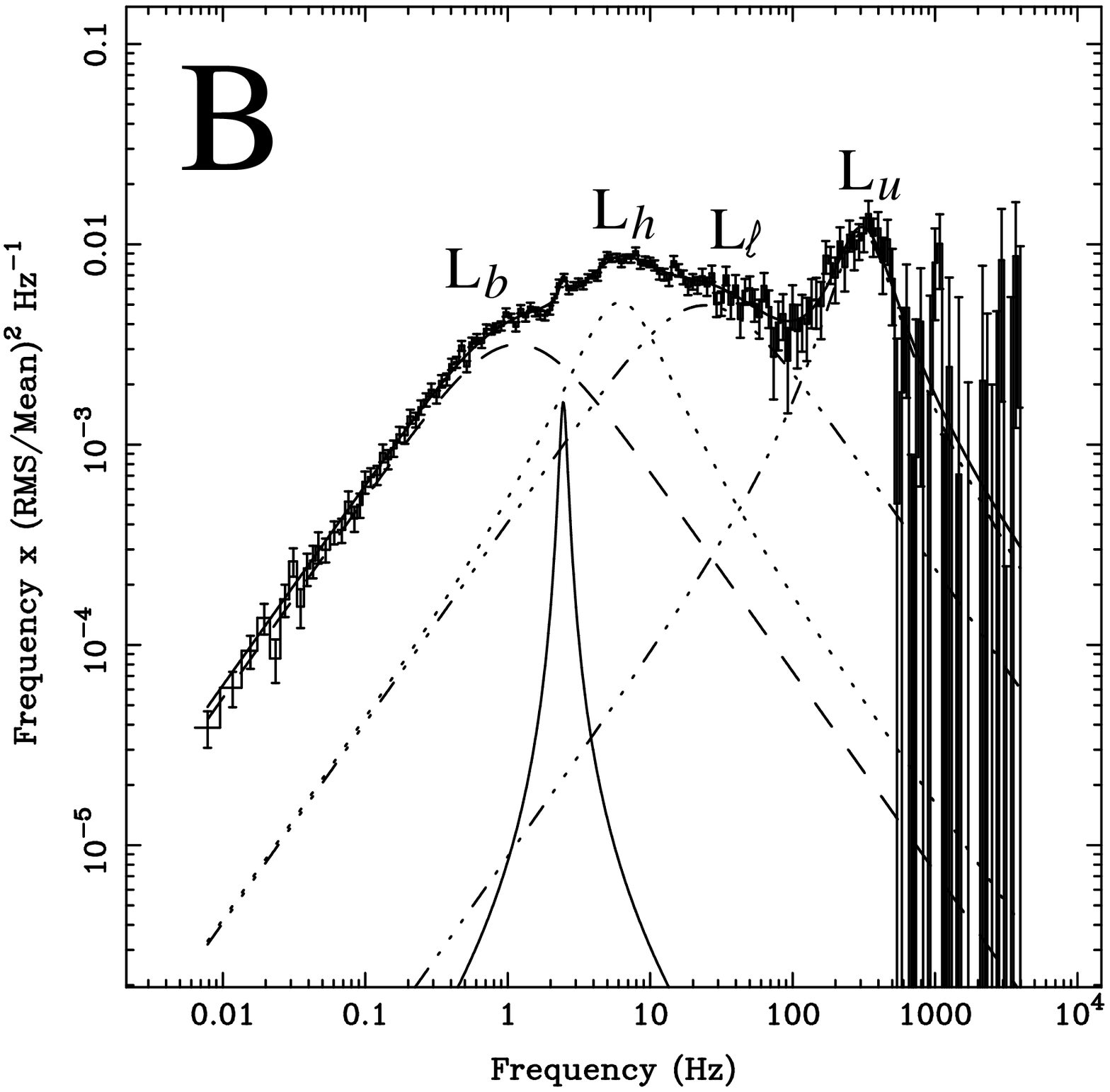} & \\
\plotone{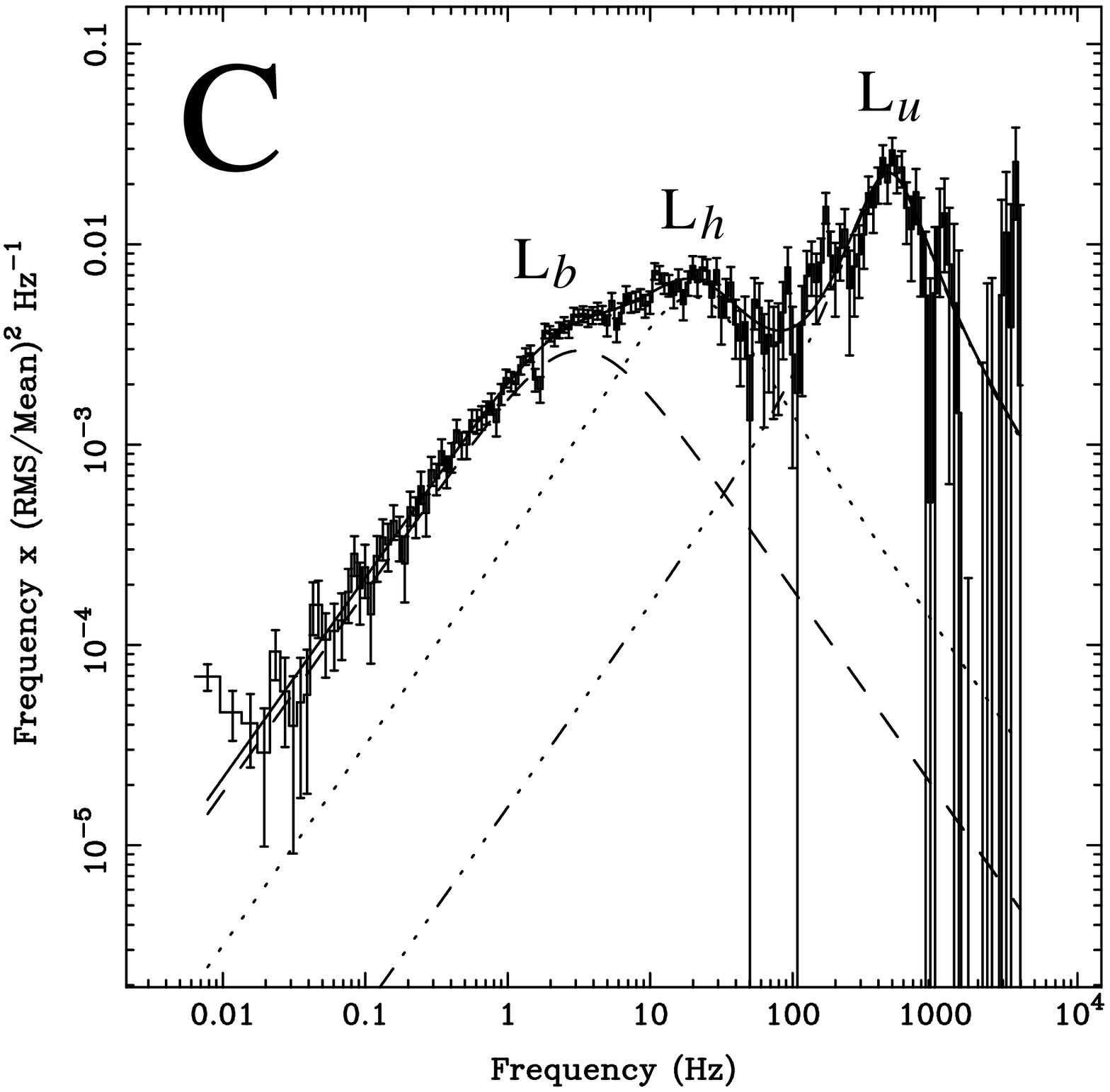} & 
\plotone{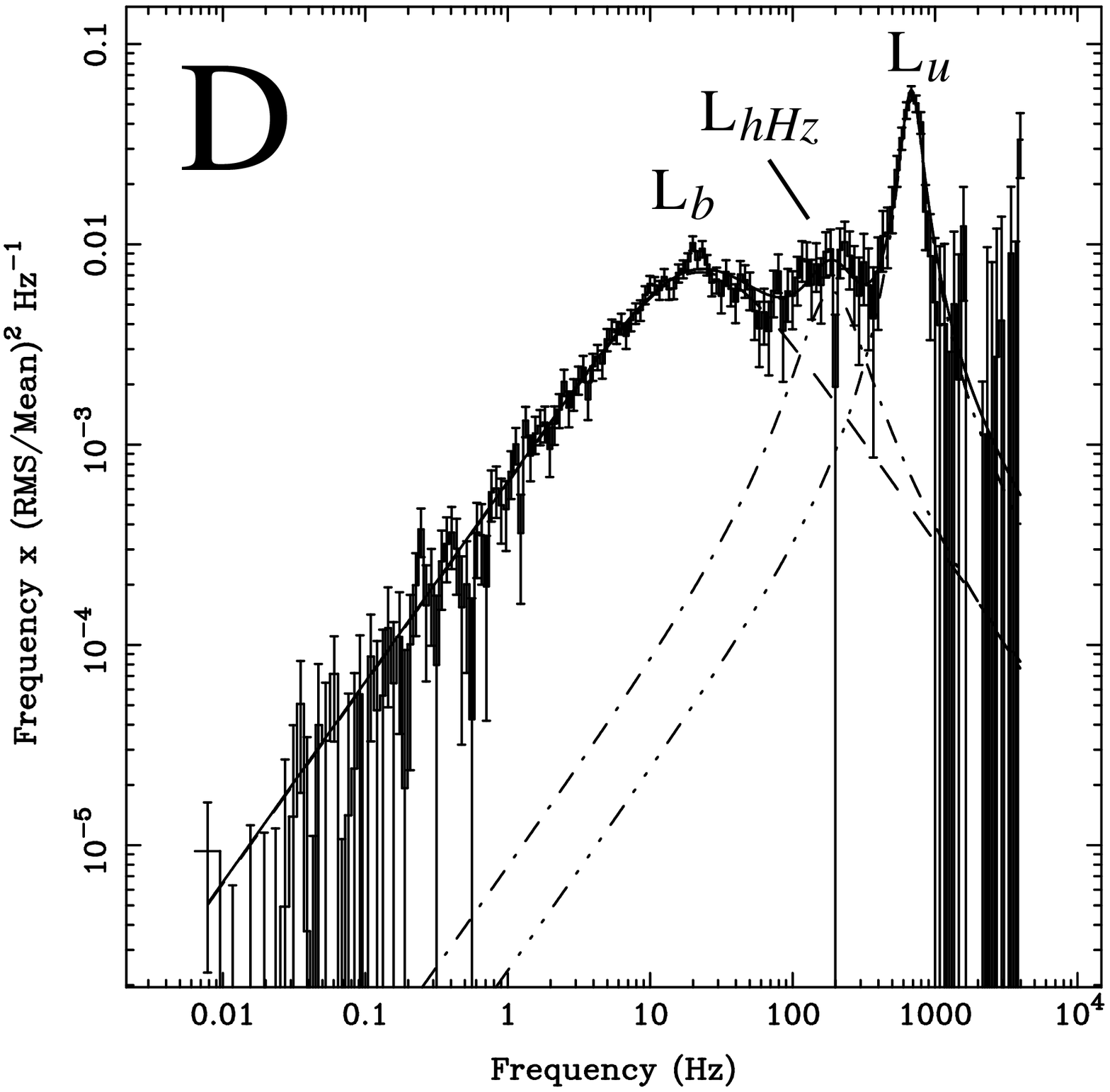} & \\
\end{tabular}
\caption{Power spectra and fit functions in the power spectral density 
times frequency representation (see \S 2) for 4U 1608--52. The different 
lines mark the individual Lorentzian components of the fit. The dashed 
lines mark both L$_{b}$ and L$_{b2}$, the dotted lines
L$_h$, the dash--dotted line L$_{hHz}$, 
and the dash--dot--dot--dotted line L$_\ell$ and L$_u$. 
In intervals F to J a power--law is included to fit the VLFN; this 
power--law is indicated with a solid line. A narrow Lorentzian found in 
interval B is indicated with a solid line. The power spectral components 
are also indicated in the plots. Interval letters are indicated.}
\label{fig.powspec_1608}
\end{figure}
\clearpage

\begin{figure}
\figurenum{7}
\epsscale{0.45}
\begin{tabular}{ccc}
\plotone{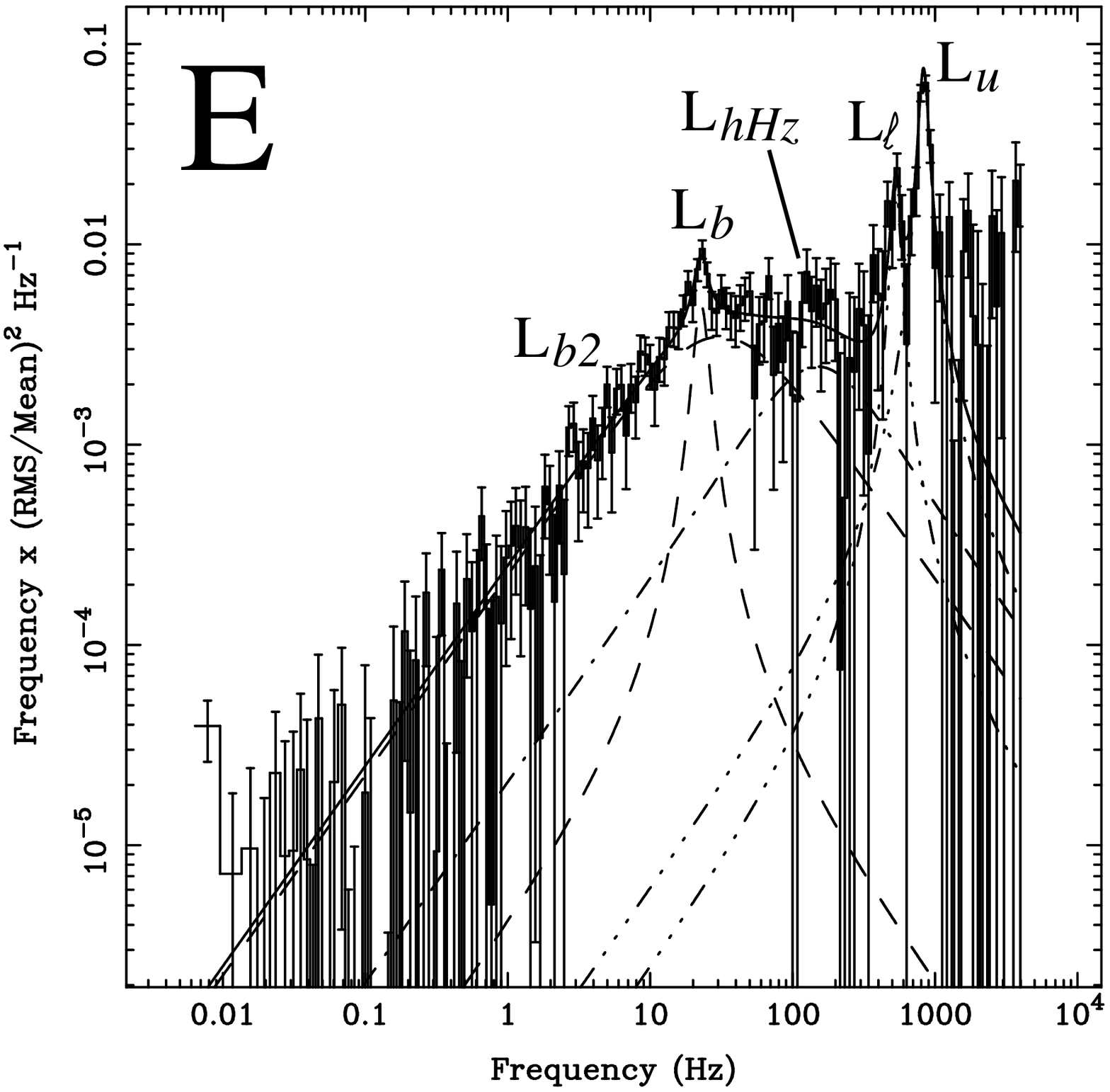} & 
\plotone{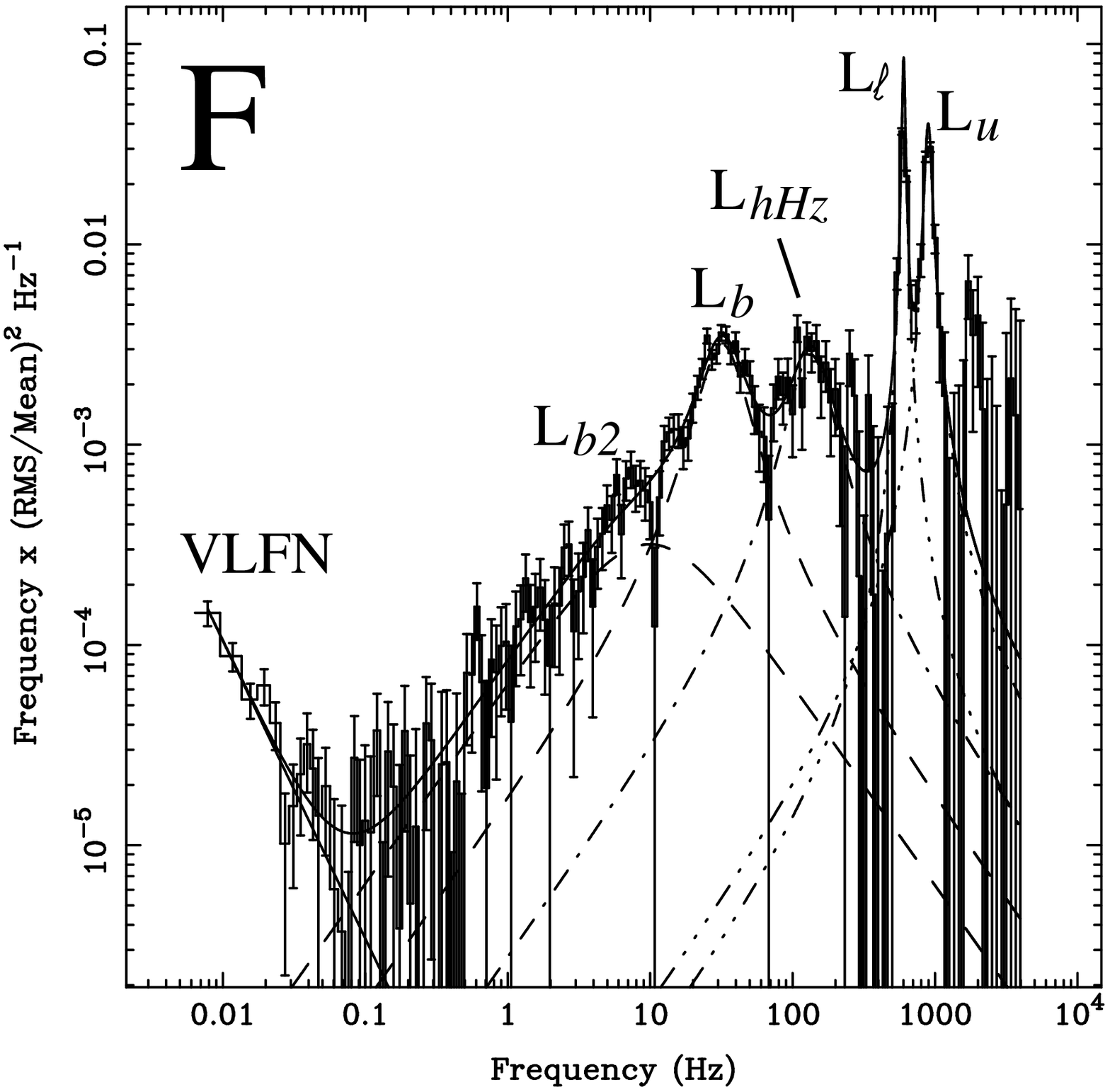} & \\
\plotone{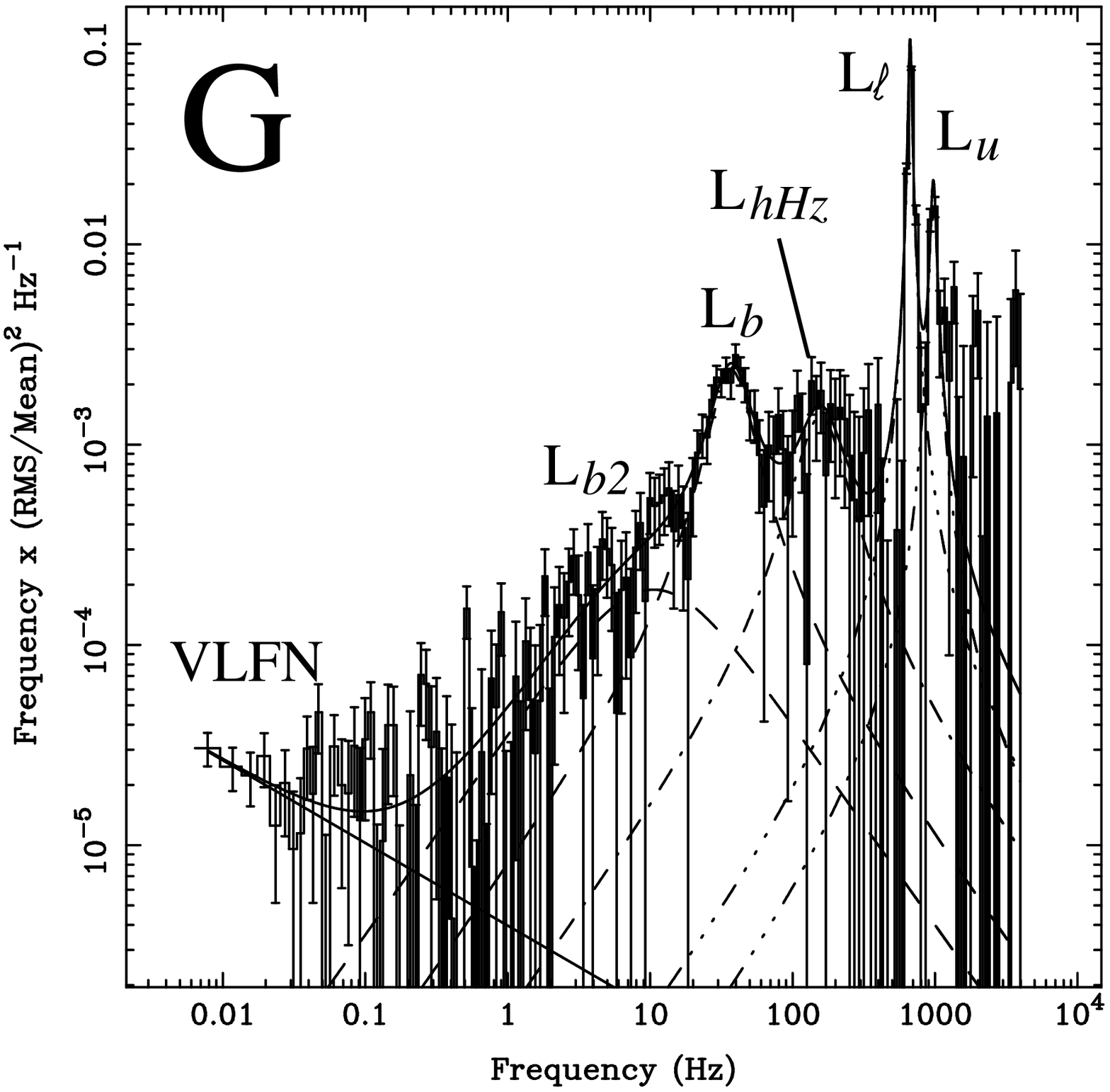} & 
\plotone{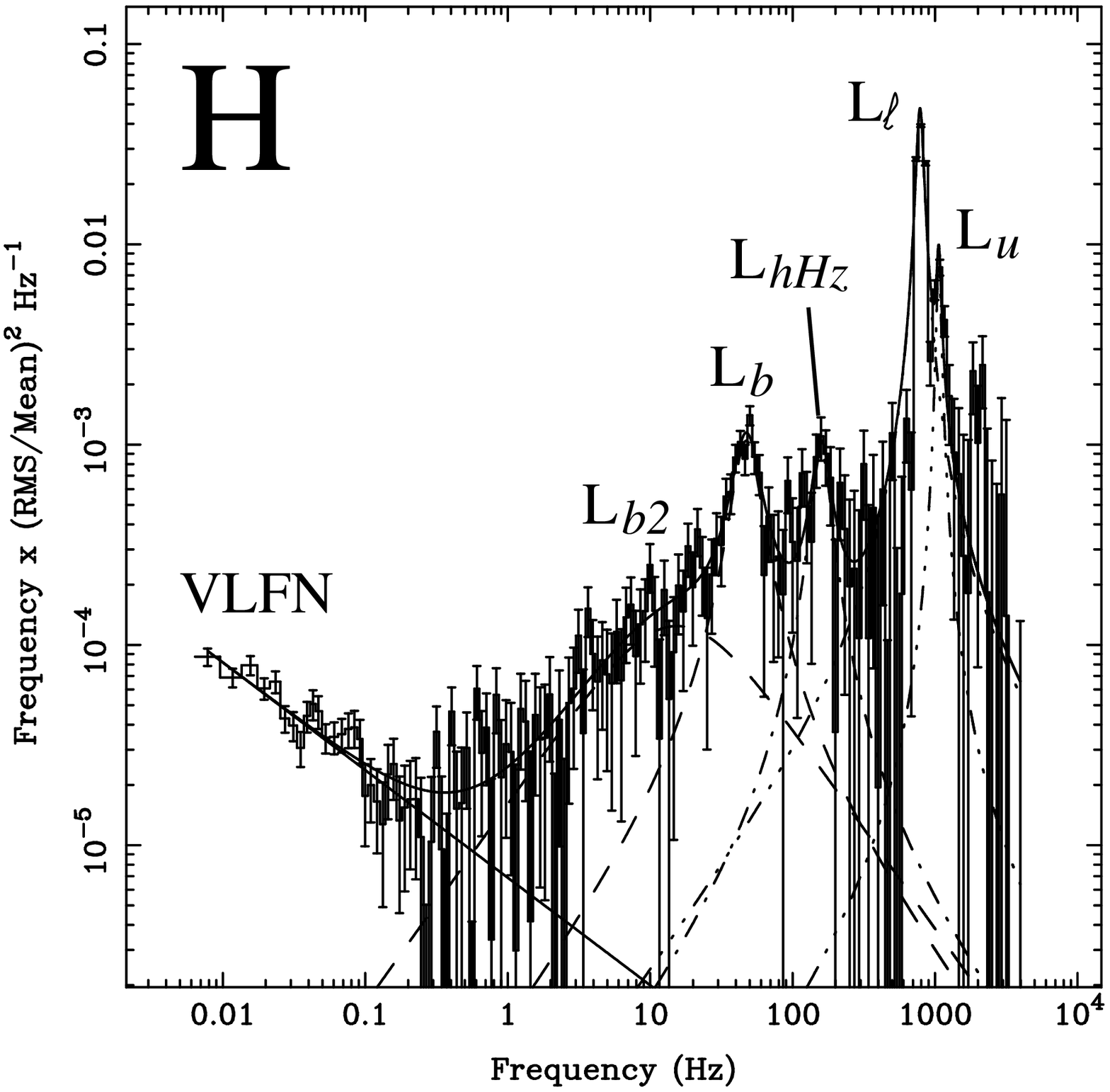} & \\
\end{tabular}
\caption{Continued}
\end{figure}
\clearpage

\begin{figure}
\figurenum{7}
\epsscale{0.45}
\begin{tabular}{ccc}
\plotone{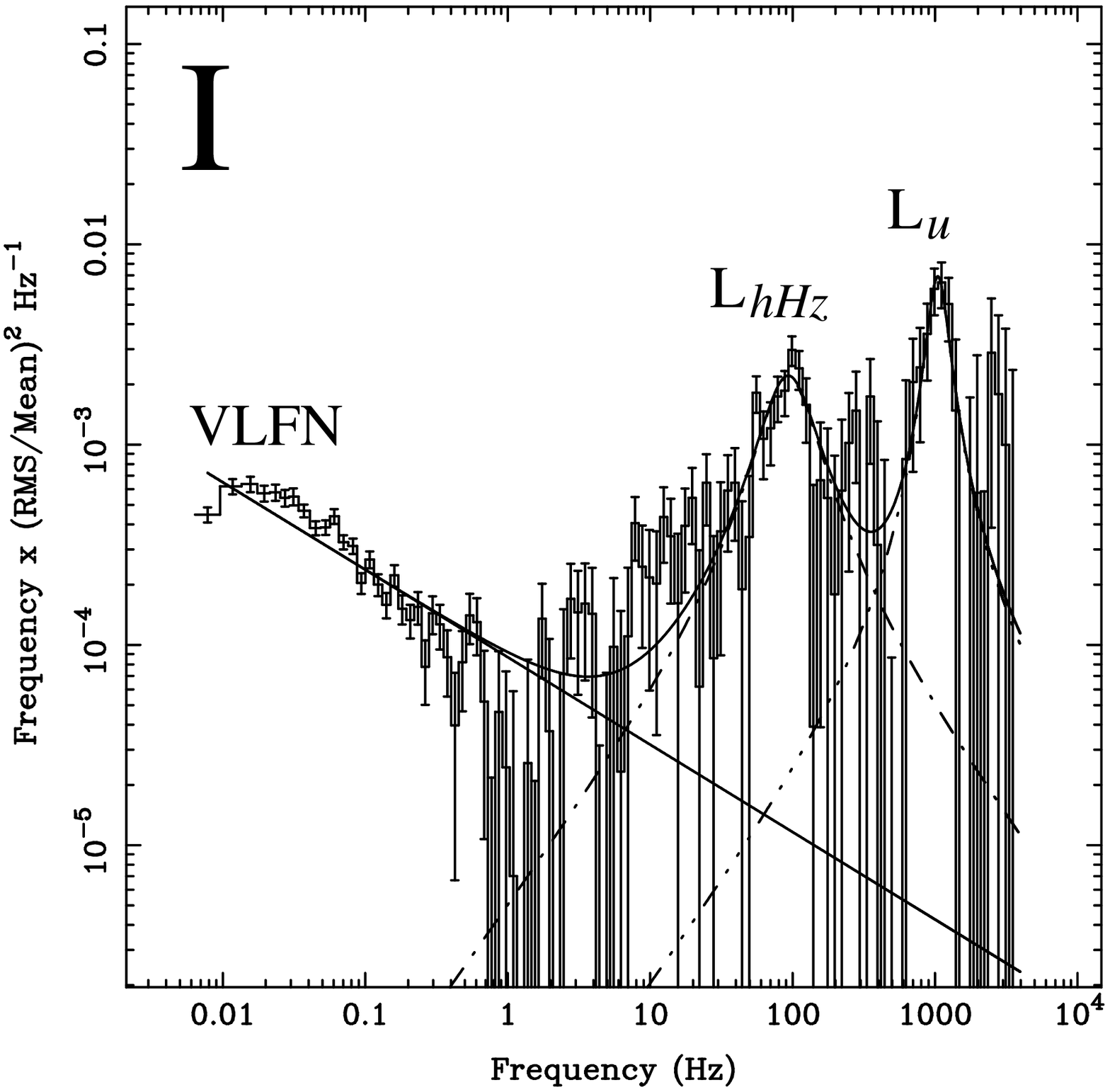} & 
\plotone{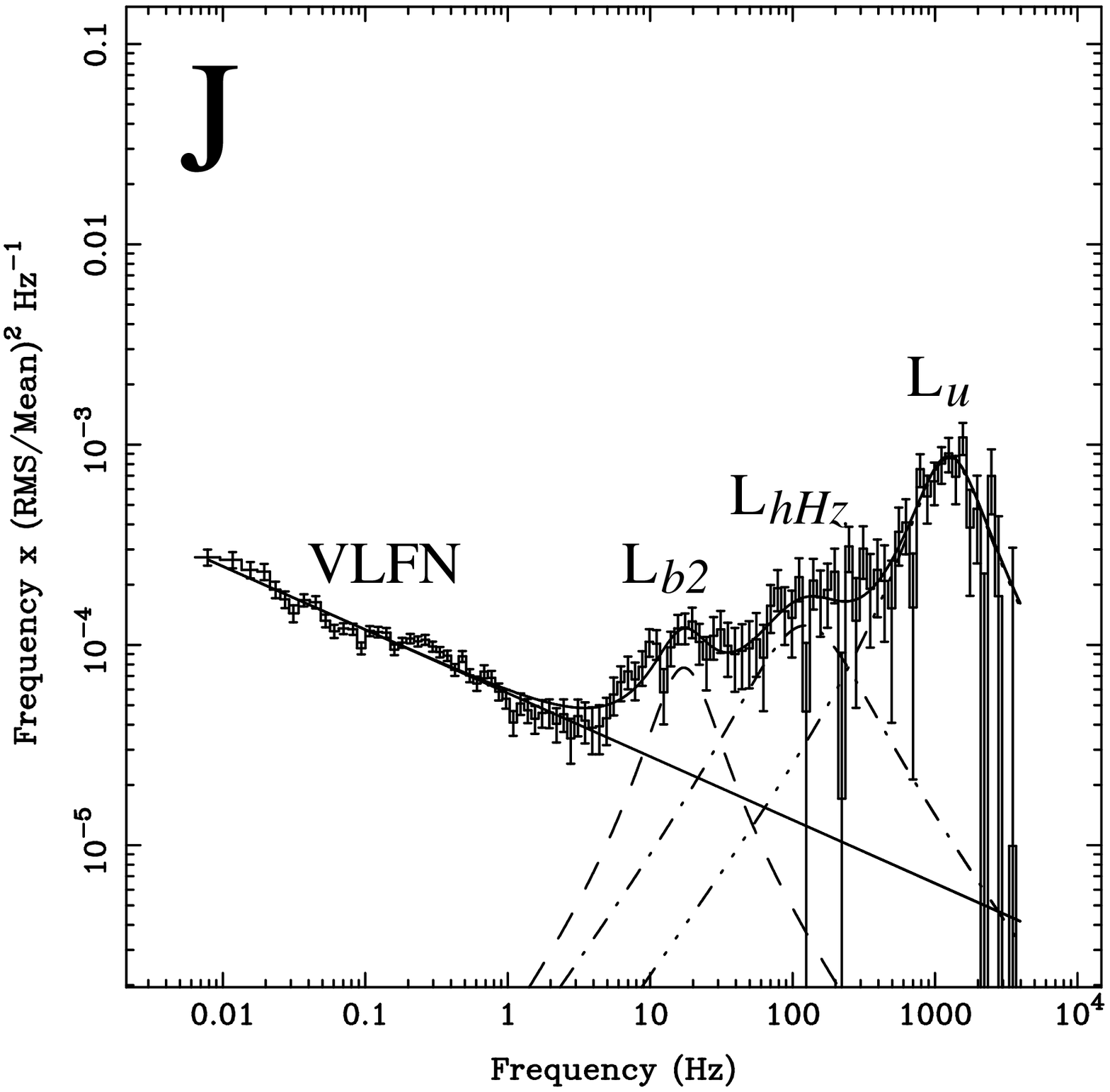} & \\
\end{tabular}
\caption{Continued}
\end{figure}
\clearpage

\begin{figure}
\figurenum{8}
\epsscale{0.8}
\plotone{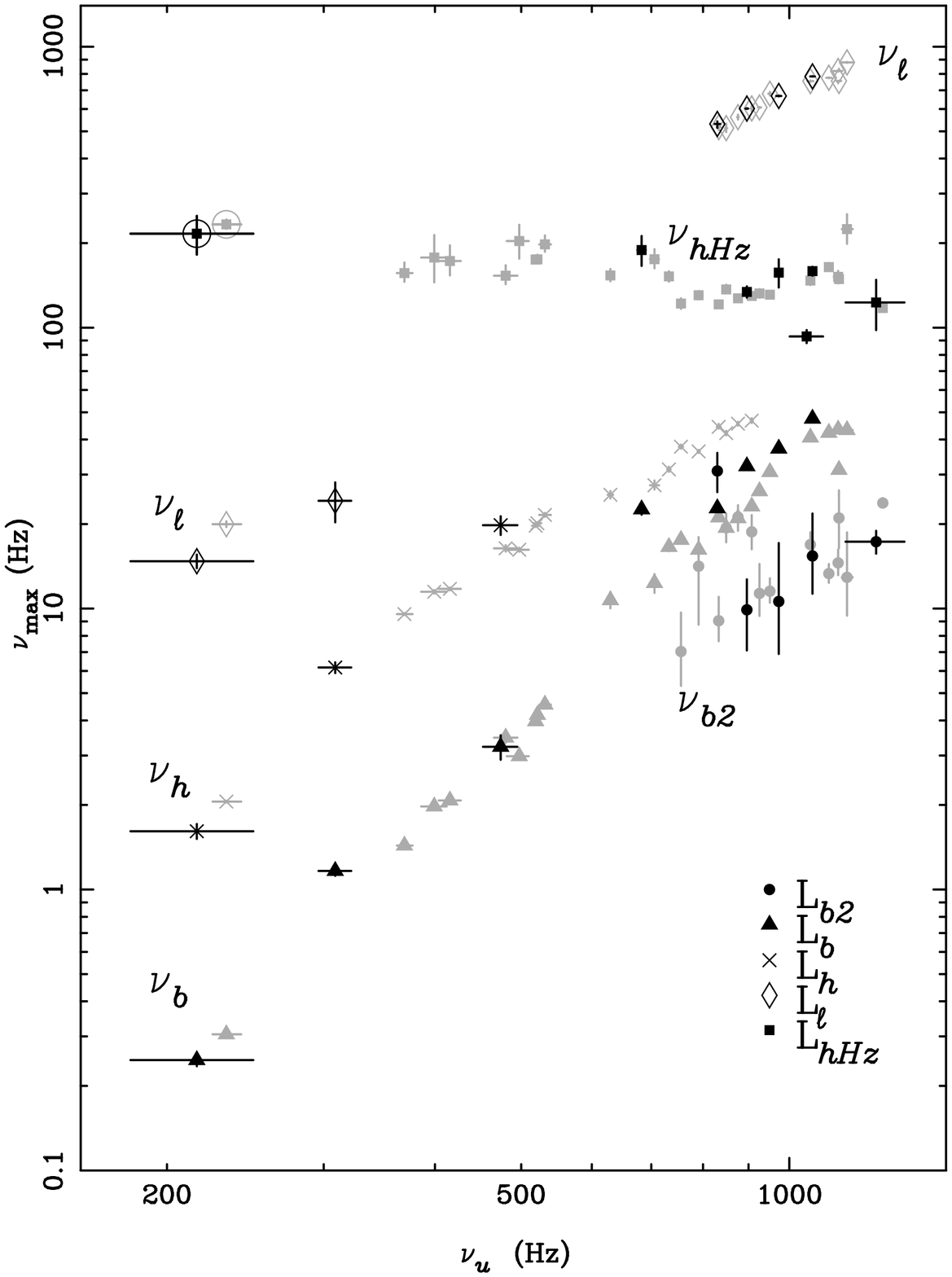}
\caption{\scriptsize
Correlations between the characteristic frequencies 
($\equiv \nu_{\rm max}$) of the several Lorentzians used to 
fit the power spectra of 4U 1608--52 and the 
$\nu_{\rm max}$ of the Lorentzian identified as the upper 
kilohertz QPO, plotted together with the results for 4U 1728--34 and 4U 0614+09 
of \citet{vstr02}. The black points 
mark the results for 4U 1608--52, the grey points the results for 4U 1728--34 and 
4U 0614+09. 
The solid dots mark L$_{b2}$, the triangles L$_{b}$, the 
crosses L$_h$, the squares L$_{hHz}$ and the diamonds L$_\ell$.
The two points with the big circles on the very left 
are from 4U 0614+09 and interval A of 4U 1608--52, for which the fourth Lorentzian 
can be identified as either the upper kilohertz QPO or the 
hectohertz Lorentzian based on its frequency (but probably is the upper kilohertz QPO; 
see \S \ref{sec.detailed_ps}). We use the parameters of 
this Lorentzian both for the upper kilohertz QPO and for the hectohertz 
Lorentzian.}
\label{fig.freq_freq}
\end{figure}

\begin{figure}
\figurenum{9}
\epsscale{0.75}
\plotone{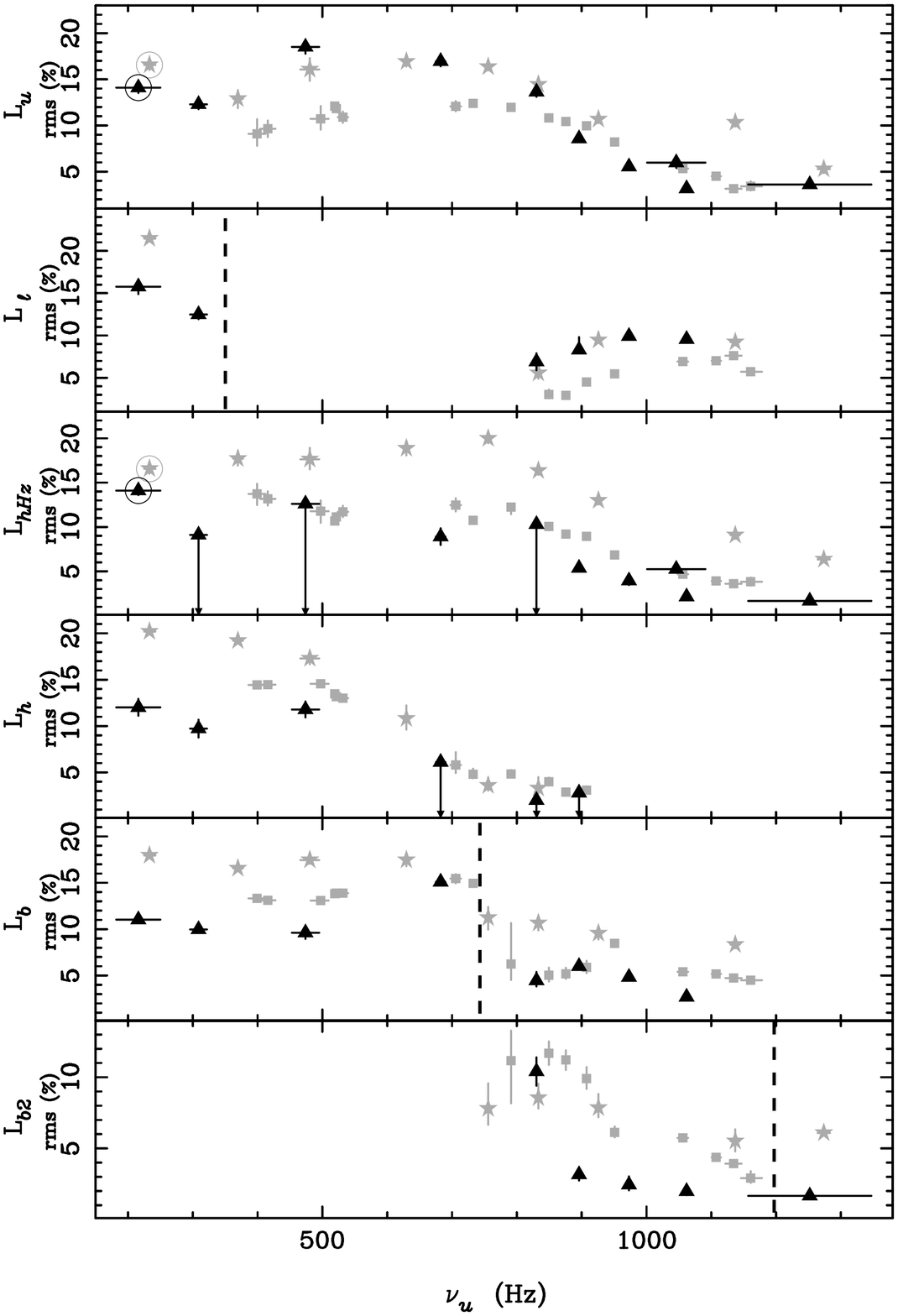}
\caption{\scriptsize The fractional rms amplitude of all components plotted versus the 
$\nu_{\rm max}$ of the upper kilohertz QPO. The black triangles mark our results for 4U 1608--52, the 
grey stars the results from 4U 0614+09, and the grey squares the results from 4U 1728--34 
\citep{vstr02}. Arrows indicate 95 \% confidence upper limits (see \S \ref{sec.comb_ps}).
The two points with the big circles on the very left 
are from 4U 0614+09 and interval A of 4U 1608--52, for which the fourth Lorentzian 
can be identified as either the upper kilohertz QPO or the 
hectohertz Lorentzian based on its frequency (but see \S \ref{sec.detailed_ps}). We use the parameters of 
this Lorentzian both for the upper kilohertz QPO and for the hectohertz 
Lorentzian.
The dashed line in the second panel marks the point below which 
L$_\ell$ is a broad low--frequency Lorentzian (see \S \ref{sec.comb_ps}).
The dashed line in the fifth panel marks the ``transformation'' 
of L$_{b}$ from band--limited noise component into a QPO and the simultaneous appearance of L$_{b2}$
(see \S \ref{sec.comb_ps}). 
The dashed line in the sixth panel marks the point above which 
a broad Lorentzian is present for which it is unclear 
whether it is L$_{b}$, L$_{b2}$ or a new component (see \S \ref{sec.detailed_ps}).
}
\label{fig.rms_all}
\end{figure}

\begin{figure}
\figurenum{10}
\epsscale{0.7}
\plotone{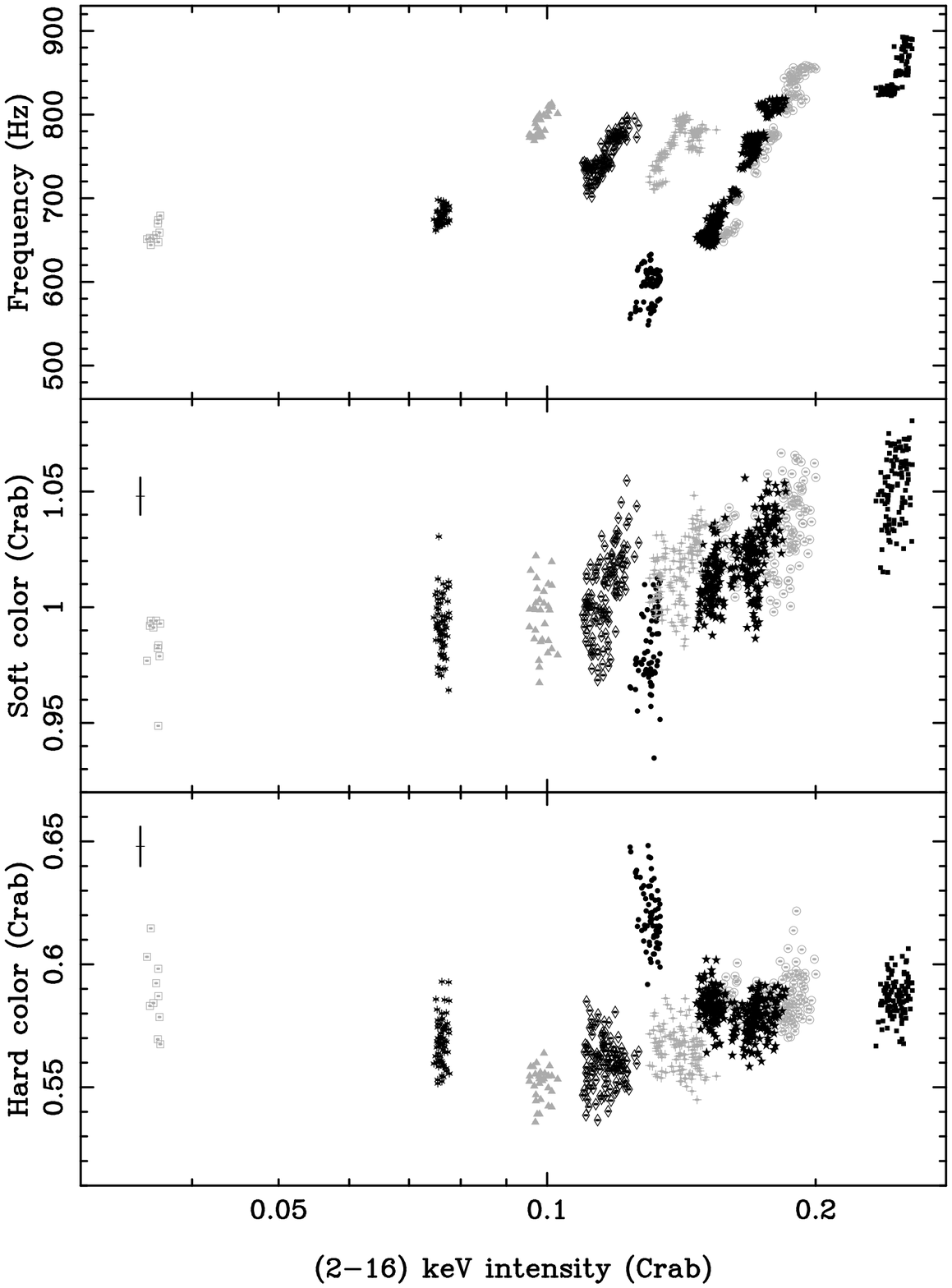}
\caption{The frequency of the lower kilohertz QPO (top panel), hard color (bottom panel) and soft color 
(middle panel) plotted versus 
2.0--16.0 keV intensity. Intensity and colors are in units of Crab (see \S \ref{sec.obs}).
The kilohertz QPO frequencies are from \citet{mendez01}. Similarly to what  
was done in \citet{mendez01} we only include data for which both kilohertz QPOs are detected 
simultaneously and therefore  the lower kilohertz QPO is identified unambiguously. 
The symbols and the alternating black/grey represent the parallel tracks in the intensity versus  
frequency diagram that are continuous in time and only have gaps due to earth occultations of 
the source ($\sim 2500$ s). Typical soft and hard color error bars are shown in the middle and bottom panel.
The errors in the intensity are smaller then the size of the symbols.}
\label{fig.mendez_rate}
\end{figure}

\begin{figure}
\figurenum{11}
\epsscale{0.8}
\plotone{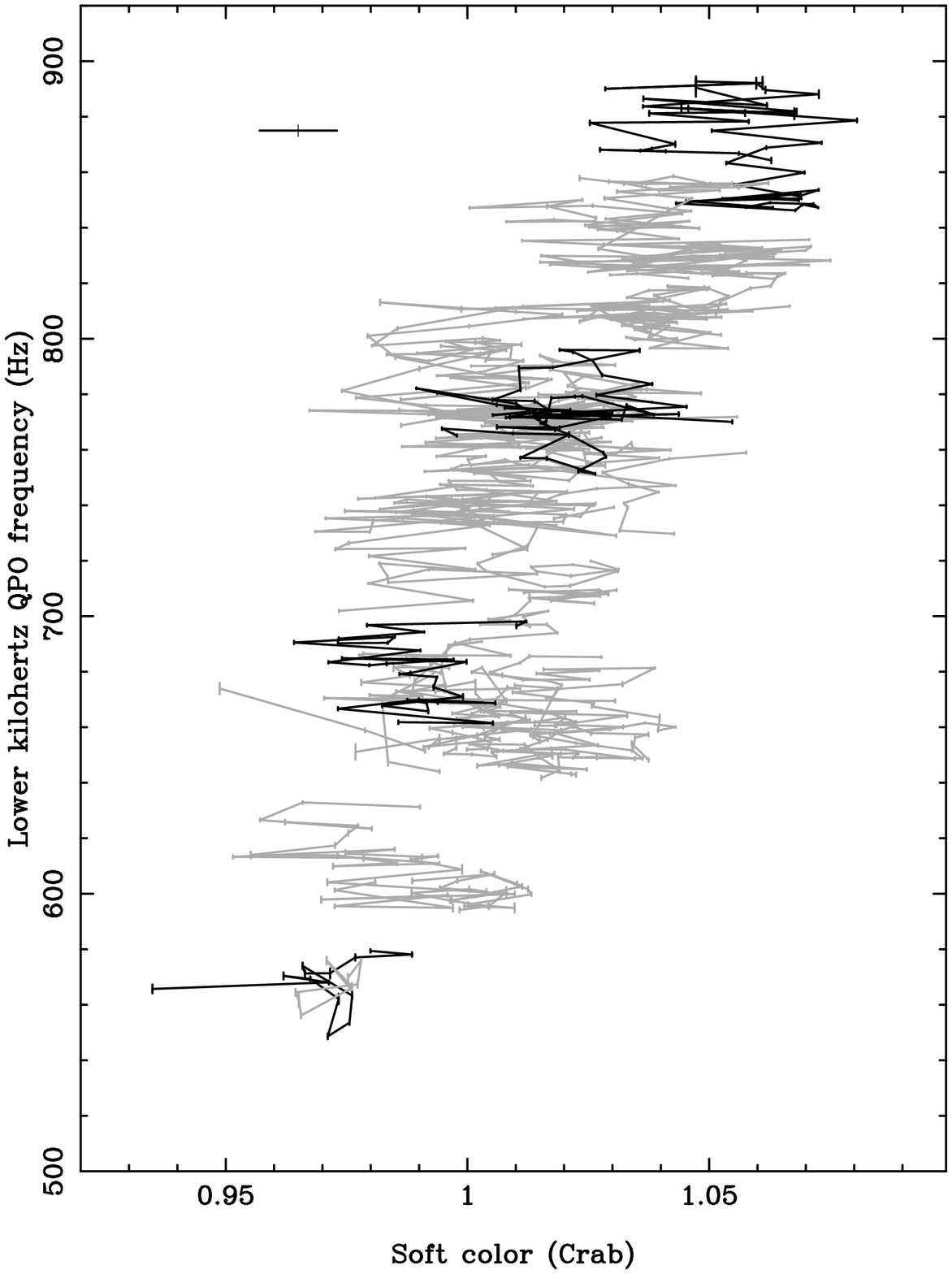}
\caption{The frequency of the lower kilohertz QPO plotted versus soft color. The kilohertz QPO frequencies are 
from \citet{mendez01}. Lines connect data from each individual satellite orbit. For clarity four representative 
individual orbits are highlighted in black. Soft color is in units of Crab (see \S \ref{sec.obs}). 
A typical soft color error is shown.}
\label{fig.mendez_sc}
\end{figure}

\begin{figure}
\figurenum{12}
\epsscale{0.8}
\plotone{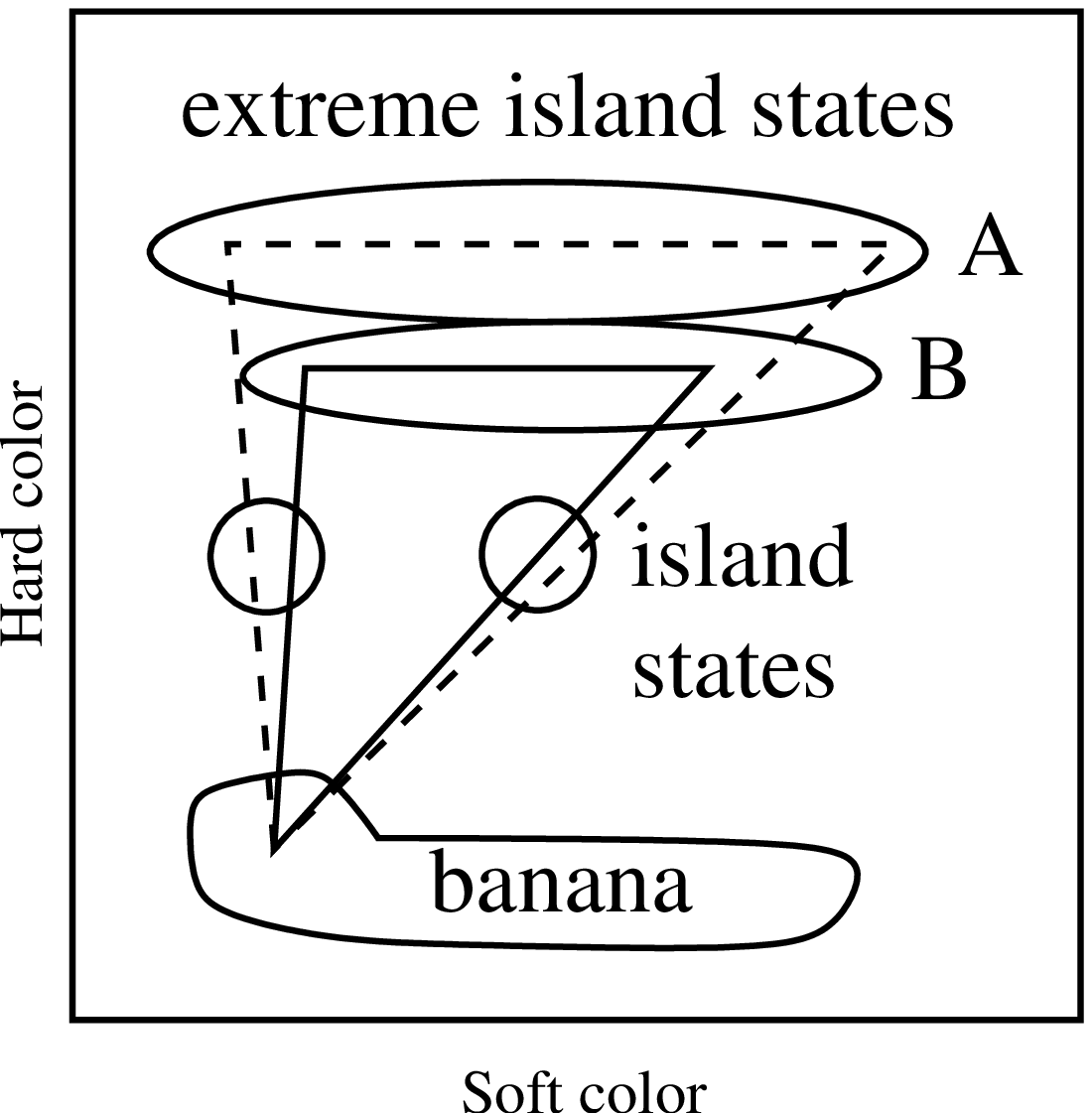}
\caption{A schematic view of the triangular shaped transition between the banana and the extreme island states.}
\label{fig.schema}
\end{figure}

\begin{figure}
\figurenum{13}
\epsscale{0.8}
\plotone{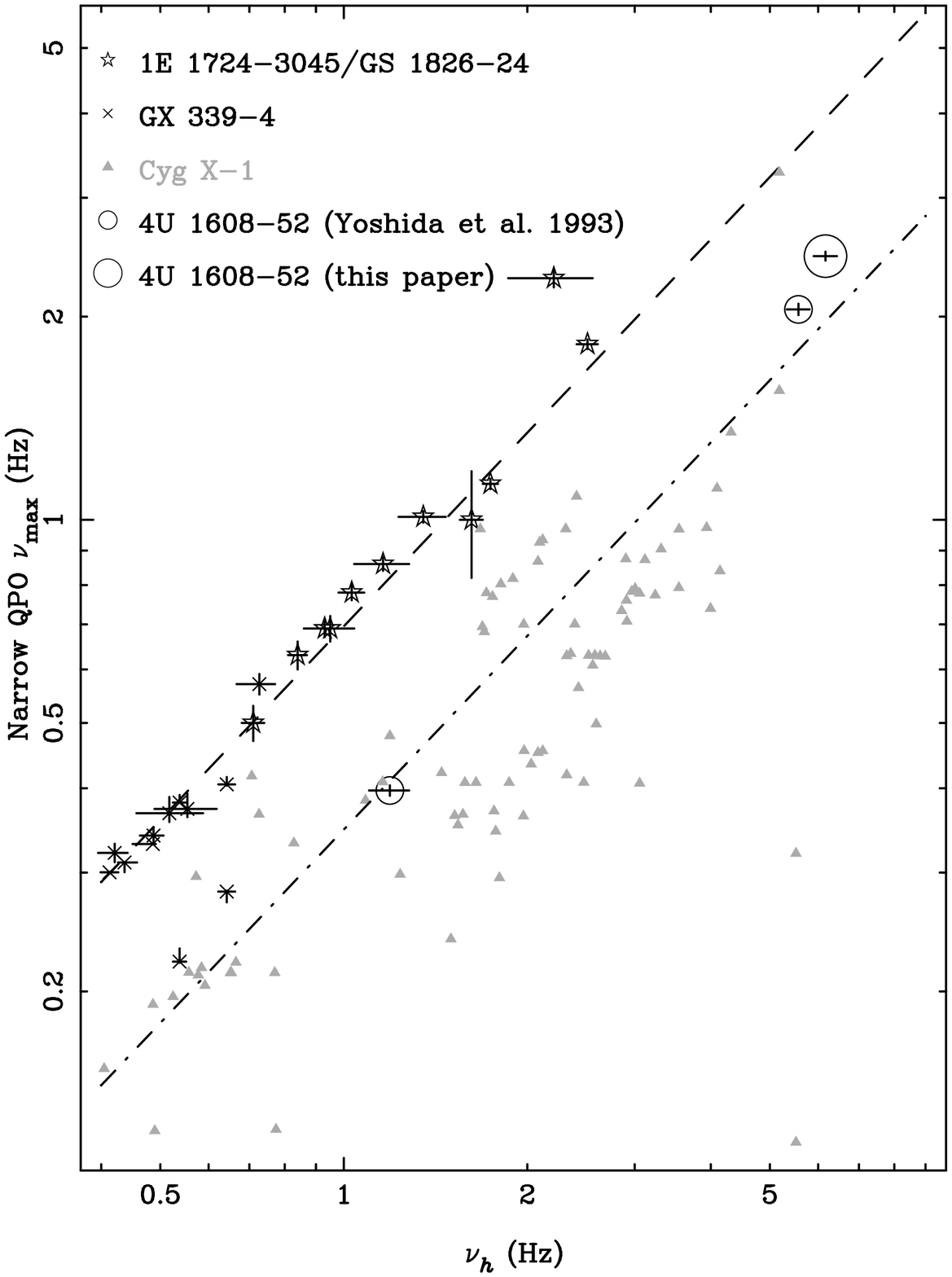}
\caption{Characteristic frequency of L$_{h}$ versus that of the narrow low--frequency QPO. The dashed line 
indicates a power law fit to the low--luminosity bursters 1E 1724--3045 and GS 1826--24, and the BHC GX 339--4. 
The dash--dotted line is a power law with a normalization half of that of the dashed line.}
\label{fig.narrowqpo}
\end{figure}

\begin{figure}
\figurenum{14}
\epsscale{0.8}
\plotone{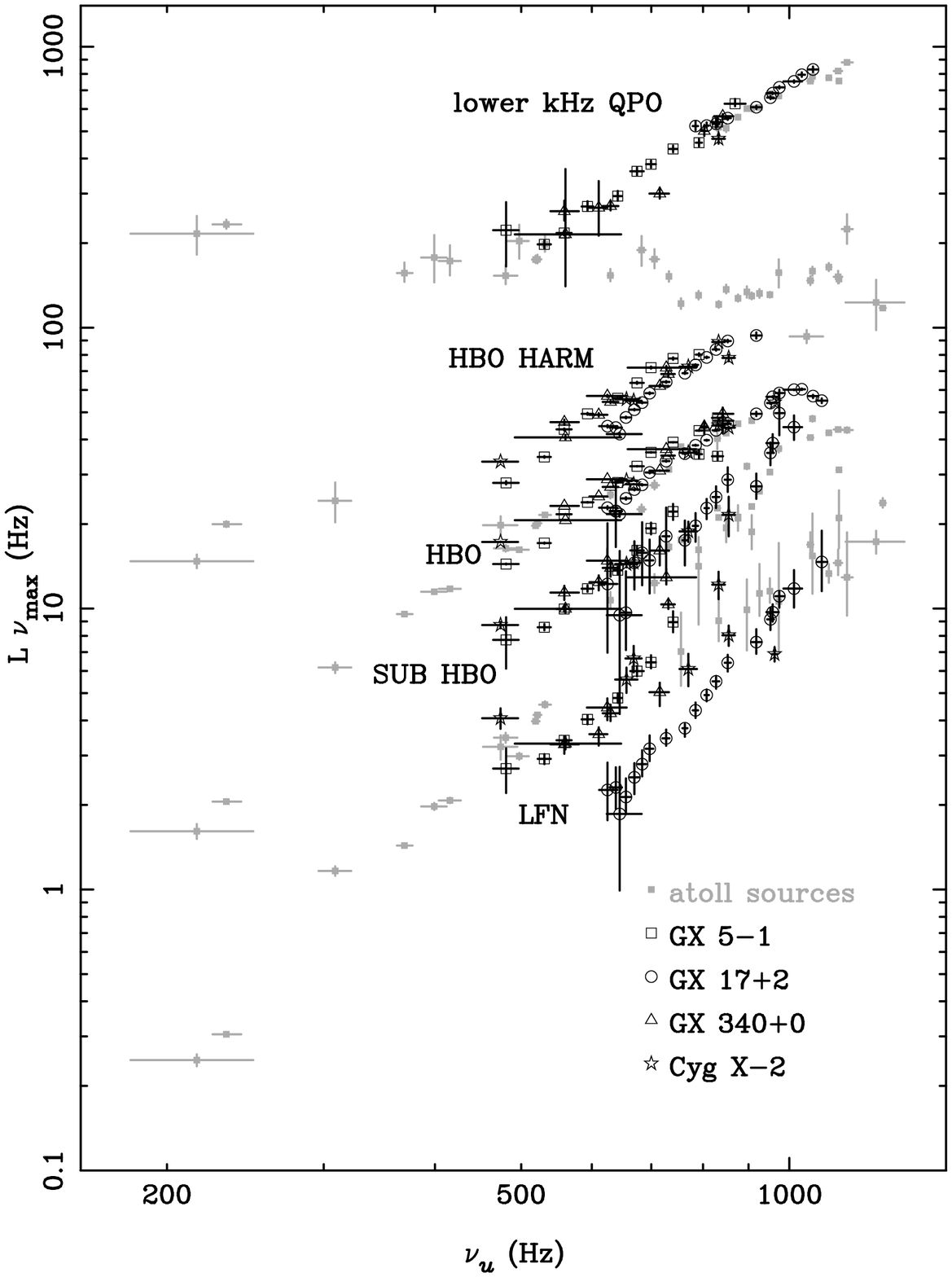}
\caption{Characteristic frequencies (see \S \ref{sec.disc} for definitions) of the different power spectral components 
versus $\nu_{\rm max}$
of the upper kilohertz QPO for atoll and Z sources. The grey symbols are the results of the multi--Lorentzian 
fit to the atoll sources 4U 1608--52, 4U 0614+09, and 4U 1728--34 of Figure \ref{fig.freq_freq}.
The black symbols represent the Z sources; GX 5-1 \citep{jonker02}, 
GX 340+0 \citep{jonker00}, GX 17+2 \citep{homan02}, and Cyg X--2 \citep{kuz02}. For the Z sources we include, 
as indicated in the plot, the lower kilohertz QPO, the Low--Frequency Noise (LFN), the Horizontal Branch 
Oscillations (HBO), and the harmonic and sub--harmonic of the HBO.
}
\label{fig.atoll_z}
\end{figure}

\begin{figure}
\figurenum{15}
\epsscale{0.8}
\plotone{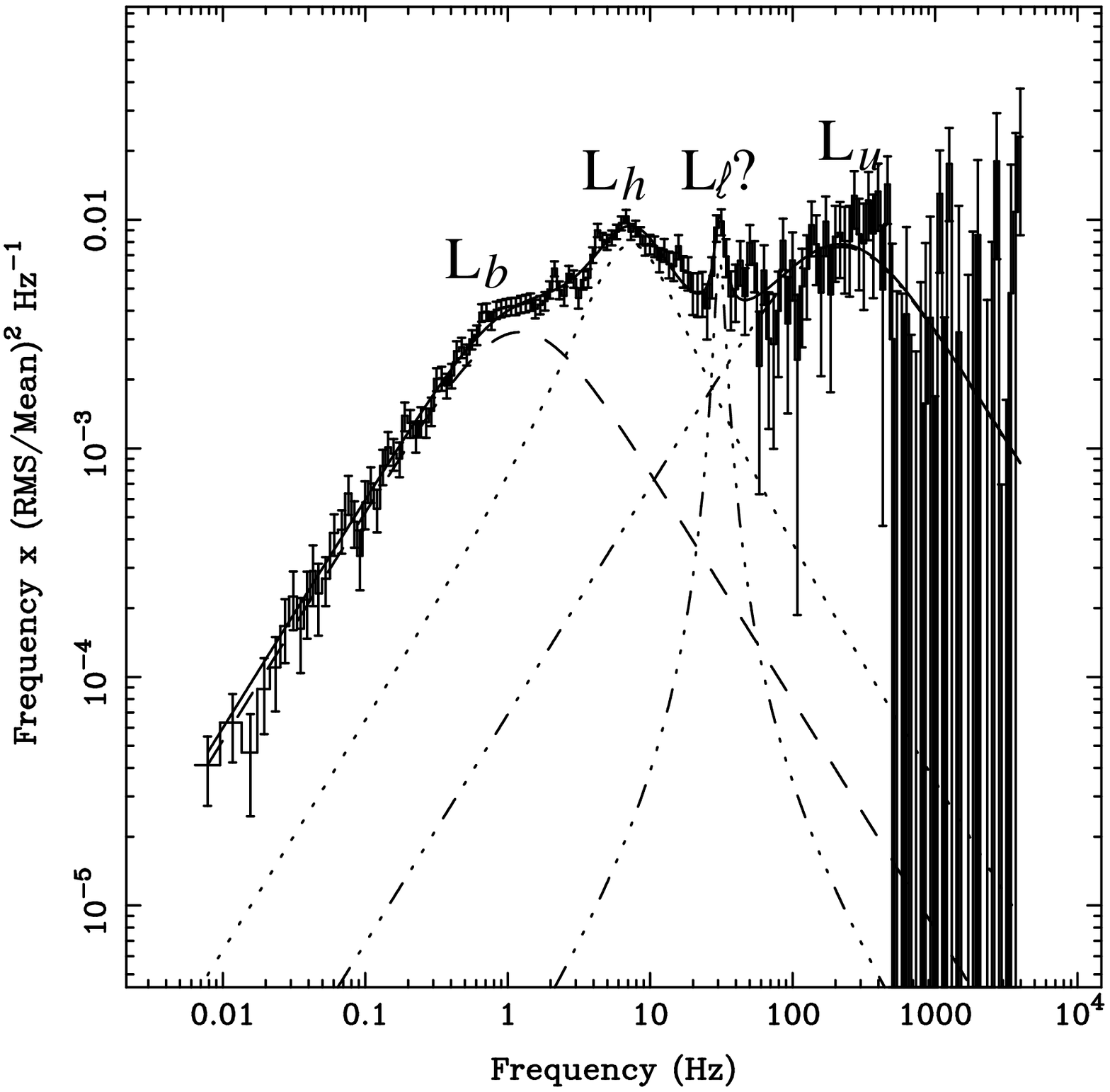}
\caption{Power spectrum of 2001 November 23 in the power spectral density times frequency representation (see \S 2) for 4U 1608--52.
This power spectrum is very similar to that of interval B (see Figure \ref{fig.powspec_1608}), except that 
instead of the broad peak at $\sim 25$ Hz, a narrow ($Q = 5.5$) QPO appears at 30 Hz.}
\label{fig.nupnu_2001}
\end{figure}

\begin{figure}
\figurenum{16}
\epsscale{0.8}
\plotone{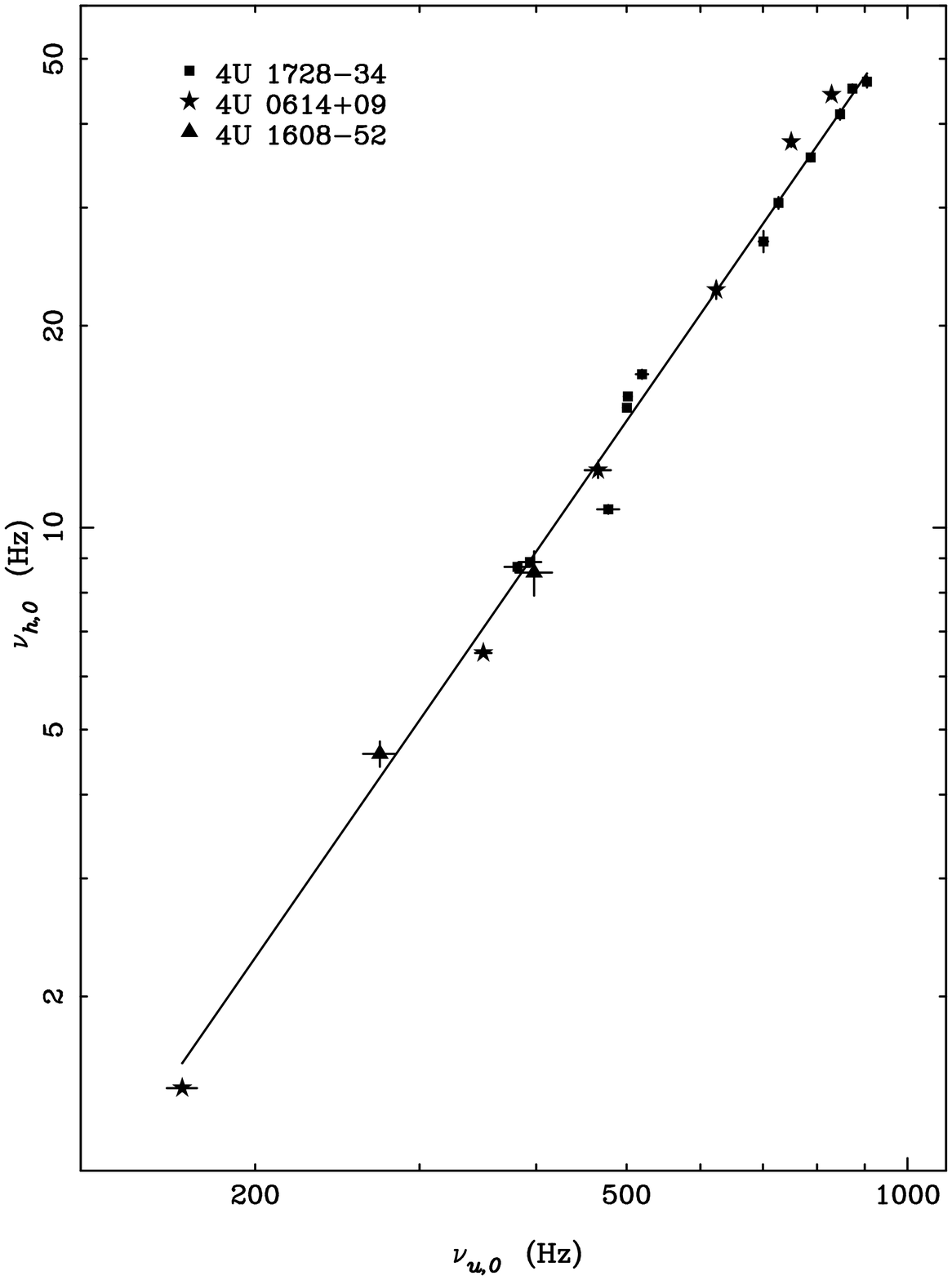}
\caption{The centroid frequency of L$_h$ versus the centroid frequency of L$_u$. 
The function shown is a power law with index 2.01.}
\label{fig.powerlawfit}
\end{figure}


\begin{thebibliography}{}

\bibitem[{Barret \& Olive}(2002)]{barret02}
	Barret, D., \& Olive, J., 2002, ApJ, 576, 391

\bibitem[{Belloni et al.}(1997)]{bel97}
        Belloni, T., van der Klis, M., Lewin, W. H. G. van Paradijs, J.,
        Dotani, T., Mitsuda, K., Miyamoto, S., 1997, A\&A, 322, 857

\bibitem[{Belloni et al.}(2002){Belloni, Psaltis \& van der Klis}]{bpk02} 
        Belloni, T., Psaltis, D., van der Klis, M., 2002, ApJ, 572, 392

\bibitem[{Berger et al.}(1996)]{berg96}
	Berger, M., et al., 1996, ApJ, 469, L13

\bibitem[{Berger \& van der Klis}(1998)]{berg98}
	Berger, M., \& van der Klis, M., 1998, A\&A, 340, 143

\bibitem[{Chakrabarty et al.}(2003)]{chakrabarty03}
	Chakrabarty, D., Morgan, E. H., Muno, M. P., Galloway, D. K., 
	Wijnands, R. A. D., van der Klis, M., Markwardt, C. B., 2003, in preparation

\bibitem[{Christian \& Swank}(1997)]{christian97}
	Christian, D. J., \& Swank, J.H., 1997, ApJS, 109, 177

\bibitem[{Crary et al.}(1996)]{crary96}
	Crary, D. J., et al. 1996, ApJ, 462, L71

\bibitem[{Di Salvo et al.}(2001)]{disalvo01}
        Di Salvo, T., M\'endez, M., van der Klis, M., Ford, E.,
        Robba, N.R., 2001, ApJ, 546, 1107 

\bibitem[{Di Salvo et al.}(2003){Di Salvo, M\'endez \& van der Klis}]{disalvo03}
        Di Salvo, T., M\'endez, M., van der Klis, M., 2003, A\&A, in press (astro-ph/0304090)

\bibitem[{Ford et al.}(1997)]{ford97}
	Ford, E. C., Kaaret, P., Tavani, M., Barret, D.,
 	Bloser, P., Grindlay, J., Harmon, B.A.,
 	Paciesas, W. S., Zhang, S.N., 1997, ApJ, 475, L123

\bibitem[{Ford et al.}(1998){Ford, van der Klis \& Kaaret}]{fordetal98}
	Ford, E. C., van der Klis, M., \& Kaaret, P. 1998, ApJ, 498, L41

\bibitem[{Ford \& van der Klis}(1998)]{ford98}
	Ford, E. C., \& van der Klis, M., 1998, ApJ, 506, L39

\bibitem[{Hasinger et al.}(1990)]{hasinger90}
	Hasinger, G., van der Klis, M., Ebisawa, K., 
	Dotani, T., Mitsuda, K., 1990, A\&A, 235, 131

\bibitem[{Hasinger \& van der Klis}(1989)]{hk89}
        Hasinger, G., \& van der Klis, M., 1989, A\&A, 225, 79

\bibitem[{Homan et al.}(2002)]{homan02}
	Homan, J., van der Klis, M., Jonker, P. G., Wijnands, R., 
	Kuulkers, E., \mm, M., Lewin, W. H. G., 2002, ApJ, 568, 878

\bibitem[{Gierlinski \& Done}(2002a)]{gier02a}
	Gierlinski, M., \& Done, C., MNRAS, 2002a, 331, L47

\bibitem[{Gierlinski \& Done}(2002b)]{gier02b}
	Gierlinski, M., \& Done, C., MNRAS, 2002b, 337, 1373

\bibitem[{Grove et al.}(1994)]{grove94}
	Grove, J., et al., AIP Conference Proc., 1994, 304, p192 

\bibitem[{Jonker et al.}(2000)]{jonker00}
	Jonker, P. G., van der Klis, M., Wijnands, R., Homan, J., 
	van Paradijs., J., \mm, M., Ford, E.C., Kuulkers, E., 
	Lamb, F. K., 2000, ApJ, 537, 374

\bibitem[{Jonker et al.}(2002)]{jonker02}
	Jonker, P. G., van der Klis, M., Homan, J., \mm, M., 
	Lewin, W. H. G., Wijnands, R., Zhang, W., 2002, MNRAS, 333, 665

\bibitem[{Kuulkers et al.}(1994)]{kuulkers94}
	Kuulkers, E., van der Klis, M., Oosterbroek, T.,
 	Asai, K., Dotani, T., van Paradijs, J.,
 	Lewin, W. H. G., 1994, A\&A, 289, 795

\bibitem[{Kuznetsov}(2002)]{kuz02}
	Kuznetsov, S. I., 2002, AstL, 28, 73

\bibitem[{Lamb \& Miller}(2001)]{lamb01}
	Lamb, F. K., \& Miller, M. C., 2001, ApJ, 554, 1210

\bibitem[{Langmeier et al.}(1989){Langmeier, Hasinger \& Tr\"umper}]{lang89}
	Langmeier, A., Hasinger, G., \& Tr\"umper, J. 1989, ApJ, 340, L21

\bibitem[{Leahy et al.}(1983)]{lea83}
        Leahy, D. A., Darbro, W., Elsner, R. F.,
        Weisskopf, M. C., Kahn, S.,
        Sutherland, P. G., Grindlay, J. E., 1983, ApJ, 266, 160

\bibitem[{Lochner \& Roussel--Dupr\'{e}}(1994)]{loch94}
	Lochner, J. C., \& Roussel--Dupr\'{e}, D., 1994, ApJ, 435, 840

\bibitem[{Mauche}(2002)]{mauche02}
	Mauche, C. W., 2002, ApJ, 580, 423

\bibitem[{\mm et al.}(1997)]{mendez97} 
	\mm, M., van der Klis, M., van Paradijs, J., Lewin, W. H. G., 
	Lamb, F. K., Vaughan, B. A., Kuulkers, E., Psaltis, D. 
	1997, ApJ, 485, L37

\bibitem[{\mm et al.}(1998)]{mendez98} 
	\mm, M., van der Klis, M., Wijnands, R. A. D., Ford, E. C., 
	van Paradijs, J., Vaughan, B.A., 1998, ApJ, 505, 23

\bibitem[{\mm et al.}(1999)]{mendez99} 
	\mm, M., van der Klis, M., Ford, E.C., 
	Wijnands, R., van Paradijs, J., 1999, ApJ, 511, L49 

\bibitem[{\mm \& van der Klis}(1999)]{mendezvdk99}
	\mm, M., van der Klis, M., 1999, ApJ, 517, L51

\bibitem[{M\'endez}(2000)]{mendez00} 
	\mm, M., 2000, Proc. 19th Texas Symposium on Relativistic 
	Astrophysics and Cosmology, ed. J. Paul, T. Montmerle,
     	\& E. Aubourg (Amsterdam: Elsevier), 15 

\bibitem[{M\'endez et al.}(2001){M\'endez, van der Klis \& Ford}]{mendez01} 
	\mm, M., van der Klis, M., 
	Ford, E.C., 2001, ApJ, 561, 1016 

\bibitem[{Miller et al.}(1998){Miller, Lamb \& Psaltis}]{miller98}
	Miller, M. C., Lamb, F. K., Psaltis, D.,
	1998, ApJ, 508, 791

\bibitem[{Mitsuda et al.}(1989)]{mitsuda89}
	Mitsuda, K. H., Inoue, H., Nakamura, N., 
	Tanaka, Y., 1989, PASJ, 41, 97 

\bibitem[{Muno et al.}(2000)]{muno00} 
	Muno, M. P., Fox, D. W., Morgan, E. H., Bildsten, L. 2000, ApJ, 542, 1016

\bibitem[{Muno et al.}(2002a){Muno, Remillard \& Chakrabarty}]{muno02a}
	Muno, M. P., Remillard, R. A., Chakrabarty, D., 2002a, ApJ, 568, L35

\bibitem[{Muno et al.}(2002b)]{muno02b}
	Muno, M. P., Chakrabarty, D., Galloway, D. K., Psaltis, D., 2002b, ApJ, 580, 1048

\bibitem[{Nowak}(2000)]{n00}
        Nowak, M. A., 2000, MNRAS, 318, 361

\bibitem[{Nowak et al.}(2002){Nowak, Wilms \& Dove}]{n02}
	Nowak, M. A., Wilms, J., Dove, J. B., 2002, MNRAS, 332, 856

\bibitem[{Olive et al.}(1998)]{olive98}
	Olive, J., Barret, D., Boirin, L., Grindlay, J., Swank, J., 
	Smale, A., 1998, A\&A, 333, 942 

\bibitem[{Olive et al.}(2003){Olive, Barret \& Gierlinski}]{olive03}
	Olive, J., Barret, D., \& Gierlinski, M., 2003, ApJ, 583, 416

\bibitem[{Osherovich \& Titarchuk}(1999)]{ot99}
	Osherovich, V., \& Titarchuk, L. 1999, ApJ, 522, L113 

\bibitem[{Pottschmidt et al.}(2002)]{pott02}
	Pottschmidt, K., Wilms, J., Nowak, M.A., Pooley, G.G., Gleissner, T., 
	Heindl, W.A., Smith, D.M., Remillard, R., Staubert, R., 2002, A\&A, submitted (astro-ph/0202258)

\bibitem[{Prins \& van der Klis}(1997)]{prins97}
	Prins, S., van der Klis, M., 1997, A\&A, 319, 498

\bibitem[{Psaltis et al.}(1999){Psaltis, Belloni \& van der Klis}]{pbk99}
        Psaltis, D., Belloni, T., van der Klis, M., 1999, ApJ, 520, 262 

\bibitem[{Reig et al.}(2000)]{reig00}
	Reig, P., \mm, M., van der Klis, M., Ford, E.C., 2000, ApJ, 530, 916


\bibitem[{Reig et al.}(2003){Reig, van Straaten \& van der Klis}]{reig03}
	Reig, P., van Straaten, S., van der Klis, M.,
	2003, in preparation

\bibitem[{Stella \& Vietri}(1998)]{stella98}
	Stella, L., \& Vietri, M., 1998, ApJL, 492, L59

\bibitem[{Stella \& Vietri}(1999)]{stella99}
	Stella, L., \& Vietri, M., 1999, PhRvL, 82, L17

\bibitem[{Strohmayer et al.}(1996)]{stroh96}
	Strohmayer, T. E., Zhang, W., Swank, J. H., Smale, A., 
	Titarchuk, L., Day, C., Lee, U. 1996, ApJ, 469, L9

\bibitem[{Titarchuk \& Osherovich}(1999)]{to99}
        Titarchuk, L., \& Osherovich, V., 1999, ApJ, 518, L95

\bibitem[{Titarchuk, Bradshaw \& Wood}(2001)]{tit_brad_wood01}
        Titarchuk, L., Bradshaw, C. F., \& Wood, K., 2001, ApJ, 560, L55

\bibitem[{Titarchuk \& Wood}(2002)]{tit_wood02}
        Titarchuk, L., \& Wood, K., 2002, ApJ, 577, L23

\bibitem[{Titarchuk}(2002)]{titarchuk02}
        Titarchuk, L., 2002, ApJ, 578, L71

\bibitem[{van der Klis}(1994a)]{vdk94a}
	van der Klis, M., 1994a, ApJS, 92, 511

\bibitem[{van der Klis}(1994b)]{vdk94b}
	van der Klis, M., 1994b, A\&A, 283, 469

\bibitem[{van der Klis}(2000)]{vdk00}
	van der Klis, M., 2000, ARA\&A, 38, 717

\bibitem[{van der Klis}(2001)]{vdk01}
	van der Klis, M., 2001, ApJ, 561, 943

\bibitem[{van Paradijs et al.}(1996)]{paradijs96}
	van Paradijs, J., et al. 1996, IAU Circ., 6336 

\bibitem[{van Straaten et al.}(2000)]{vstr00}
        van Straaten, S., Ford, E. C., van der Klis, M., M\'endez, M., Kaaret, P. 
        2000, ApJ, 540, 1049

\bibitem[{van Straaten et al.}(2001)]{vstr01}
        van Straaten, S., van der Klis, M., Kuulkers, E., M\'endez, M., 
        2001, ApJ, 551, 907

\bibitem[{van Straaten et al.}(2002)]{vstr02} 
	van Straaten, S., van der Klis, M., Di Salvo, T., 
	Belloni, T., 2002, ApJ, 568, 912

\bibitem[{Warner \& Woudt}(2002)]{warner02}
	Warner, B., \& Woudt, P.A., 2002, MNRAS, 335, 84

\bibitem[{Warner \& Woudt}(2003)]{warner03}
	Warner, B., \& Woudt, P. A., 2003, to appear in the proceedings of 
	``Magnetic Cataclysmic Variables'', eds. M. Cropper \& S. Vrielmann (astro-ph/0301168)

\bibitem[{Wijnands et al.}(1996)]{wijnands96}
	Wijnands, R. A. D., van der Klis, M.,
 	Psaltis, D., Lamb, F. K., Kuulkers, E.,
 	Dieters, S., van Paradijs, J., Lewin, W. H. G.,  1996, ApJ, 469, L5

\bibitem[{Wijnands et al.}(1997)]{wijnands97}
	Wijnands, R., et al., 1997, ApJ, 490, L157

\bibitem[{Wijnands \& van der Klis}(1998)]{wk98}
        Wijnands, R., van der Klis, M., 1998, ApJ, 507, L63

\bibitem[{Wijnands et al.}(1998a)]{wijnands98a}
	Wijnands, R., et al., 1998a, ApJ, 493, L87

\bibitem[{Wijnands et al.}(1998b)]{wijnands98b}
	Wijnands, R., \mm, M.,
 	van der Klis, M., Psaltis, D.,
 	Kuulkers, E., Lamb, F. K., 1998b, ApJ, 504, L35

\bibitem[{Wijnands et al.}(1998c)]{wijnands98c}
	Wijnands, R. A. D., van der Klis, M., \mm, M., 
	van Paradijs, J., Lewin, W. H. G., Lamb, F. K., Vaughan, B.,
     	Kuulkers, E., 1998b, ApJ, 495, L39

\bibitem[{Wijnands \& van der Klis}(1999)]{wk99}
        Wijnands, R., van der Klis, M., 1999, ApJ, 514, 939

\bibitem[{Yoshida et al.}(1993)]{yoshida93}
	Yoshida, K., Mitsuda, K,,
 	Ebisawa, K., Ueda, Y.,
 	Fujimoto, R., Yaqoob, T., Done, C., 1993, PASJ, 45, 605

\bibitem[{Yu et al.}(1997)]{yu97}
	Yu, W., Zhang, S.N., Harmon, B. A.,
 	Paciesas, W. S., Robinson, C. R., Grindlay, J. E.,
 	Bloser, P., Barret, D., Ford, E. C., Tavani, M.,
 	Kaaret, P., 1997, ApJ, 490, L153

\bibitem[{Zhang et al.}(1993)]{zhang93}
        Zhang, W., Giles, A. B., Jahoda, K., Soong, Y., Swank, J. H., 
        Morgan, E. H., 1993, SPIE, 2006, 324 

\end{thebibliography}
\end{document}